\documentclass{ieeeaccess}
\usepackage{kotex}
\usepackage{cite}
\usepackage{amsmath,amssymb,amsfonts}
\usepackage[bookmarks=false]{hyperref}
\usepackage{multirow}
\usepackage{array}
\usepackage{tabularx}
\usepackage{afterpage}
\usepackage[ruled,vlined]{algorithm2e}
\usepackage{graphicx}
\usepackage{textcomp}
\usepackage{stfloats}
\usepackage{stackengine}

\usepackage{bm}
\makeatletter
\AtBeginDocument{\DeclareMathVersion{bold}
\SetSymbolFont{operators}{bold}{T1}{times}{b}{n}
\SetSymbolFont{NewLetters}{bold}{T1}{times}{b}{it}
\SetMathAlphabet{\mathrm}{bold}{T1}{times}{b}{n}
\SetMathAlphabet{\mathit}{bold}{T1}{times}{b}{it}
\SetMathAlphabet{\mathbf}{bold}{T1}{times}{b}{n}
\SetMathAlphabet{\mathtt}{bold}{OT1}{pcr}{b}{n}
\SetSymbolFont{symbols}{bold}{OMS}{cmsy}{b}{n}
\renewcommand\boldmath{\@nomath\boldmath\mathversion{bold}}}
\makeatother

\def\BibTeX{{\rm B\kern-.05em{\sc i\kern-.025em b}\kern-.08em
    T\kern-.1667em\lower.7ex\hbox{E}\kern-.125emX}}

\begin{document}
\history{Date of publication xxxx 00, 0000, date of current version xxxx 00, 0000.}
\doi{Not Assigned}

\title{Semantic Networks as Clues: A Theoretical Foundation and Process Optimization for Semantic Network Construction}
\author{\uppercase{JinWoo Ha}\authorrefmark{1} AND
\uppercase{Dongsoo Kim}\authorrefmark{2}}

\address[1]{Department of Industrial and Information Systems Engineering, Soongsil University, Seoul 06978, Republic of Korea (e-mail: realfriend@soongsil.ac.kr)}
\address[2]{Department of Industrial and Information Systems Engineering, Soongsil University, Seoul 06978, Republic of Korea (e-mail: dskim@ssu.ac.kr)}
\tfootnote{This research was supported by the G--LAMP Program of the National Research Foundation of Korea (NRF) grant funded by the Ministry of Education (No. RS-2025-25441317).}

\markboth
{J. Ha \headeretal: Semantic Networks as Clues: A Theoretical Foundation and Process Optimization for Semantic Network Construction}
{J. Ha \headeretal: Semantic Networks as Clues: A Theoretical Foundation and Process Optimization for Semantic Network Construction}

\corresp{Corresponding author: Dongsoo Kim (e-mail: dskim@ssu.ac.kr)}

\begin{abstract}
The subject matter of this paper is twofold. One is to review the theoretical foundation of a specific type of Semantic Networks (SNs) representing textual non-propositional knowledge. The other involves proposing a framework (ClueNetwork) for ranking candidate SNs generated through various Semantic Network Construction (SNC) processes for the type. In the first fold, it is clarified that the type serves as clues, not surrogates, of reality, making gold standards elusive. Then, it is discussed why this type nevertheless holds scientific legitimacy in terms of abduction. Grounded in this legitimacy, the three main stages of SNC, comprising Automatic Keyphrase Extraction (AKE), Edge Weighting (EW), and Community Detection (CD), are reviewed alongside their objectives and operations. In the second fold, evaluation criteria (comprising two established and one reformulated) for achieving the objectives are first defined and justified, followed by illustrative experiments based on the criteria. Thereafter, SNC is reformulated as a Process Optimization Problem (POP), and its global objective function that integrates the local criteria is defined and justified. Based on these, ClueNetwork is ultimately proposed.
\end{abstract}


\titlepgskip=-21pt

\maketitle

\section{Introduction}
\label{sec:introduction}

\medskip

Our current subject matter is twofold. One is to \emph{review} the foundation of Semantic Networks (SNs) as clues, and the other involves proposing a framework to \emph{optimize} SN Construction (SNC). In this work, SNC is confined to the core phase of SN Analysis (SNA) --- the Text Mining (TM) technique comprising data collection and preprocessing, SNC, and post--hoc interpretation of SN representing knowledge \cite{b1, b2}. Peers have just encountered key concepts, namely TM, SN, Knowledge Representation (KR), and a network.

\medskip

\noindent\textbf{Text Mining and Semantic Network}. Firstly,

\smallskip

\begin{itemize}
    \item TM is a range of techniques for ``extracting meaningful \emph{information}'' from textual data \cite{b1}.
    \item SN is a network representing \emph{knowledge} \cite{b3, b4}.
\end{itemize}

\smallskip

\noindent As may be noticed, a conceptual tension exists 
between these two definitions. TM extracts \emph{information}, and SNA is a TM technique. Given that information and knowledge can be defined as ``structured data,'' and ``a mix of information and some ingredient'' (e.g., understanding, experience, skills, capability, or values), respectively \cite{b5}, why is an SN said to represent \emph{knowledge} rather than information?

The answer lies in an interesting observation that while the field of TM tends to regard SNs as \emph{``clues''} \cite{b6} toward true knowledge, the field of KR sometimes regards SNs as \emph{``surrogates''} \cite{b7} of true knowledge. The aforementioned definition of SN reflects reality that discussions of SN itself have been more driven by the KR community that uses specialized SNs, also termed Knowledge Graphs (KGs) \cite{b8}, as their key instruments, rather than by the TM community, where SNA is merely one of various available tools. 

Of course, this contrast is made for convenience. We already know that many peers may belong to both communities. For now, it is sufficient to note that, within this paper, SNs are confined to \emph{clues}. Given that such clues represent knowledge in their own way (Subsection \ref{subsec:surrogates and clues} discusses this point in detail), this paper accepts the aforementioned definition of an SN only in a \emph{literal sense}, not as a surrogate.

\medskip

\noindent\textbf{Knowledge Representation}. Likewise, within this paper, the notion of KR differs slightly from the following widely accepted definition within the KR community:

\medskip

\begin{itemize}
    \item KR is the use of ``\emph{formal symbols} to represent a collection of \emph{propositions} believed by some putative agent,'' or ``the field of study concerned with'' such uses \cite{b9}.
\end{itemize}

\medskip

\noindent Herein, knowledge requires belief. Given that such belief is grounded in an agent's understanding, experience, or other ingredients, this definition of knowledge \emph{partially} harmonizes with the aforementioned definition of knowledge (i.e., information $+$ some ingredient) \cite{b5}. But why ``partially?''

\begin{table*}[b!]
\caption{Key Correspondences between Key Components for SNC.}\label{table:Table 1}
\centering
\begin{tabular}{>{\arraybackslash}m{0.9cm}|>
{\arraybackslash}m{0.6cm}|>{\arraybackslash}m{4.2cm}|>
{\arraybackslash}m{2.0cm}|>
{\arraybackslash}m{1.2cm}|>
{\arraybackslash}m{2.9cm}|>
{\arraybackslash}m{2.9cm}}
Level & Task & Role within SNC & Objective & Evaluation \newline Criterion & Input within SNC & Output within SNC \\
\hline\hline
\multirow{3}{0.9cm}{Local \newline stages} & AKE & Facilitating \emph{sparse} SN visualization in terms of \emph{vertices} & Keyness & $h\text{F}_1$ & Textual dataset & Document--Term--Matrix
\\
& EW & Facilitating \emph{sparse} SN visualization in terms of \emph{edges} & Interpretability & $\text{RI}$ & Document-Term-Matrix & Raw SN before CD
\\
& CD & Facilitating the capture of \emph{local topics} inherent in a given textual dataset & Distinctiveness & $Q$ \cite{b21} & Raw SN before CD & Final SN after CD
\\ \hline
Global \newline process & SNC & Providing \emph{clues} toward true knowledge of a given textual dataset & Knowledge representation & $J$ & Textual dataset & Final SN after CD
\\
\end{tabular}
\end{table*}

While any proposition can be information\footnote{The proposition ``this paper has been accepted without revisions'' is sadly false, yet it can be informative, indicating that its agent believes a falsehood.}, any information is \emph{not necessarily} a proposition. Information is defined as ``structured data'' \cite{b5}, yet non--propositionally structured data also exists in the world (Appendix \ref{Appendix A}). This paper selects a specific type of SN that represent textual non--propositional information and knowledge, as opposed to typical KGs. Subsection \ref{subsec:surrogates and clues} also discusses this point in detail. For now, it is sufficient to note that, within this paper, KR should not exclude the representation of non-propositional knowledge.

\medskip

\noindent\textbf{Network}. Revisiting the definition of KR \cite{b9}, knowledge can be represented by ``formal symbols.'' SNC is the process of representing knowledge by constructing a network as a symbolic system, and a network is

\medskip

\begin{itemize}
    \item a graph--based representation composed of a system’s \emph{components} and their \emph{interactions}, which are represented as \emph{vertices} and their \emph{edges}, respectively \cite{b10}.
\end{itemize}

\medskip

\noindent However, if a network is something represented by such graphical symbols of vertices and edges, why, then, do we not simply call it a graph? Even though a graph as a \emph{data structure} is necessarily used to construct any network, the \emph{system} represented possesses its own inherent properties. Because those properties constrain the graph's topology, it is only then legitimately distinguished as a network, not a mere graph \cite{b7}.

\medskip

\noindent\textbf{Network Science.} Another noteworthy point is that remarkably diverse systems or phenomena across nature \cite{b11}, technology \cite{b12, b13}, and society \cite{b14}--\cite{b16} have been represented as networks. Consequently, the discovery of principles, which govern specific networks, has been achieved through \emph{inductive reasoning} \cite{b17}: That is, architectures of networks from diverse domains have proven to be similar, and such findings have been elevated to ``universal organizing principles'' \cite{b17} through empirical studies. As a discipline in which such principles, also termed ``predictive models'' \cite{b18}, converge, network science has acquired its own universality and has become the foundation for studying networks.

Therefore, any network is constrained not only by its system but also by network science. We have also relied heavily on the established percolation theory in network science to facilitate SNC optimization. Subsubsection \ref{subsubsec:robustness improvement} details it.

\medskip

To summarize, our subject matter is about SNC. For a textual dataset, the lower bound of the expectation of SNC is the extraction of meaningful information, and the upper bound is KR. The selected type of SN represents non--propositional knowledge. Such SNs are symbolic systems, governed by universal organizing principles of networks.

What, then, does the selected type look like, and how is it constructed? First, it represents \emph{keyphrases} extracted from a given textual dataset as vertices and \emph{weights of semantic relatedness} between them as edges. In Subsection \ref{subsec:surrogates and clues}, we discuss that this type has trade-offs with the typical type of KG, which represents \emph{entities} as vertices and their \emph{relations} as edges. For now, it is sufficient to note that the selected type can be a better choice as a \emph{clue} rather than a \emph{surrogate}.

Next, when constructing such a clue, three primary stages of Automatic Keyphrase Extraction (AKE), Edge Weighting (EW), and Community Detection (CD) are typically undertaken (Subsections \ref{subsec:automatic keyphrase extraction}--\ref{subsec:community detection}) \cite{b19}. As representing all vertices and edges increases cognitive load during SN interpretation, AKE and EW facilitate SN sparsification by retaining only keyphrases as vertices and only significant interactions as edges \cite{b20}. CD also clusters vertices into distinct communities \cite{b21, b22}, thereby capturing local topics.

Corresponding to these roles, the stages aim to achieve respective objectives of keyness, interpretability, and distinctiveness. The degrees of belief in achieving these objectives can also be measured, and $h\text{F}_{1}$, percolation--theory--based $\text{RI}$, and a community scoring function \cite{b21} are used as those criteria in this paper. Subsections \ref{subsec:automatic keyphrase extraction}--\ref{subsec:community detection} clarify the definitions of these stages and objectives, and Subsection \ref{subsec:local evaluation criteria} justifies the choices of the criteria. For now, it is sufficient to note the correspondences of these components (Table \ref{table:Table 1}).

Beyond reviewing these components, our final aim is to propose a framework to \emph{optimize} SNC. What, then, is the background of this proposal? It is as follows:

AKE, semantic relatedness estimation (referred to as edge weighting within this paper), and community detection are research areas that can independently exist \emph{even outside} SNC. As to be reviewed in Subsections \ref{subsec:automatic keyphrase extraction}--\ref{subsec:community detection} and \ref{subsec:local evaluation criteria}, each area encompasses available methods and evaluation protocols. So, any researcher conducting SNC should systematically select an appropriate method from a rich set of alternatives at each stage. However, many applications still rely on a limited number of conventional pipelines\footnote{They typically involve Term Frequency (TF) or TF--IDF \cite{b23} for AKE, Co--occurrence Frequency (CF) for EW, and the Clauset--Newman--Moore (CNM) \cite{b24} or Louvain \cite{b25} algorithms for CD.} \cite{b19}.

Whether SNs constructed in such manners guarantee optimal outcomes remains an unaddressed question. This limitation stems from the fact that while evaluating methods within each stage is straightforward, ranking resulting SNs in an integrative context remains challenging\hypertarget{joint confidence}{.}

Against this backdrop, Subsection \ref{subsec:cluenetwork} presents our SNC framework facilitating such \emph{ranking}. What, then, is the criterion for determining that one SN is \emph{more reliable} than another? Crucially, our notion of an optimal SN is distinct from some external \emph{gold standard} assumed in KG refinement or KG alignment. Rather, the notion aligns with some SN that achieves the \emph{highest joint confidence} for the local objectives (i.e., keyness, interpretability, and distinctiveness), given available method combinations across the stages.

\begin{figure*}[b!]
    \centering
    \includegraphics[width=\textwidth]{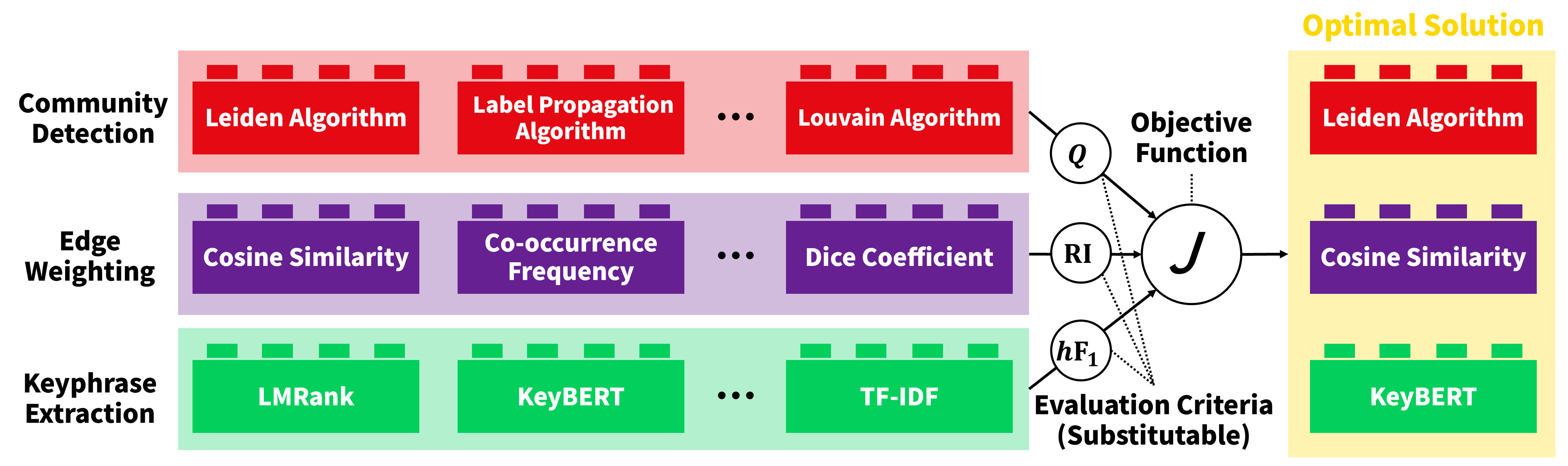}
    \caption{\textbf{SNC Depicted as a Brick--Stacking Problem. Optimal construction is achieved through the systematic selection and combination of bricks. Likewise, SNC requires the systematic selection and integration of methods. This illustration is inspired by the BERTopic documentation \href{https://maartengr.github.io/BERTopic/algorithm/algorithm.html}{(link)}.}}
    \label{Figure 1}
\end{figure*}

Within this paper, such confidence is measurably defined as a global objective function $J$. Subsection \ref{subsec:justification of j} justifies its form. For now, it is sufficient to refer to Fig. \ref{Figure 1} illustrating the simplified version of our framework \emph{ClueNetwork} (its full version is detailed in Subsection \ref{subsec:cluenetwork}). Herein, SNC is depicted as a Process --- also can be termed a pipeline, policy, or brick--stacking --- Optimization Problem (POP):

First, choices of local methods should be evidence--based. At each stage, candidate methods are evaluated by using a stage--specific criterion. Next, the gathered evaluation results are integrated into $J$. Finally, some optimal combination of methods that maximizes $J$ is identified for a textual dataset. This paper contributes to the body of knowledge in reaching the framework as follows:

\medskip

\noindent\textbf{Theoretical Foundation}. Section \ref{sec:theoretical foundation} addresses the following Research Questions (RQs) to ensure a \emph{theoretical} foundation for SNs as clues: How does the selected type of SN represent knowledge (\textbf{RQ1})? When is the type, rather than typical KGs, recommended to be constructed (\textbf{RQ2})? 
What are the definitions of the SNC stages and their objectives (\textbf{RQ3})? Which stage--specific methods have been selected for the Proof of Concept (PoC) of \emph{ClueNetwork} (\textbf{RQ4})?

\medskip

\noindent\textbf{Empirical Record}. Section \ref{sec:local evaluation} holds value as an \emph{empirical} record. By providing illustrative experiments, which stage--specific selected methods yield high performance (\textbf{RQ6}) is presented. Before it, the section answers how the local evaluation criteria ($h\text{F}_1$, $\text{RI}$, and $Q$) are defined and justified (\textbf{RQ5}).

\medskip

\noindent\textbf{ClueNetwork}. In response to how evaluation results across the SNC stages are integrated to identify the optimal SN (\textbf{RQ7}), Section \ref{sec:global optimization} presents our framework \emph{ClueNetwork}. Crucially, it facilitates \emph{ranking} candidate SNs for a textual dataset. The section also answers how \emph{ClueNetwork}'s objective function $J$ is defined and justified (\textbf{RQ8}). \emph{ClueNetwork} is more than integrative, as it incorporates two innovative methodological components, namely $\text{RI}$ and a veracity pretest for edge weighting measures. They are built upon the established percolation theory and the Matthews correlation coefficient \cite{b26}, respectively, yet with a touch of novelty.

\medskip

Finally, only after Section \ref{sec:discussion} carefully discusses not only such contributions but also \emph{ClueNetwork}'s limitations does Section \ref{sec:conclusions} conclude this paper.

\section{Theoretical Foundation}
\label{sec:theoretical foundation}

\medskip

This review section comprises four subsections. Subsection \ref{subsec:surrogates and clues} first reviews several philosophical topics to distinguish the selected type of SN from typical KGs, thereby addressing \textbf{RQs 1} and \textbf{2}. Next, Subsections \ref{subsec:automatic keyphrase extraction} through \ref{subsec:community detection} review the SNC stages along with their respective objectives and considerations. Each subsection also summarizes the philosophies of selected methods that optimize the corresponding local objective. Thereby, Subsections \ref{subsec:automatic keyphrase extraction} through \ref{subsec:community detection} themselves serve as the answers to \textbf{RQs 3} and \textbf{4}.

\medskip

Before that, we would like to clarify that collecting or granular clustering as many local methods as possible is beyond our scope. We only need some \emph{materials} for the PoC of \emph{ClueNetwork}. Accordingly, we have selected relatively accessible and implementable local methods and have provided a simplified (not granular) taxonomy of methods for each SNC stage. Instead, we have cited some specialized review papers focusing on granular clustering for interested readers.

\begin{table*}[b!]
\caption{Four Central Theses, Formulated by \cite{b28}, of Scientific Realism.}\label{table:Table 2}
\centering
\begin{tabular}{>{\arraybackslash}m{1.0cm}|>
{\arraybackslash}m{4.9cm}|>
{\arraybackslash}m{10.5cm}}
No. & Core Stance & Full Principle \\
\hline\hline
Thesis 1 & Confidence for the \emph{correspondence} between scientific terms and reality & Even when scientific theories mention unobservable terms (e.g., dark matter), such terms should be regarded as \emph{referring} to real things. \\ \hline
Thesis 2 & Confidence for the \emph{verifiability} of whether scientific theories align with reality & As scientific theories are frequently ``confirmed as approximately true by ordinary scientific evidence,'' interpreted based on ``methodological standards'' \cite{b28}, they are \emph{verifiable}. \\ \hline
Thesis 3 & Confidence for scientific \emph{progress} & Science progresses largely through ``successively more accurate approximations to the truth'' \cite{b28}. That is, later scientific theories are established largely ``by standing on the shoulders of giants'' \cite{b29} (i.e., ``the knowledge embodied in previous'' scientific theories \cite{b28}). \\ \hline
Thesis 4 & Confidence for the \emph{independence} of reality & Reality exists regardless of whether scientific theories aiming to explain it are true or false. That is, reality remains ``largely independent of our thoughts or theoretical commitments'' \cite{b28}. \\
\end{tabular}
\end{table*}

\subsection{Surrogates and Clues}\label{subsec:surrogates and clues}

\medskip

To begin with, there is a straightforward fact. If any type of SN aims at KR, who constructs them most actively? Certainly, scientists (including engineers who consider scientific principles). Science is, by its very nature, an activity dedicated to true knowledge. Indeed, it is defined as

\medskip

\begin{itemize}
    \item ``the organized and systematic enterprise that gathers \emph{knowledge} about the world and condenses the knowledge into \emph{testable} laws and principles'' \cite{b27}.
\end{itemize}

\medskip

Hence, if we can comprehend scientists' common sense, we may be able to recommend appropriate types of SNs for them. Let us consider the following.

\medskip

\noindent\textbf{Scientific Realism} \cite{b28} is a doctrine embodying the theses presented in Table \ref{table:Table 2}. According to them, reality exists \emph{independently} of human minds (Thesis 4), scientific theories (sets of ``testable laws and principles'' \cite{b27}) \emph{correspond} to reality (Thesis 1), and such correspondences are \emph{verifiable} (Thesis 2). Scientific realism aligns with the common sense of most scientists. For example, this journal \emph{IEEE Access} asks reviewers, ``Does the paper contribute to the \emph{body} of knowledge?'' It presupposes that such a body \emph{independently} exists and that reviewers can \emph{verify} whether submissions contribute to it.

\medskip

\noindent\textbf{Reality}. Throughout those theses, \emph{reality} serves as an independent criterion to verify knowledge. What, then, is reality? 

\medskip

\begin{itemize}

    \item Reality is the domain of ``that which \emph{there is}. (...) \emph{How much} of it there is forms the subject of ontology'' \cite{b30}.

\end{itemize}

\medskip

\noindent Here, one might ask, ``Wait, we already encounter everyday beings. Why, then, should we single out reality?'' For most philosophers, existence is far from a trivial condition. For instance, Plato, one of the earlier realists who deeply contemplated reality, believed that only \emph{fixed} (constant, uniform, and immutable) things purely exist \cite{b31}. Plato referred to such things as ``the Forms'' \cite{b31}--\cite{b34}. Whether we agree with Plato's claim or not, it warrants consideration\hypertarget{banana}{.}

In a corner of our laboratory, a banana is rolling around, seemingly in existence. If we ignore it for two months while focusing on this re--revision, will it still be a banana? It is likely to become some horrific organic matter, not a banana anymore. Most everyday things are \emph{not fixed}. Plato referred to such transient things as ``rolling around between what is not and what purely is'' \cite{b32}. Numerous philosophers have spent the last 2,400 years debating what belongs to reality.

\medskip

\noindent\textbf{The Forms and Gold Standards}. Apart from that debate, why is reality desired as an \emph{independent} criterion of knowledge? To scientists, transient things remain too uncertain to serve as criteria. Claiming that their knowledge is true based on the knowledge itself could be circular reasoning. If there is something that exists fixedly and independently, it becomes far easier to say, ``Look! My knowledge harmonizes with reality, so it is true.'' Interestingly, a similar approach is observed in KG \emph{refinement} and KG \emph{alignment}, often performed by knowledge engineers. Herein,

\medskip

\begin{itemize}
    \item KG refinement consists of KG completion and KG error detection; the former is the task of ``adding missing knowledge'' to a KG, and the latter is the task of ``identifying and removing errors'' in a KG \cite{b35}.
    \item KG alignment, also termed entity alignment, is ``the task of finding'' \cite{b36} and ``linking entities sharing the same identity'' \cite{b37} across multiple KGs.
\end{itemize}

\medskip

The concerns of KG refinement and KG alignment are whether KGs are \emph{complete and correct} \cite{b35} and whether KGs are \emph{properly} integrated \cite{b37}, respectively. To address these concerns, knowledge engineers rely largely on independent criteria\footnote{For example, KGs deemed to possess higher qualities than target KGs, common knowledge bases (e.g., Wikidata \cite{b38}), or domain experts.} termed ``\emph{gold} standards'' \cite{b35, b39}, subsets of which are sometimes classified as ``partial gold standards,'' ``silver standards,'' or ``human (post--hoc) judgment'' \cite{b35} depending on their qualities or formats. Does this setup look familiar?

\begin{figure*}[b!]
    \centering
    \includegraphics[width=\linewidth]{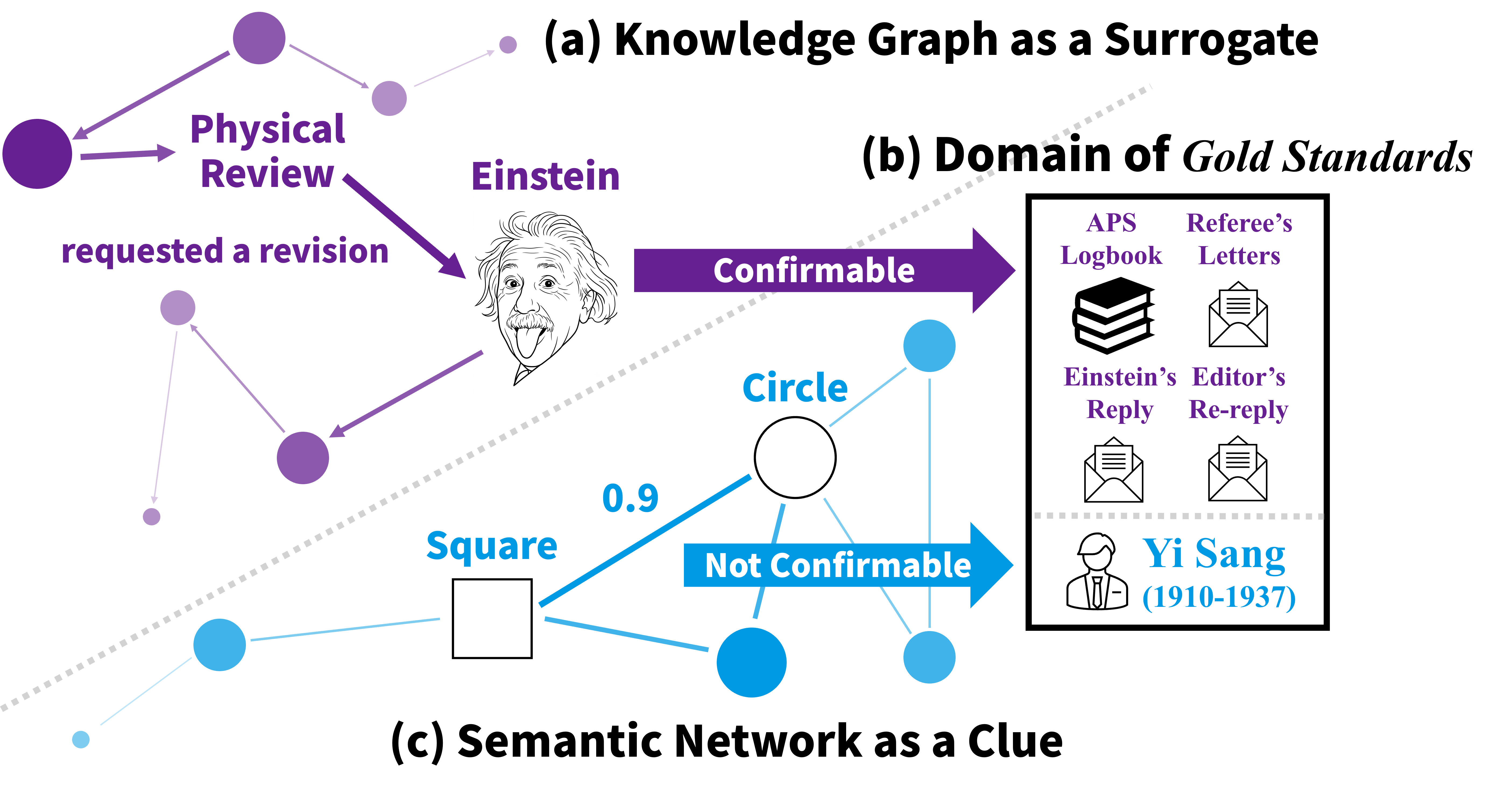}
    \caption{\textbf{The Physical Review's decision is confirmable through records, whereas the meaning of (Square, $0.9$, Circle) cannot be queried to Yi Sang.}}
    \label{Figure 2}
\end{figure*}

Yes, it is Plato again. He believed that the reason an everyday thing seems somehow beautiful is that it \emph{partially shares} the Form of Beauty \cite{b31}. Likewise, the reason a KG is deemed to represent true knowledge partially is that the KG \emph{partially reflects} a gold standard. Plato believed that as the Forms are fixedly being, contemplating them is to gain true knowledge \cite{b32, b33}. Similarly, as gold standards are fixed, \emph{comparing} them to KGs is deemed to determine the veracity of the KGs.

\medskip

\noindent\textbf{Surrogates and Propositions}. As such, whether welcome news or not, the Platonic way of thinking somehow resembles the way KGs operate. To many peers, KGs are something that, by approximating true knowledge, ultimately become its ``\emph{surrogates}'' \cite{b7}. What are such KGs likely to be composed of to make it easier to verify them? Certainly, \emph{propositions}, as they are ``declarative sentences, believed by some putative agent, that can be \emph{true or false, right or wrong}'' \cite{b9}.

Indeed, typical KGs consist of subject--predicate--object structures, also termed \emph{triplets}, that can be intuitively transformed into propositions. Let us suppose there is a KG of physics (Fig. \ref{Figure 2}--a). It can include the triplet (Physical Review, requested a revision, Einstein), from which ``some putative agent'' \cite{b9} (we) can easily infer that the triplet represents the proposition, ``The \emph{Physical Review} requested a revision from A. Einstein.'' And it is very confirmable (Fig. \ref{Figure 2}--b), because the historical records of the review process exist \cite{b40}.

Likewise, when \cite{b8} defined a KG as ``a graph of data intended to accumulate'' and represent knowledge, many peers likely regarded such knowledge as ``a collection of propositions'' \cite{b9}. And there are reasons. With the resurgence of AI over the past few decades, studies leveraging KGs as \emph{references} used by AI--based systems to achieve something innovative are increasing. Given that such systems often exhibit a \emph{black--box} nature, making even their references something whose ``true or false, right or wrong'' \cite{b9} and meanings are ambiguous might amount to opening \emph{Pandora's box}. To many peers, KGs should be ``surrogates'' \cite{b7} that reduce room for variances from AI--based systems.

\medskip

\noindent\textbf{Case of Square Circle}. However, humanity has taken on the role of ``some putative agent'' \cite{b9} long before AI. Let us suppose an instance of the selected SN type regarding the poetry of Yi Sang, who is a Korean writer\hypertarget{unlike KGs}{.}

Although the SN (Fig. \ref{Figure 2}--c) can include the triad (Square, $0.9$, Circle), because $0.9$ is not a predicate, the triad cannot be \emph{immediately} converted into a proposition. Only through circuitous reasoning do we say, ``The triad \emph{implies} that the collocation of square and circle occurs in the dataset with a degree of $0.9$,'' yet the triad \emph{itself} does not represent any proposition. This derivation requires \emph{intervention} from the agent (us), as the collocation is far from everyday contexts.

\begin{figure}[t!]
    \centering
    \includegraphics[width=0.8\linewidth]{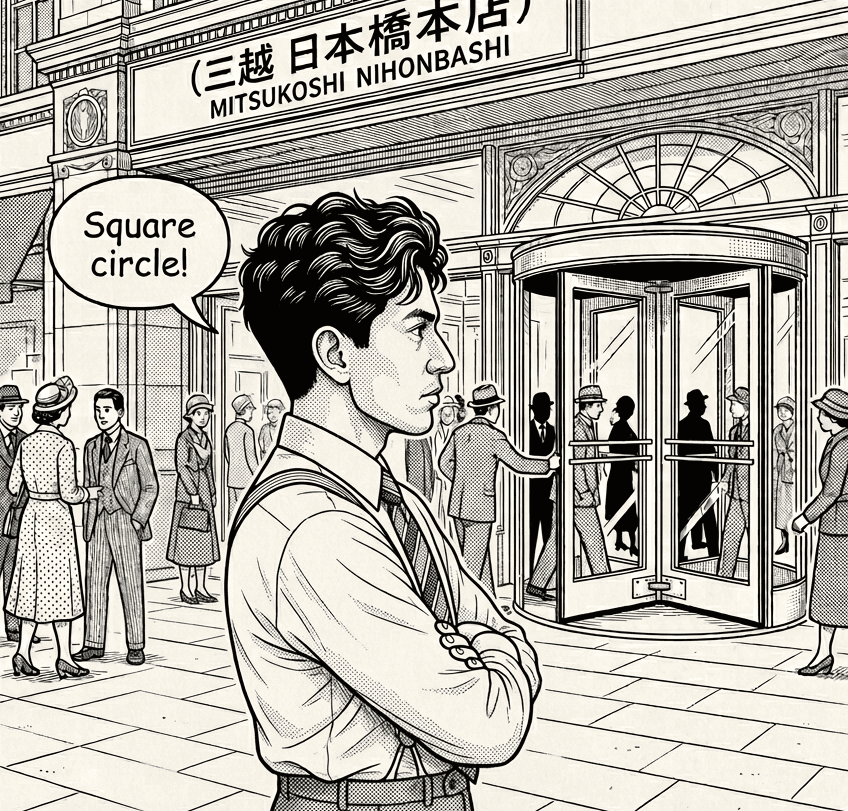}
    \caption{\textbf{Yi Sang likely envisioned the square circle upon seeing a revolving door. This figure was generated by using \cite{b41}.}}
    \label{Figure 3}
\end{figure}

In fact, the square circle refers to a \emph{revolving door} of a department store (Fig. \ref{Figure 3}), given that our literature teachers taught us so. The teaching was possible because Yi Sang researchers had reached a \emph{consensus} based on several clues. Specifically, Yi Sang was an architect; the objects in the poem \href{https://www.poetryfoundation.org/poetrymagazine/poems/148686/au-magasin-de-nouveautes}{(link)} resemble the features of the department store \href{https://cp.mistore.jp/global/en/nihombashi.html}{(link)}\hypertarget{Yi Sang}{;} the title is ``\emph{At the Department Store}'' \cite{b42}. However, these are not Yi Sang himself. He returned to stars in 1937 (Fig. \ref{Figure 2}--b).

That is, in certain cases, reaching gold standards is \emph{extremely difficult}. Far more datasets than expected are context--dependent, trendy, or too large in volume to be read. Their gold standards are largely \emph{not yet ready}.

As many researchers cannot afford to wait for such criteria, they strive to produce possible explanations \emph{at the moment} by adding their expertise--based \emph{beliefs} as ingredients to available clues. Whether true or not, such explanations harmonize with the definition of knowledge (information $+$ some ingredient) by \cite{b5}. The attitude of pursuing more coherent explanations also aligns with Thesis $3$ reserved until now of scientific realism (Table \ref{table:Table 2}). As long as such explanations are under researchers' responsibility, there is no reason to exclude the explanations from the domain of knowledge.

\medskip

\noindent\textbf{Semantic Networks as Clues}. Then, can (Square, $0.9$, Circle) \emph{itself} be considered knowledge? Yes, it can. First, as it is ``structured data'' \cite{b5}, it qualifies at least as information. Second, it is also knowledge, given that this structuring was only possible because Yi Sang's \emph{belief} --- that the revolving door resembles the square circle --- had been embedded within the data. Third, as discussed in Subsections \ref{subsec:automatic keyphrase extraction} through \ref{subsec:community detection}, the triad already reflects the \emph{beliefs} of particular AKE and EW algorithms regarding which phrases are representative and worthy of being connected.

In short, some data producers, some local SNC algorithms, and some interpreters are all ``some putative agents'' \cite{b9}. Given that \emph{beliefs} intervene at least at the SNC phase\footnote{Certain data might be mere noise. And even if an SN is constructed, it might be left uninterpreted. However, without SNC, the SN cannot exist.} among the three stages (production, SNC, and interpretation), triads of the selected type of SN should be considered to represent knowledge. Therefore, we are now ready to answer \textbf{RQ1} (``How does the selected type represent knowledge?'').

\emph{Based on} information constructors' beliefs (i.e., local SNC algorithms' philosophies), it represents knowledge by \emph{implying} data producers' beliefs as non--propositional triads. To convert such triads into propositions, interpreters intervene \emph{only after} SNC. As a system of such triads, the selected type can be considered as a \emph{clue}, in that it can serve as the starting point for interpreters to derive \emph{hypotheses} (propositions not verified yet) regarding what is implied.

Although their KR ways differ, both the selected type and typical KGs belong to the SN family, as they do KR \cite{b3, b4}. They merely diverged in the late 1980s when some researchers in the Netherlands designed a type of SN characterized by ``a limited set of relations'' \cite{b43}, as KGs. To avoid confusion and for brevity, the selected type and typical KGs are hereafter referred to as SNs and KGs, respectively.

\medskip

\noindent\textbf{Abduction and Exploratory Research}. Here, one might ask, ``Okay, SNs do KR. However, if such knowledge cannot be compared to some gold standards, would it not violate Thesis 2 and thus fail to qualify as \emph{scientific} knowledge?'' It does not. First, Thesis 2 (Table \ref{table:Table 2}) simply states that ``\emph{methodological} standards'' \cite{b28} should be used for scientific verification. It does not restrict them to \emph{gold} standards. Second, although such \emph{methodological} standards vary, they are fundamentally certain combinations of \emph{deduction}, \emph{induction}, and \emph{abduction}, termed ``the three types of reasoning'' \cite{b44}.

As presented in Table \ref{table:Table 3} and according to \cite{b44}, deduction is the reasoning that ``proves that something must be'' by deriving ``\emph{necessary} consequences'' from given propositions; induction is the reasoning that determines the \emph{actual} degree to which something operates through observations; abduction is the \emph{only} reasoning that ``forms an \emph{explanatory} hypothesis'' that ``suggests that something may be.''

\begin{table}[h!]
\caption{Examples of Deduction, Induction, and Abduction. This table is secondarily cited from \cite{b45}, and the examples have been altered by us.}\label{table:Table 3}
\centering
\begin{tabular}{>{\arraybackslash}m{2.4cm}|>
{\arraybackslash}m{2.4cm}|>
{\arraybackslash}m{2.4cm}}
Deduction & Induction & Abduction \\
\hline\hline
Rule: \emph{All} PhDs in this lab get grants. & Case: \emph{These} PhDs are in this lab. & Rule: All PhDs in this lab get \emph{grants}. \\ \hline
Case: These PhDs are in this lab. & Result: These PhDs get grants. & Result: These PhDs get grants. \\ \hline
Result: \emph{These} PhDs get grants. & Rule: \emph{All} PhDs in this lab get grants. & Case: These PhDs are in this \emph{lab}.
\\ \hline
\multicolumn{3}{m{7.2cm}}{Each cell in the final row represents a derived proposition.} \\
\end{tabular}
\end{table}

Most gold--standard--based KG evaluations are certain combinations of deduction and induction. A typical combination for arbitrary propositions $a, b, c, \text{ and } d$ is as follows, where the propositions shared by both reasonings are presented without parenthetical annotations:

\medskip

\begin{itemize}
    \item Rule (Deduction): The gold standard says $a,b,c,\text{ and }d$.
    \item Case: The KG claims $\neg a, b, c, \text{ and }\neg d$.
    \item Result: The KG yields false, true, true, and false for $a, b, c, \text{ and } d$, respectively.
    \item Rule (Induction): The KG's $\text{F}_{1}$ score is $0.67$.
\end{itemize}

\medskip

\noindent Herein, interestingly, knowledge engineers (i.e., modern scientists) do not insist on deduction. However, Plato regarded deduction as the superior intellectual activity \cite{b33}. That is, while science operates similarly to the Platonic way in some aspects, it is obviously not always the case. 

Scientists also legitimately and frequently employ certain combinations of abduction and induction. For example, \cite{b46} derived a hypothesis through SNA--based abduction that Reddit\footnote{Reddit is a well-known social media platform. r/Republican and r/democrats are its subreddits (discussion forums).} users with different political orientations share views on which events and politicians warrant discussion:

\medskip

\begin{itemize}
    \item Rule: Submissions from r/Republican and r/democrats share linguistic patterns when sharing views on what warrants discussion.
    \item Result (Abduction): \textbf{Wow}! Collected submissions from r/Republican and r/democrats share linguistic patterns.
    \item Case: The collected submissions share views on what warrants discussion.
    \item Result (Induction): The differences between the scores of words in the collected submissions are not statistically significant.
    \item Rule: The submissions from r/Republican and r/democrats share linguistic patterns when sharing views on what warrants discussion.
\end{itemize}

\medskip

Herein, \cite{b46} only performed their work up to the hypothesis (Case) generation level. Nevertheless, it was legitimately published in a scientific journal, because the generated hypothesis can be tested through induction at \emph{any time}, thereby providing a new \emph{starting point} for research. Such abduction originates from a \emph{wow moment} when a researcher encounters a ``surprising fact'' \cite{b47}, leading to a hypothesis by filtering that fact through the researcher's existing belief (Rule).

That is, science progresses not only through gold standards or hypothesis testing, but also through such \emph{exploratory} research \cite{b48}. Knowledge \emph{reserved} for verification is not equivalent to \emph{unverifiable} knowledge. Even Yi Sang's square circle cannot be definitively assumed to remain indefinitely as literary knowledge rather than scientific knowledge, as his undiscovered diaries might come to light someday. Accordingly, we are now ready to answer \textbf{RQ2} (``When are SNs, rather than KGs, recommended to be constructed?''):

In works requiring verification--oriented KR, KGs are superior to SNs. Conversely, SNs are recommended for works focused on hypothesis--generation--oriented KR or where gold standards are lacking. While KGs should be employed in works requiring propositions for AI, SNs can be considered when researchers need the power of open interpretation to move beyond predefined propositions. As such, there are simply trade--offs between KGs and SNs.

\subsection{Automatic Keyphrase Extraction}\label{subsec:automatic keyphrase extraction}

\medskip

In the previous subsection, we reviewed the philosophical foundation of SNs. Now, we can transition to the review of the individual stages required for constructing such SNs. This subsection is the first of Subsections \ref{subsec:automatic keyphrase extraction} through \ref{subsec:community detection}, which themselves constitute the comprehensive answers to \textbf{RQ3} (``What are the definitions of the SNC stages and their objectives?'') and \textbf{RQ4} (``Which stage--specific methods have been selected for the PoC of \emph{ClueNetwork}?'').

\subsubsection{What is AKE, and Why Does It Matter?}\label{subsubsec:what is ake, and why does it matter}

\medskip

\noindent\textbf{Keyphrase Extraction (KE)} is ``a fundamental subtask of Natural Language Processing (NLP)'' \cite{b49}, and also frequently serves as an important subtask across diverse applications of Text Mining (TM) and Information Retrieval (IR\hypertarget{definition of ake}{)} \cite{b50}. According to \cite{b50} and \cite{b51}, KE is defined as

\medskip

\begin{itemize}
    \item the \emph{automatic} extraction of \emph{phrases that} ``best represent'' \cite{b50} or ``concisely summarize'' \cite{b51} a document.
\end{itemize}

\medskip

\noindent As shown above, relevant papers (e.g., \cite{b52}--\cite{b56}) often emphasize that KE is performed \emph{automatically}, thereby also referring to KE as Automatic KE (AKE). Why? Given that the volume of available documents today typically exceeds what users can thoroughly read and analyze, KE as an automatic rather than manual subtask is feasible \cite{b50}.

Meanwhile, in the above definition, a phrase refers to a textual unit composed of one or more tokens \cite{b49, b50}. As a phrase is distinguished from larger textual units (e.g., documents, paragraphs, or sentences), it is also referred to as a ``lexical unit'' \cite{b51} or simply a ``term'' \cite{b49}.

\medskip

\noindent\textbf{Keyness}. From the definition of AKE, we can also deduce the definition of \emph{keyness}. Since keyphrases have already been defined as ``phrases \emph{that} best represent or concisely summarize a document,'' the property described by this relative clause is precisely what constitutes keyness in terms of AKE\hypertarget{definition of max keyness}{:}

\medskip

\begin{itemize}
    \item Keyness is the property \emph{whereby} a document is ``best represented'' \cite{b50} or ``concisely summarized'' \cite{b51}.
\end{itemize}

\medskip

Phrases possessing this property are legitimately prefixed with \emph{key--}. However, according to \cite{b50}, keyness is linked to at least ten different properties (for which please see \cite{b50}), rendering further specification of its meaning ``elusive.'' Considering its scope and for brevity, this paper certainly does not undertake that work. Instead, Subsubsection \ref{subsubsec:how does AKE work} reviews different philosophies of keyness embedded within the selected AKE algorithms, in light of the observation by \cite{b50} that interpretations of keyness are application--dependent.

\medskip

\noindent\textbf{Necessity in SNC}. AKE is also essential in SNC because no SN can be constructed without vertices. One might ask, ``Okay, but why should SNs represent keyphrases as vertices? KGs represent entities without any problems.'' A KG is graph--structured data intended to accumulate and represent knowledge, whose vertices and edges are ``entities of \emph{interest}'' and their relations \cite{b8}. Herein, any KG assumes an \emph{interest}, whereas, given that an SN is directly extracted from a textual \emph{dataset}, entities of specific interest are largely not predefined. That is, KGs represent knowledge of specialized \emph{interests} \cite{b8}, whereas SNs represent knowledge of \emph{texts} \cite{b2}. In terms of SNC, what matters is not which vertices accurately represent an interest, but rather which vertices ``best represent'' \cite{b50} texts. This is why keyphrases are employed as vertices, and AKE serves as an irreplaceable stage in SNC.

\medskip

\noindent\textbf{Maximization of Keyness}. Ultimately, the objective of AKE is to \emph{comprehensively} extract phrases that possess keyness for a document. Within this paper, this objective is termed ``the \emph{maximization} of keyness.'' Such maximization is specifically and typically regarded as maximizing the \emph{alignment} of a set of extracted phrases with the ``\emph{golden} set of keyphrases'' \cite{b49}. The actual degree of such alignment can also be quantified and evaluated by using a specific criterion (e.g., $\text{F}_{1}$, $p\text{F}_{1}$, or $h\text{F}_{1}$) as explained in Subsection \ref{subsubsec:harmonic F1 Score}.

\medskip

\noindent\textbf{Scientific Realism Revisited}. As may be noticed, while \emph{ClueNetwork} does not assume the existence of a gold SN for a textual dataset, AKE assumes that of gold keyphrases. That is, by reflecting Theses 1, 2, and 4 (Table \ref{table:Table 2}), AKE serves as an anchor rendering \emph{ClueNetwork} scientifically realistic. This reflection is feasible, given that constructing sets of gold keyphrases is easier than constructing a gold SN (Subsubsection \ref{subsubsec:handling lack of gold keyphrases}). Furthermore, the definition of a clue is ``a piece of evidence leading to \emph{discovery}'' \cite{b57}. To scientists, discovery often signifies finding true knowledge consistent with reality. Accordingly, technologically sound AKE should enable the clues (SNs) to encompass some pieces of reality.

\subsubsection{How Does AKE Work?}\label{subsubsec:how does AKE work}

\medskip

\noindent\textbf{Beliefs as Ingredients}. AKE typically operates in a manner where an AKE algorithm $x$ formulates a proposition, such as ``The keyness scores of candidate phrases $a, b, c$, and the rest for a document $d$ are $1.0, 0.9, 0.8$, and below, respectively,'' based on its belief (interpretation of keyness). Herein, the definition of a belief is a judgment that ``\emph{the world is one way} and not another'' \cite{b9}. That is, $x$ judges the world to be one where $a, b, \text{ and } c$ possess higher keyness.

\medskip

\noindent\textbf{Ranking Candidates}. As may be noticed, ranking candidates is central to AKE, given that their priorities ultimately matter. Here, a relevant consideration arises. Some AKE algorithms use lower--is--better scoring (e.g., \cite{b58} and \cite {b59}), and their resulting scores often follow skewed distributions that can complicate data analysis. Hence, $ClueNetwork$ normalizes candidate scores to a $0$ to $1$ scale under the higher--is--better convention. Specifically, \eqref{Eq 1}--a is applied to original higher--is--better scores, and \eqref{Eq 1}--b to original lower--is--better scores.

\begin{equation} \label{Eq 1}
\begin{aligned}
\begin{cases}
    \frac{r-r_{min}+\epsilon}{r_{max}-r_{min}+2\epsilon}\cdots(a) \\ \\
    \frac{r_{max}-r+\epsilon}{r_{max}-r_{min}+2\epsilon}\cdots(b)
    
\end{cases}
\end{aligned}
\end{equation}

\medskip

Herein, for a document, $r$ is the rank of a candidate's score (higher rank numbers indicate higher original scores); $r_{max}$ and $r_{min}$ are the maximum and minimum ranks, respectively; $\epsilon=10^{-8}$ is a small constant handling corner cases where only tied candidates exist (i.e., $r_{max} = r_{min}$).

\medskip

\noindent\textbf{Construction of 2--mode Matrix}. In terms of SNC, although each term's final score is required to represent only keyphrases \emph{across} a dataset as vertices \cite{b20}, AKE is typically performed on \emph{respective} documents \cite{b49}. Hence, to facilitate computing each term's final score by aggregating its normalized individual scores, a 2--mode matrix is constructed, with rows, columns, and cells corresponding to windows, terms, and scores, respectively. Herein, a window is a textual range (e.g., document, paragraph, sentence, or $n$--gram) depending on the purpose, but it is typically set to a document.

That is, in SNC, as a result of document--wise AKE, an initial Document--Term Matrix ($\mathbf{DTM} \in \mathbb{R}^{m \times n}$) is typically built, where $m$ and $n$ are the numbers of documents and terms, respectively. Herein, $\text{Score}(t_{i,\text{ }h})$ is the normalized score of the $i$\textsuperscript{th} term $t_{i}$ for the $h$\textsuperscript{th} document, and the term's final score is obtained by summing its normalized individual scores:

\begin{equation} \label{Eq 2}
\text{Final Score}(t_i)=\sum_{h=1}^{m}\text{Score}(t_{i,\text{ }h})=\sum_{h=1}^{m}\mathbf{DTM}_{h,\text{ }i}
\end{equation}

\medskip

\noindent\textbf{Simplified Taxonomy of Unsupervised Algorithms}. What algorithms exist to score terms under respective philosophies of keyness? Numerous relevant algorithms (e.g., \cite{b52}--\cite{b56} and \cite{b58}--\cite{b83}) have been proposed with benchmarking (i.e., gold--standard--based) experiments \cite{b49, b51}. According to \cite{b51}, such algorithms can be categorized as supervised or unsupervised ones. As the latter, which requires no training data, is more widely employed, this paper focuses on it. It can be further categorized as follows \cite{b49, b51}:

\begin{figure}[h!]
    \centering
    \includegraphics[width=\linewidth]{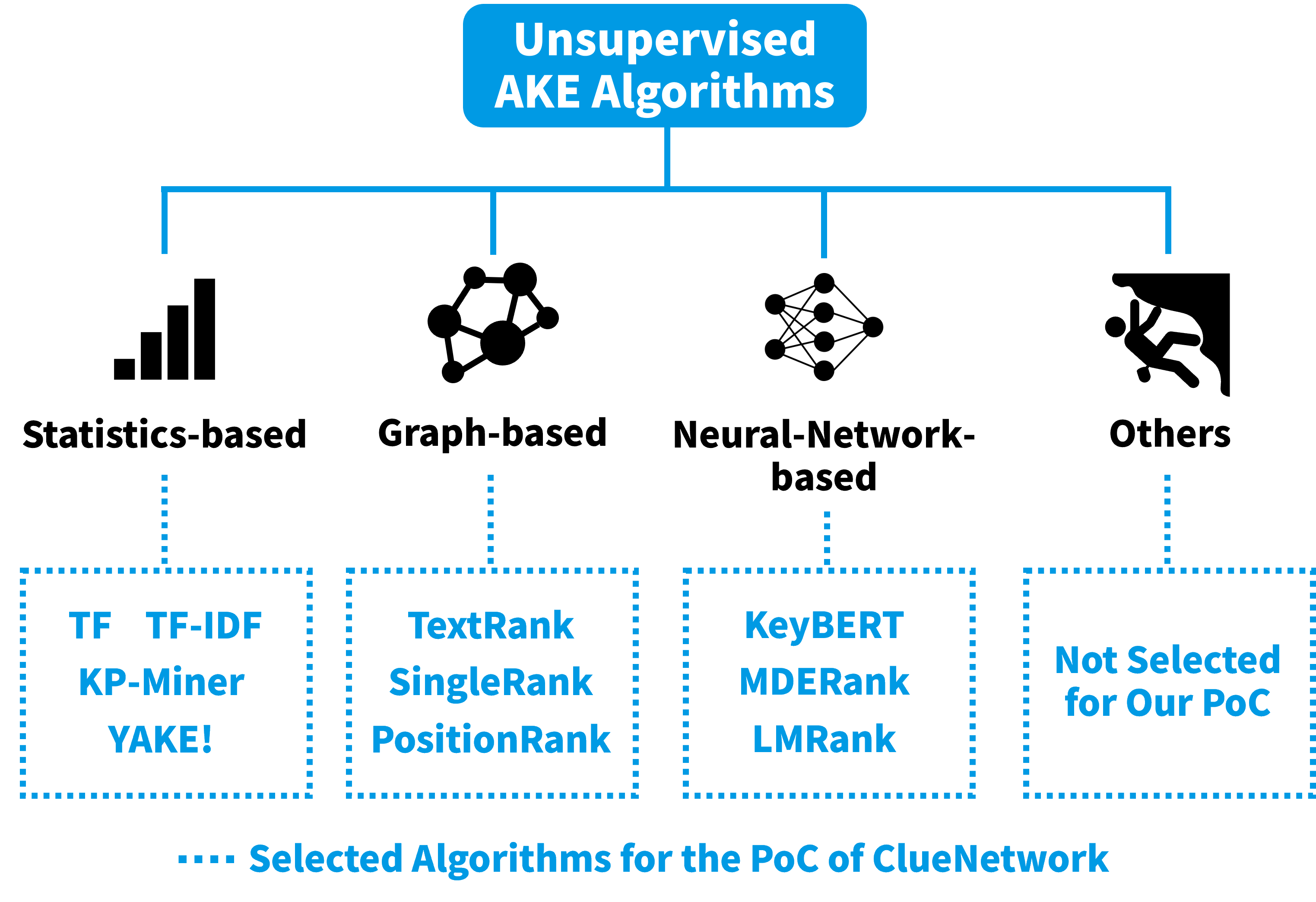}
    \caption{\textbf{Simplified Taxonomy of Unsupervised AKE Algorithms.}}
    \label{Figure 4}
\end{figure}

First, statistics--based algorithms exploit statistical features of textual units (e.g., \cite{b23}, \cite{b55}, \cite{b58}, \cite{b60}, and \cite{b61}). For instance, TF and TF--IDF \cite{b23} regard a \emph{higher frequency} of a term as a necessary condition for keyness. Their successors, such as KP--Miner \cite{b61} and YAKE! \cite{b58}, also incorporate other features, such as \emph{positional or linguistic} ones.

Second, graph--based algorithms exploit a graph, which represents textual units as vertices and their interactions as edges, inherent in a document (e.g., \cite{b52}--\cite{b54} and \cite{b62}--\cite{b76}). Interestingly, these algorithms remind us that graphs are prevalent in the world (Section \ref{sec:introduction}), as they fundamentally view \emph{influential vertices} as possessing keyness.

Third, Neural--Network--based (NN--based) algorithms explicitly execute neural models to capture \emph{contextual semantics} of a document. Although their frameworks and philosophies of keyness diverge (e.g., \cite{b54}, \cite{b75}, \cite{b77}, and \cite{b78}), two prominent trends are observed. One trend (e.g., \cite{b59} and \cite{b79}--\cite{b81}) is to compute similarity scores between vector representations (also termed embeddings), extracted by a Pretrained Language Model (PLM), of a document and those of candidate phrases. The other (e.g., \cite{b82} and \cite{b83}) is to extract keyphrases by feeding a document and an appropriate prompt into a PLM or a Large Language Model (LLM).

Fourth, most AKE algorithms are somewhat hybrid or idiosyncratic. Can a graph--based algorithm leveraging a PLM also be NN--based (e.g., \cite{b75})? If an algorithm does not execute a neural network but references third--party static word embeddings, can it be NN--based (e.g., \cite{b65} and \cite{b69})? Are alternative categories required for algorithms that obtain word embeddings through matrix factorization (e.g., \cite{b84}) or leverage clustering (e.g., \cite{b60}) or game theory (e.g., \cite{b56})?

These corner cases illustrate that the boundaries between the three main categories are often blurred. Hence, Fig. \ref{Figure 4} should be regarded as a convenient abstraction. Readers interested in a more meticulous classification may consult \cite{b51}, an AKE--specialized review paper. For insights into relatively recent trends, readers may also consult \cite{b49}.

\medskip

Readers interested in specific frameworks or parameters of the selected AKE algorithms may consult the original references. For brevity, the following paragraphs provide only high--level summaries of the algorithms selected based on whether they significantly reflect their category characteristics, whether their Python implementations are available, and whether their reported performances are competitive \cite{b49, b51}, \cite{b81, b82}, \cite{b85, b86}. Parameters associated with specific descriptions of the algorithms have been denoted within parentheses with distinct symbols, abbreviated as ``pa.'':

\medskip

\begin{itemize}
    \item For instance, ``KP--Miner boosts IDF \underline{(pa. $\sigma$ and $\alpha$)}.'' 
\end{itemize}

\medskip

\noindent\textbf{Term Frequency (TF) and TF--Inverse Document Frequency (TF--IDF)}. TF is solely concerned with how many times a term occurs in a document. That is, TF regards a higher frequency of a term as a \emph{necessary and sufficient} condition for keyness. Given that using only TF may extract common terms that cannot represent local topics in a document, TF--IDF \cite{b23} incorporates Inverse Document Frequency (IDF) to counterbalance TF. IDF penalizes common terms and compensates for terms that are rare across a textual dataset yet frequently occur within certain subsets of the dataset. The philosophy underlying TF--IDF is to view both TF and IDF as the \emph{necessary} conditions for keyness.

\medskip

\noindent\textbf{KP--Miner} \cite{b61} introduces four improvements to TF--IDF. First, for a document, KP--Miner does not regard certain terms as candidates if they occur fewer times than a specified ``least allowable seen frequency (pa. \emph{lasf})'' \cite{b61}, or first occur after a specified number of words (pa. \emph{cutoff}). Second, KP--Miner prevents the underestimation of multigram terms. Third, it boosts IDF (pa. $\sigma$ and $\alpha$). Fourth, it weights terms that first occur in early positions (pa. $\text{P}_{f}$). In short, KP--Miner believes not only unigrams but also multigrams possess keyness if they occur above a threshold at the beginning of a document, where key information can be concentrated.

\medskip

\noindent\textbf{YAKE!} \cite{b58} rewards a term when it occurs more frequently than average, in many sentences, in early sentences, or in capitalized forms. Conversely, YAKE! penalizes a term when its left [right] neighbors are diverse, indicating a higher likelihood of being a common term (pa. $\text{window}$), or when its unigrams do not form a meaningful multigram but rather end or begin with a stopword (e.g., ``doctor \emph{of}'' instead of ``doctor \emph{of} philosophy''). Given that YAKE! collects all these features, it marks a singularity that statistics--based AKE algorithms encountered limitations in adhering to frequentism.

\medskip

\noindent\textbf{Selected Graph--based Algorithms}. TextRank \cite{b62}, a pioneer of graph--based algorithms, builds a graph representing lexical units as vertices. The graph links two vertices if they co--occur in at least one sliding window frame\footnote{Given a document ``GPUs go brrr'' and a specified window size of $2$, the sliding window frames are ``\textbf{[GPUs go]} brrr'' and ``GPUs \textbf{[go brrr]}.''} (pa. $\text{window}$). Then, TextRank runs PageRank \cite{b87} that recursively updates each vertex's score until it converges (pa. $\delta$). PageRank rewards a vertex if it has \emph{many edges} or is linked to such \emph{influential} vertices. Only vertices whose final scores rank in the top $T$ percent are retained (pa. $T$). Each candidate's score is determined by summing its constituent vertices' scores.

SingleRank \cite{b63} resembles TextRank but differs in that edge weights are determined by \emph{co--occurrence frequencies} of lexical units across sliding window frames rather than being binary. PositionRank \cite{b71} also resembles SingleRank but differs in that PositionRank rewards a vertex if it occurs in \emph{early positions} in a document. In short, all three algorithms believe influential terms possess keyness, but SingleRank and PositionRank believe such terms are frequently co--occurring terms. PositionRank further believes such terms should occur at the beginning of a document.

\medskip

\begin{figure}[h!]
    \centering
    \includegraphics[width=\linewidth]{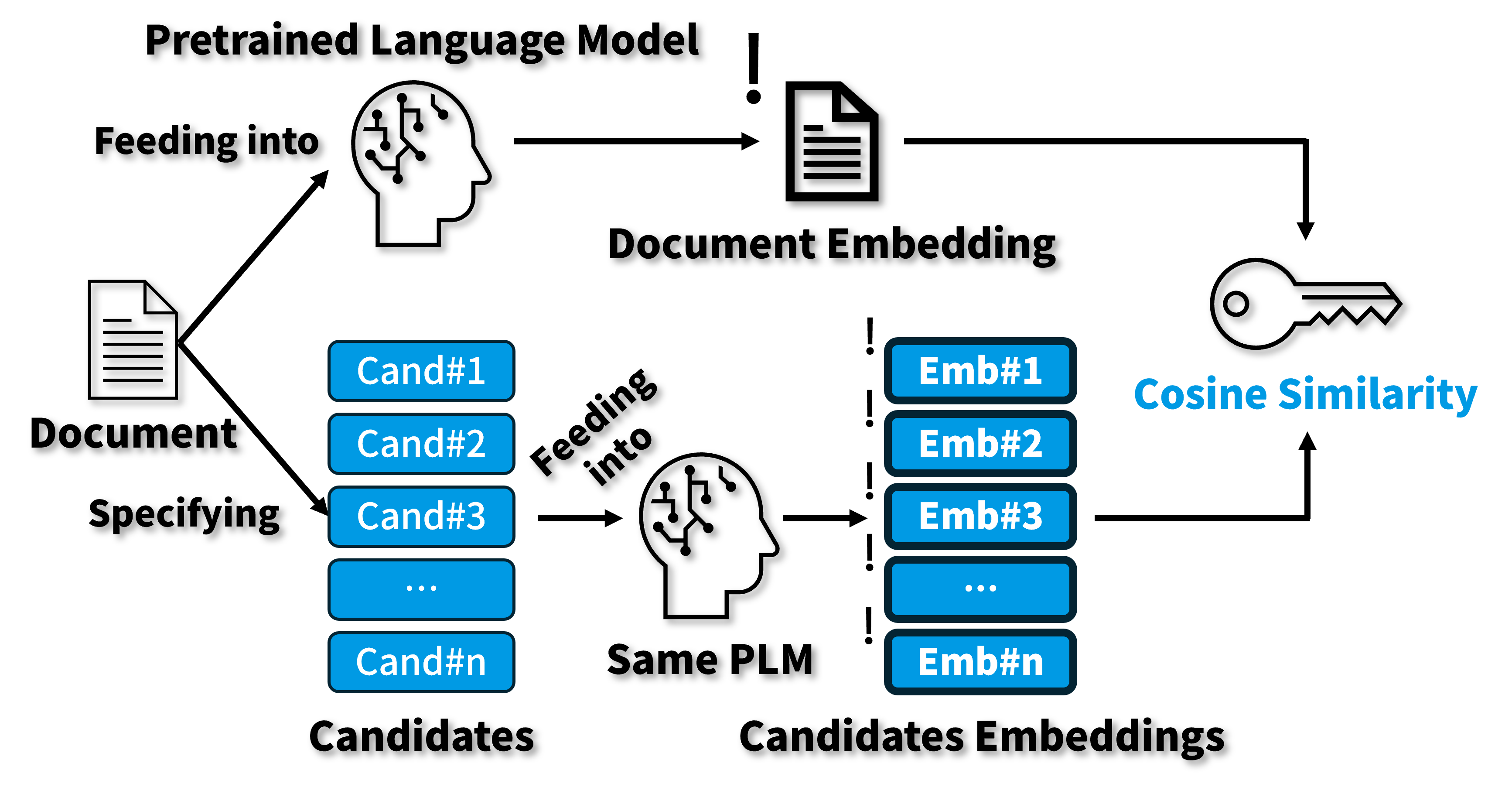}
    \caption{\textbf{Overall Workflow of KeyBERT.}}
    \label{Figure 5}
\end{figure}

\noindent\textbf{Selected NN--based Algorithms}. Firstly, KeyBERT (Fig. \ref{Figure 5}) \cite{b80} is a very simple NN--based algorithm.

\noindent It feeds a document and corresponding candidate terms into a PLM, thereby obtaining a document embedding and candidate embeddings. Then, it computes the cosine similarity between the document embedding and each candidate embedding, selecting candidates with the \emph{highest similarity} values as keyphrases. ``BERT'' \cite{b88} in KeyBERT's name is not necessary. While a lightweight BERT variant is typically employed as a PLM backbone in KeyBERT, a recent State--Of--The--Art (SOTA) model such as harrier--oss--v1--27b \cite{b89} can also be employed, provided that a high--end GPU is available.

\begin{figure}[h!]
    \centering
    \includegraphics[width=\linewidth]{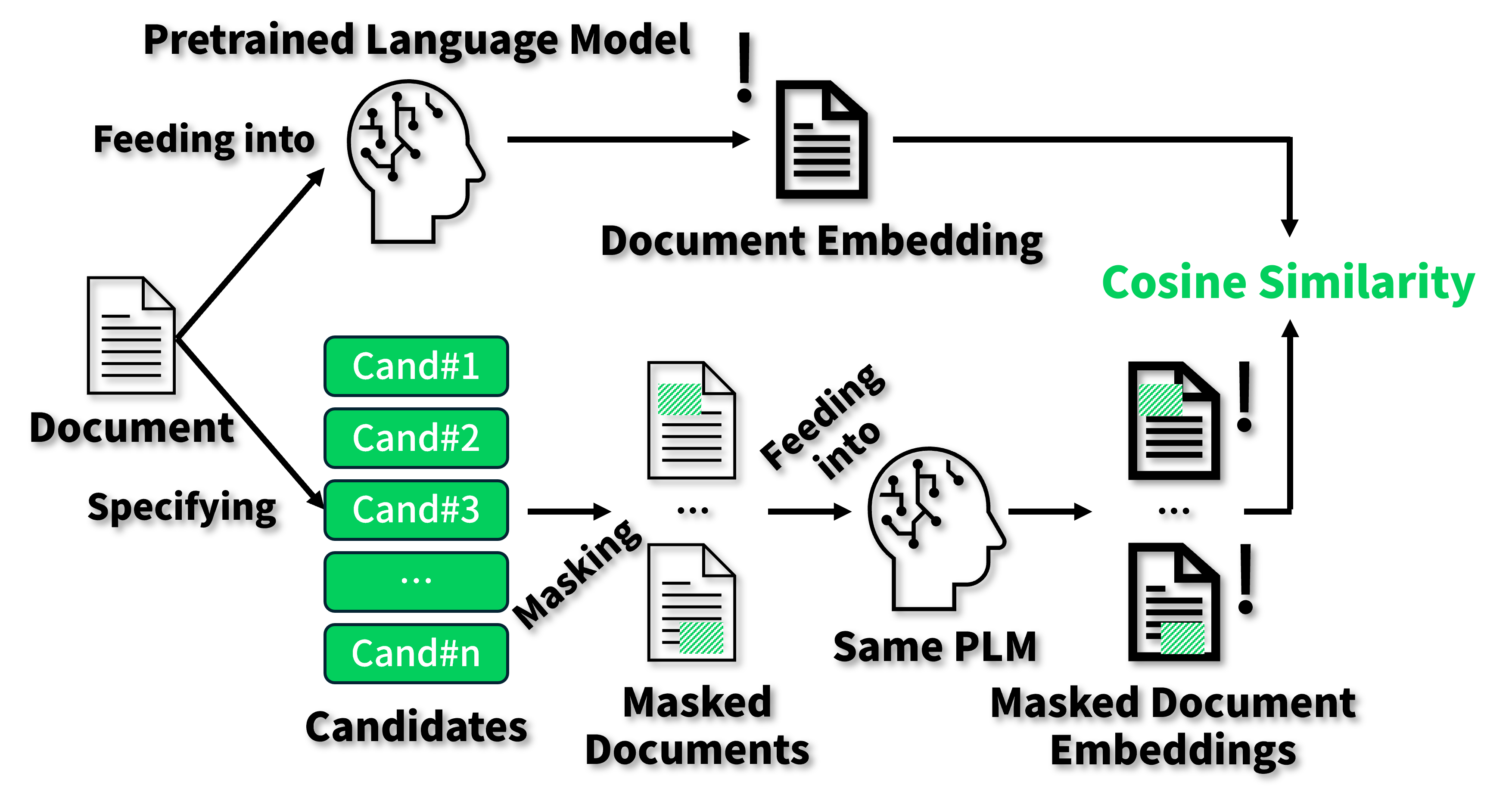}
    \caption{\textbf{Overall Workflow of MDERank.}}
    \label{Figure 6}
\end{figure}

In contrast, MDERank (Fig. \ref{Figure 6}) \cite{b59} believes direct similarity computation between a document and each candidate isolates candidates from the document's full context. Hence, MDERank computes the similarity between the embedding of the original document and each embedding of a masked version of the document, in which the corresponding candidate is masked with the special token \texttt{[MASK]}. Here, MDERank uses lower--is--better scoring, given that if a candidate is a keyphrase and its masking causes a significant loss of the \emph{full context}, the similarity score decreases.

Meanwhile, LMRank \cite{b81} resembles KeyBERT but differs in several aspects. First, LMRank avoids employing a BERT variant, typically pretrained through Masked Language Modeling (MLM) \cite{b88}. Instead, LMRank employs MPNet \cite{b90} as an alternative PLM backbone. MLM can learn positional information of tokens, but fails to capture dependencies between masked tokens \cite{b88, b90}. Hence, MPNet believes a BERT variant struggles to adapt when encountering a complex textual context \cite{b90}. Second, although MPNet has a shorter input limit of 384 tokens \cite{b90} compared with the typical 512--token limit of BERT variants \cite{b88}, LMRank circumvents this constraint by segmenting a long document and computing its final embedding through average pooling.

Third, LMRank runs meticulous preprocessing to select appropriate candidates, such as syntactic dependency parsing between terms, removing trivial terms, restricting candidates to noun phrases (pa. $keeps\_noun\_adjs$), and handling overlapping terms by retaining only the most representative ones (pa. $deduplicate$). Fourth, LMRank can also boost similarity scores of candidates that occur in \emph{early positions} in a document (pa. $positional\_feature$ and $\mu$).

In short, these NN--based algorithms believe terms closely aligned with their document's \emph{context} within a high--dimensional embedding space possess keyness. 

\subsection{Edge Weighting}\label{subsec:edge weighting}

\medskip

\subsubsection{What is EW, and Why Does It Matter?}\label{subsubsec:what is ew, and why does it matter}

\medskip

\noindent\textbf{Edge Weighting (EW)}. We now suggest that peers approach EW, employed as the second stage in SNC, without considering its definition too abstractly. It \emph{literally} refers to assigning specific weights to edges while constructing a network. Depending on what kind of system a network represents, an edge weight can signify, for instance, ``the average number of interactions per day'' \cite{b15} between two employees in a workplace, or ``the number of meetings that two individuals jointly attended'' \cite{b16}. In SNC, an edge weight is confined to \emph{semantic relatedness} between two keyphrases (vertices).

\medskip

\noindent\textbf{Semantic Relatedness} is defined as a \emph{human--perceivable connection} between two \emph{concepts} \cite{b91, b92}. Herein, concepts subtly differ from words (terms)\footnote{In the field of NLP, terms, words, phrases, and lexical units often have blurry boundaries. This paper also uses them interchangeably.}. Concepts refer to \emph{word senses}, and a word (e.g., doctor) can convey different word senses (e.g., doctor as a Ph.D. versus a medical doctor) depending on \emph{semantic context} (e.g., academic versus medical) \cite{b93}--\cite{b95}. According to \cite{b92, b93}, humans can perceive two words (ultimately, two concepts) as connected if there exists at least one semantic or mere lexical relation between them, as detailed in Appendix \ref{Appendix B}. When such relations are somehow \emph{typed}, they are termed ``\emph{explicit}'' \cite{b92, b94, b95} relations.

Semantic similarity also exists, but it is a mere special aspect of semantic relatedness \cite{b91}, \cite{b94}--\cite{b97}. Criteria for distinguishing them are presented in Appendix \ref{Appendix B}. More relevantly, semantic relatedness can be \emph{measured} as a weight between $0$ and $1$, depending on the perceived degree of such a connection \cite{b95}. Here, semantic relatedness is also regarded as the opposite of semantic \emph{distance} \cite{b91, b93, b96, b97}\hypertarget{definition of ew}{.}

In short, EW in SNC is equated to \emph{measuring (estimating)} semantic relatedness weights between extracted keyphrases. Indeed, semantic relatedness is also defined as ``\emph{how much} connection humans perceive between two concepts'' \cite{b91}. Hereafter, it is referred to as relatedness.

\medskip

\noindent\textbf{Necessity in SNC}. A structure where only unlinked keyphrases float is a mere word cloud. Edges are essential to form a meaningful SN. However, even an SN representing only $50$ vertices can represent up to $\binom{50}{2}=1,225$ edges. Representing all such edges imposes an excessive cognitive load on an interpreter. Hence, only significant edges can be retained by omitting or visually thinning zero-- or low--weighted relatedness \cite{b2, b20}. Given that EW provides the criteria for such SN sparsification, it is necessary in SNC.

\medskip

\noindent\textbf{Maximization of Interpretability}. Seemingly, the objective of EW may be to maximize the alignment of estimated weights with golden relatedness values. Indeed, the literature (e.g., \cite{b91, b92, b94}--\cite{b96}) has established gold--standard--based evaluation protocols that typically compare estimates of relatedness measures with gold annotations using Pearson’s correlation coefficient or Spearman’s rank correlation \cite{b98}. Such annotations are typically confined to weights between \emph{predefined or generic} words for the ``intrinsic evaluation'' \cite{b95} of relatedness measures\hypertarget{many edges}{.}

In contrast, before AKE, which keyphrases are extracted remains \emph{veiled}. Even a small textual dataset easily reaches above $1,000$ candidates. Should a gold standard comprising $\binom{1000}{2} = 499,500$ weights then be prepared? Impractical. Hence, \emph{ClueNetwork} circumvents elusive gold edge weights by applying the presumption of innocence to qualified relatedness measures (Subsubsection \ref{subsubsec:separate penguin from orca}), and then evaluating whether they construct \emph{semantic--percolation--friendly} SN topologies \emph{under} their philosophies of relatedness. So, we define the objective of EW in SNC as ``maximization of interpretability'' (Subsubsection \ref{subsubsec:robustness improvement}).

\subsubsection{How Does EW Work?}\label{subsubsec:how does ew work}

\medskip

\noindent\textbf{Distributional Hypothesis}. Herein, ``applying the presumption of innocence to relatedness measures'' means that if some measures adhering to the distributional hypothesis \cite{b99, b100} ensure \emph{minimal veracity}, they are evaluated in terms of \emph{interpretability} rather than veracity. The hypothesis states that semantically close words tend to co--occurr in close contexts \cite{b92, b95, b96, b99, b100} and a word's sense is determined by its neighbors \cite{b95, b99, b100}. More relevantly, ``close contexts'' signify context windows (in this paper, documents), and more modernly, ``distributions'' signify vector representations of words (in this paper, $\mathbf{DTM}$ columns).

As some measures based on the hypothesis facilitate weighting word relatedness without considering explicit relations, they are also considered to capture ``\emph{implicit}'' word relations \cite{b94}. Although termed a ``hypothesis,'' the hypothesis and its descendants function as de facto \emph{axioms} in the field of NLP (e.g., \cite{b101}--\cite{b104}). \emph{ClueNetwork} also accepts it as such, questioning the selected EW measures: ``Do you follow the distributional hypothesis?'' (Subsubsection \ref{subsubsec:separate penguin from orca})

\medskip

\begin{figure}[b!]
    \centering
    \includegraphics[width=\linewidth]{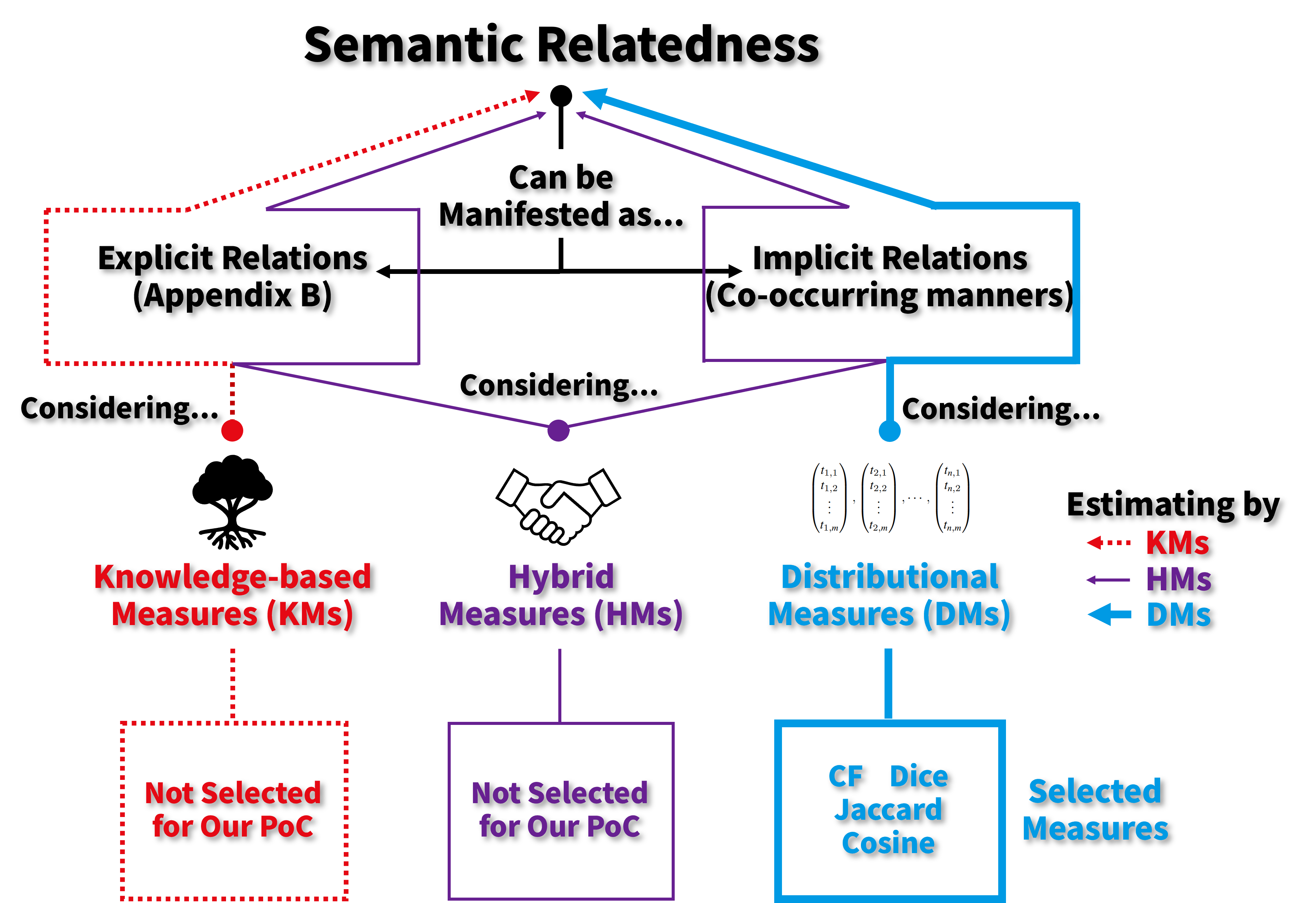}
    \caption{\textbf{Simplified Taxonomy of Semantic Relatedness Measures.}}
    \label{Figure 7}
\end{figure}

\noindent\textbf{Beliefs as Ingredients and Simplified Taxonomy of Semantic Relatedness Measures}. Looking across the literature \cite{b91}--\cite{b97}, there are three main sides of semantic relatedness measures (Fig. \ref{Figure 7}). First, the knowledge--based side estimates relatedness by exploiting some knowledge bases \cite{b93, b95}, in which \emph{explicit} word relations are encoded \cite{b92}. Next, the distributional side is precisely the side following the \emph{distributional hypothesis} \cite{b91, b92, b95}.

Finally, the hybrid side believes estimating relatedness requires considering not only explicit relations but also \emph{implicit} relations favored by the distributional side \cite{b95, b96}. Readers interested in which measures belong to these sides may consult the specialized review papers \cite{b92, b93}, \cite{b95}--\cite{b97}.

A more relevant point is that, in terms of SNC, a relatedness measure $x$ as ``some putative agent'' \cite{b9} also injects beliefs as ingredients into an SN, such as ``Given their relations, the degree of connection between terms $a$ and $b$ is $0.7$.'' That is, $x$ judges the world to be one where the edge weight is $0.7$ depending on its philosophy of relatedness. 

For theoretical impartiality, this paper avoids definitive commentary on the \emph{general} performances of these sides. They can be better or worse depending on downstream tasks. Nonetheless, only several simple distributional measures have been selected for the PoC of \emph{ClueNetwork}. As discussed in the case of the square circle, target datasets for SNC are largely \emph{context--dependent}. We have a concern that the other sides consider more generic explicit relations. Given that the distributional side is adaptive to domains, languages, and lexicons \cite{b92, b96}, it appears suitable for SNC.

\medskip

\noindent\textbf{Selected EW Measures}. The selected semantic relatedness measures, hereafter referred to as EW measures, are presented in Table \ref{table:Table 4}. Given that these are well--established simple measures, we believe readers can sufficiently grasp their underlying philosophies by examining the presented formulas and components. Accordingly, only aspects most relevant to the remainder of this paper are briefed here.

Co--occurrence Frequency (CF) believes the larger the \emph{intersection} of individual document sets in which two keyphrases occur, the greater the relatedness. Dice Coefficient (Dice) \cite{b105} and Jaccard Index (Jaccard) \cite{b106} are similar to CF, yet Dice and Jaccard apply \emph{normalization} denominators to avoid the scaling effect of individual set sizes. Because Dice and Jaccard share a \emph{monotonically increasing} relationship, ranks of their resulting adjacency matrix entries are always identical for an SN. This paper refers to CF, Dice, and Jaccard as \emph{discrete (frequency--based)} measures.

Minkowski Distance \cite{b98} measures a point--to--point distance between two keyphrases in a vector space derived from a $\mathbf{DTM}$. Among infinite forms of Minkowski, this paper considers only the popular Euclidean Distance (Euclidean). Cosine Similarity (Cosine) \cite{b98} believes the larger the dot product between two keyphrase vectors, the greater their relatedness. Here, when an entry in either vector at any dimension is $0$, that dimension contributes $0$ to the dot product.

Document Mover's Distance (DMD) is a variant of Earth Mover's Distance (EMD) \cite{b107}. EMD is depicted as the minimum total cost of transporting dirt piles from a supply area to a demand area \cite{b108}. Likewise, DMD is defined as the minimum total cost of moving the document--level scores of one keyphrase vector to another. Such a cost is derived by identifying the optimal transport plan matrix, in which individual flow amounts are defined. In short, DMD believe the lower the semantic transport cost between two keyphrases, the greater their relatedness. This paper refers to Minkowski, Cosine, and DMD as \emph{vector--based} measures.

\begin{table*}[t!]
\caption{Formulas of Selected Edge Weighting Measures}\label{table:Table 4}
\centering
\renewcommand{\arraystretch}{2}
\begin{tabular}{>{\arraybackslash}m{4cm}|>
{\arraybackslash}m{6.9cm}|>{\arraybackslash}m{5.5cm}}
\hline
Measure & Formula & Meanings of Components \\
\hline\hline
Co--occurrence Frequency & $\text{CF}(t_{i},t_{j})=|\mathbf{O}(t_{i}) \cap \mathbf{O}(t_{j})|$ & \begin{itemize}
    \item $t_{y}$: $y$\textsuperscript{th} term.
    \item $\mathbf{O}(t_{y})$: Set of windows where $t_y$ occurs.
    \item $|\dots|$: Size of a given set.
\end{itemize} \\ \hline
Dice Coefficient \cite{b105} & $\text{Dice}(t_{i},t_{j})=\frac{2|\mathbf{O}(t_{i}) \cap \mathbf{O}(t_{j})|}{|\mathbf{O}(t_{i})| + |\mathbf{O}(t_{j})|}$ & \begin{itemize}
    \item As defined in Co--occurrence Frequency.
\end{itemize} \\ \hline
Jaccard Index \cite{b106} & $\text{Jaccard}(t_{i},t_{j})=\frac{|\mathbf{O}(t_{i}) \cap \mathbf{O}(t_{j})|}{|\mathbf{O}(t_{i}) \cup \mathbf{O}(t_{j})|}$ & \begin{itemize}
    \item As defined in Co--occurrence Frequency.
\end{itemize} \\ \hline
Minkowski Distance \cite{b98} & $\text{Minkowski}(\mathbf{t}_{i},\mathbf{t}_{j})=\bigg[\sum_{k=1}^{m}|t_{i,k}-t_{j,k}|^{p}\bigg]^{1/p}$ & \begin{itemize}
    \item $\mathbf{t}_y$: $y$\textsuperscript{th} column vector of a given $\mathbf{DTM}$.
    \item $m$: The number of rows of the $\mathbf{DTM}$.
    \item $t_{y,k}$: $k$\textsuperscript{th} entry of $\mathbf{t}_y$.
    \item $|\dots|$: Absolute value of a given scalar.
    \item $p$: When $p = 1$, this measure corresponds to the Manhattan distance (cf. $\text{L}_{1}$ norm), and when $p = 2$, this measure becomes the Euclidean distance (cf. $\text{L}_{2}$ norm).
\end{itemize} \\ \hline
Cosine Similarity \cite{b98} & $\text{Cosine}(\mathbf{t}_{i},\mathbf{t}_{j})=\frac{\mathbf{t}_{i}\cdot\mathbf{t}_{j}}{\Vert \mathbf{t}_{i} \Vert_{2} \Vert 
\mathbf{t}_{j} \Vert_{2}}$ & \begin{itemize}
    \item $\mathbf{t}_{i}\cdot\mathbf{t}_{j}$: Dot product between $\mathbf{t}_{i}$ and $\mathbf{t}_{j}$.
    \item $\Vert \dots \Vert_{2}$: $\text{L}_{2}$ norm of a given vector.
\end{itemize} \\ \hline
\multirow{4}{4cm}{Document Mover's Distance \cite{b107}} & & \multirow{4}{5.5cm}{\begin{itemize}
    \item $\mathbf{P}$: Transportation plan matrix.
    \item $\mathbf{P}\geq0$: All entries of $\mathbf{P}$ are non--negative.
    \item $\mathbf{P}_{kl}$: Flow amount from $t_{i,k}$ to $t_{j,l}$.
    \item $\mathbf{d}_x$: $x$\textsuperscript{th} row vector of the $\mathbf{DTM}$.
    \item $\text{Euclidean}(\mathbf{d}_{k}, \mathbf{d}_{l})$: Physical transport distance from $t_{i,k}$ to $t_{j,l}$.
\end{itemize}} \\
& $\text{DMD}(\mathbf{t}_{i}, \mathbf{t}_{j})=\underset{\mathbf{P}\geq0}{\text{min}}\sum_{k,l=1}^{m}\bigg[\mathbf{P}_{kl}\cdot\text{Euclidean}(\mathbf{d}_{k}, \mathbf{d}_{l})\bigg],$ & \\ 
& $\text{subject to } \sum_{l=1}^{m}{\mathbf{P}_{kl}}=t_{i,k} \text{ and } \sum_{k=1}^{m}{\mathbf{P}_{kl}}=t_{j,l} \;\ \forall \;\ k,l$ & \\ 
& & \\ \hline
\end{tabular}
\end{table*}

To adhere to the higher--is--closer convention, although Euclidean and DMD are termed ``distances,'' \emph{ClueNetwork} normalizes their resulting values by ``a suitable inverse function'' \cite{b93} $1-\text{dist}(\mathbf{t}_i, \mathbf{t}_j) / \text{max}(\text{dist}(\mathbf{t}_x, \mathbf{t}_y))$. Herein, $\text{dist}(\mathbf{t}_i, \mathbf{t}_j)$ is the distance between two keyphrases. Meanwhile, although the range of Cosine is between $-1$ and $1$, in \emph{ClueNetwork}, Cosine yields only non--negative edge weights within a $0$ to $1$ range, as all $\mathbf{DTM}$ entries are non--negative (cf. \eqref{Eq 1}). 

\medskip

\noindent\textbf{Construction of 1--mode Matrix}. How, then, do the selected measures specifically work within SNC? When constructing a raw SN, a Term--Term Matrix ($\mathbf{TTM} \in \mathbb{R}^{l \times l}$) whose rows and columns are both dimensioned by $l$ keyphrases selected based on \eqref{Eq 2} for SN sparsification, is built first. Here, the vector--based or discrete measures intervene in distinct manners, respectively. When undertaking such a task with a vector--based measure, $\mathbf{DTM}$ columns are first regarded as vector representations of keyphrases. Then, the measure estimates relatedness weights between these vectors. It results in a $\mathbf{TTM}$ whose entries are weights of possible edges.

When constructing such an adjacency matrix $\mathbf{TTM}$ using a discrete measure, a $\mathbf{DTM} \in \mathbb{R}^{m \times l}$ is first converted into a binary matrix $\mathbf{B}$ whose entries simply indicate whether each keyphrase is present ($1$) or absent ($0$) in each document. Then, a raw $\mathbf{TTM}$ is built by computing $\mathbf{B}^{\top}\mathbf{B}$ whose entries are CFs between keyphrases. Finally, these entries are updated by applying the measure, thereby resulting in the final $\mathbf{TTM}$.

Meanwhile, whether directionality is assigned to edges is also a consideration. When representing an undirected SN, a symmetrical $\mathbf{TTM}$ is first built, and then keyphrases are linked with undirected edges derived from either the Strictly Upper Triangular Matrix ($\mathbf{SUTM}$) or the Strictly Lower Triangular Matrix ($\mathbf{SLTM}$). When representing a directed SN, both matrices are used, as they encode distinct directions. Whether undirected or directed, the main diagonal (i.e., loop edges) is typically excluded from SNC. In practice, undirected SNs are more widely used \cite{b19} due to a few constraints (Appendix \ref{Appendix C}). This paper also focuses on undirected SNs, and potential extension to directed SNs is left for future work.

\subsubsection{Separate Penguin from Orca}\label{subsubsec:separate penguin from orca}

\medskip

\noindent\textbf{Distributional Hypothesis Revisited}. Although Minkowski is familiar and DMD appears calculative, we should avoid jumping to the next stage too hastily. Let us suppose that a swimming penguin is positioned at $(0, 0.5)$ on a maritime coordinate plane, and an orca at $(0.5, 0)$ spots the penguin.

\begin{figure}[h!]
    \centering
    \includegraphics[width=\linewidth]{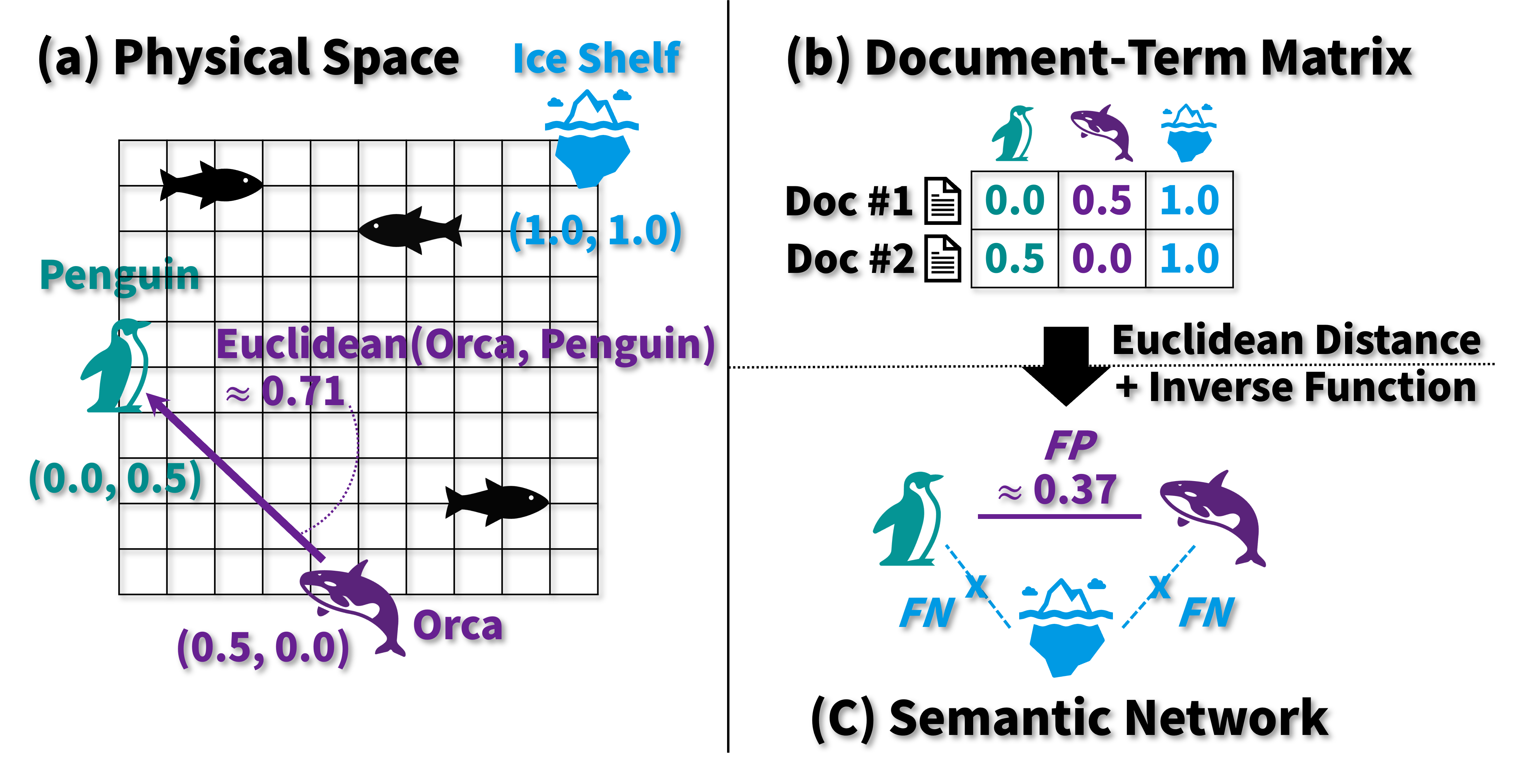}
    \caption{\textbf{Behaviors of Euclidean Distance in Different Worlds.}}
    \label{Figure 8}
\end{figure}

Given that $(0, 0)$ is a relative origin in that physical space, we can \emph{meaningfully} state that their Euclidean distance is approximately $0.71$ (Fig. \ref{Figure 8}--a). In contrast, in a $\mathbf{DTM}$ (Fig. \ref{Figure 8}--b), zero signifies that a keyphrase does not occur in a document. Given that there are no documents in which \emph{penguin} and \emph{orca} co--occur (i.e., \emph{penguin} is safe), the edge weighted by Euclidean between \emph{penguin} and \emph{orca} is a false positive (Fig. \ref{Figure 8}--c). Given that \emph{ice shelf} co--occurs with both \emph{penguin} and \emph{orca} once, the edges omitted by Euclidean are false negatives.

Such false edges certainly violate not only intuition but also the distributional hypothesis. What causes such violations? It is the component $|t_{i,k}-t_{j,k}|^{p}$ in Minkowski. It yields a non--zero value even when only one of $t_{i,k}$ and $t_{j,k}$ has a small non--zero value. DMD incorporating Euclidean \cite{b107} is also not free from this problem. That is, Minkowski is vulnerable to noise in spaces (particularly, sparse high--dimensional spaces) where zero is \emph{meaningless}. This vulnerability has been theoretically \cite{b109} and empirically \cite{b109, b110} proven. Readers interested in it may consult Appendix \ref{Appendix D} or \cite{b109, b110}.

Indeed, Euclidean or EMD are more commonly considered in the field of computer vision \cite{b12, b13, b107, b108} rather than NLP. Why? In dense 2-- or 3--dimensional spaces, zero can be \emph{meaningful} \cite{b109, b110}. There are countless measures in the world. Manually confirming which ones are suitable for SNC can be challenging for peers. Hence, \emph{ClueNetwork} incorporates the following pretest to question candidate measures: ``Do you follow the distributional hypothesis?''

\medskip

\noindent\textbf{Minimal Veracity Pretest}. Its basic concept is to filter out liar measures by quantifying anti--false edge abilities of candidate measures. Such an ability is defined as the Matthews Correlation Coefficient \cite{b26, b11} of an EW measure:

\begin{equation} \label{Eq 3}
\begin{array}{r}
\text{MCC} = \frac{\text{TP}\times\text{TN}-\text{FP}\times\text{FN}}{\sqrt{(\text{TP}+\text{FP})(\text{TP}+\text{FN})(\text{TN}+\text{FP})(\text{TN}+\text{FN})}}.
\end{array}
\end{equation}

\medskip

\noindent Herein, for a $\mathbf{DTM}$, ``$\text{TP}$ (True Positive)'' is the number of $\mathbf{TTM}$ entries correctly set as non--zero; ``$\text{TN}$ (True Negative)'' is the number of $\mathbf{TTM}$ entries correctly set as zero; ``$\text{FP}$ (False Positive)'' is the number of $\mathbf{TTM}$ entries falsely set as non--zero; ``$\text{FN}$ (False Negative)'' is the number of $\mathbf{TTM}$ entries falsely set as zero.

For a textual dataset, when two keyphrases do not co--occur in any document, a \emph{veracious} measure must set their corresponding $\mathbf{TTM}$ entry to zero. When they co--occur in at least one document, the measure must set the entry to a non--zero value. Such passers attain the maximum $\text{MCC}$ of $1$, whereas failures attain less than $1$ and are filtered out. When the denominator of $\text{MCC}$ is zero, special criteria are applied:

Specifically, when all ground truths (actual co--occurrences) are non--zero (or zero) and predictions perfectly match them, $\text{MCC}=1$ is legitimately assigned to a given measure. In contrast, when the measure sets all entries as non--zero (or zero) regardless of actual co--occurrences, $\text{MCC}=0$ is assigned, given its lack of discriminative power.

Why is the choice of MCC justified? MCC penalizes or rewards all components of the confusion matrix \cite{b112}. Thereby, it is suitable for detecting bad measures that link \emph{penguin} with \emph{orca} or separate both from \emph{ice shelf}. Here, the well--established MCC is certainly not our creation. However, the idea of applying it to verify whether EW measures adhere to \emph{reality} (i.e., the distributional hypothesis) constitutes a touch of novelty. In Subsubsection \ref{subsubsec:results of ew}, this pretest is performed.

\subsection{Community Detection}\label{subsec:community detection}

\medskip

\subsubsection{Why Does CD Matter, and What is it?}\label{subsubsec:why does cd matter, what is it}

\medskip

\noindent\textbf{Necessity in SNC}. Whether Community Detection (CD) is essential for SNC can be an open question. About 37\% of applications are satisfied with \emph{raw} SNs \cite{b19}. However, when an interpreter encounters a raw SN with many vertices and edges in its unpartitioned state, a reaction like ``Let me see. \textbf{Uhhh...} how am I supposed to interpret this?'' can escape. Here, social scientists often view a community in an SN as something akin to a ``topic'' \cite{b113, b114} in topic modeling that \emph{facilitates} interpretation \cite{b2}. Computer scientists also often believe that a community reveals its members and their interactions \emph{distinct} from the outside \cite{b115}. Accordingly, \emph{ClueNetwork} leverages CD to alleviate such uhhh moments\hypertarget{definition of cd}{.}

\medskip

\noindent\textbf{Community Detection (CD)}. For a network $\mathbf{G}(\mathbf{V}, \mathbf{E})$, where $\mathbf{V}$ and $\mathbf{E}$ represent the sets of vertices and edges, respectively, a community $U_{i} \in \text{the set of communities } \mathbf{U}$ is a group whose member vertices ``satisfy the \emph{condition}'' \cite{b115} of having stronger (denser) ``intra--group'' \cite{b117} connections and weaker (sparser) ``inter--group'' \cite{b117} connections \cite{b21, b115, b116, b117}. And CD is ``to design a mapping'' \cite{b116} (or ``partition'' \cite{b118}) that regards every $v_{j} \in \mathbf{V}$ as a member of at least\footnote{``At least'' indicates that certain works assume overlapping communities. For convenience, the current PoC assumes that $U_i \cap U_j = \emptyset \;\ \forall \;\ i \neq j$.} one community \cite{b116}\hypertarget{definition of max distinctiveness}{.}

\medskip

\noindent\textbf{Maximization of Distinctiveness}. That is, satisfying the above condition across all $v_{i} \in \mathbf{V}$ is equated to the objective of CD. Many criteria, termed Community Scoring Functions (CSFs), for evaluating such objective achievement have been accumulated \cite{b119}. In applications like SNC where gold--standard communities are elusive, internal CSFs such as Triad--Participation--Ratio (TPR) \cite{b120}, Conduction \cite{b121}, or Modularity $Q$ \cite{b21, b24} can be considered \cite{b115}.

As these CSFs interpret the objective in different ways, \cite{b120} first classified such objective achievement into four properties (separability, density, cohesiveness, and clustering coefficient) for gold--standard--based meta--evaluation. Then, \cite{b120} demonstrated that while TPR aligned best with density, cohesiveness, and clustering coefficient, and Conduction best with separability, Modularity $Q$ did not. For \emph{brevity}, these four properties are hereafter \emph{collectively} referred to as distinctiveness --- as the objective ultimately says that communities should be somehow distinct. Accordingly, the objective is also referred to as the maximization of distinctiveness.

Meanwhile, $Q$ is used not only as a CSF but also as an objective function that certain CD algorithms aim to maximize. \cite{b122} pointed out that when a network is sufficiently large, a partition yielding a higher $Q$ may merge genuinely distinct small communities into a single larger community. Hence, \cite{b122} proposed the Constant Potts Model (CPM) as an alternative objective function. Readers interested in specific definitions of TPR, Conduction, the four properties, or CPM may consult the original references\hypertarget{justification of q}{.}

\medskip

\noindent\textbf{Material Is All We Need}. In short, we already know $Q$ is not the most ideal CSF. We nevertheless have employed it in the PoC of \emph{ClueNetwork}. Why? The PoC is just a PoC. Its true aim is to demonstrate that if a researcher chooses a stage--specific evaluation criterion, SNC pipelines can be ranked \emph{within} its rule. We have only employed $Q$ as material to achieve this very aim. That is, while considering alternatives is important in future applications, it is outside the current scope. Nonetheless, we considered using the combination of TPR and CPM. Given that some implementations of the selected modularity--based algorithms support $Q$ but not CPM, we have fixed the criterion to $Q$ for a fair comparison.

\medskip

\noindent\textbf{Modularity} $-1 \leq Q \leq 1$ \cite{b21, b24} can be \emph{mathematically} defined as the basic form \eqref{Eq 4}--a or the long form \eqref{Eq 4}--b:

\begin{equation} \label{Eq 4}
\begin{array}{r}
Q=\frac{1}{2m}\sum_{ij}\left[ \mathbf{A}_{ij}-\frac{k_{i}k_{j}}{2m} \right]\delta(c_{i},c_{j}) \cdots(a) \\ \\
= \sum_{l}\left[\frac{\sum_{ij}{\mathbf{A}_{ij}\delta(c_{i},l)\delta(c_{j},l)}}{2m}-\frac{\sum_{i}k_{i}\delta(c_{i},l)}{2m}\frac{\sum_{j}k_{j}\delta(c_{j},l)}{2m}\right]\cdots(b) 
\end{array}
\end{equation}

\medskip

\noindent For an undirected weighted graph $\mathbf{G}$, the meanings of the components in \eqref{Eq 4}--b are also presented in Table \ref{table:Table 5}.

\begin{table}[h!]
\caption{Meanings of Components in \eqref{Eq 4}}\label{table:Table 5}
\centering
\begin{tabular}{>{\arraybackslash}m{1.3cm}|>
{\arraybackslash}m{6.3cm}}
Component & Meaning \\
\hline\hline
$l$ & Individual community within $\mathbf{G}$. \\ \hline
$m$ & Total sum of edge weights in $\mathbf{G}$. \\ \hline
$\mathbf{A}$ & Adjacency matrix of $\mathbf{G}$. \\ \hline
$\delta$ & Kronecker delta function that returns $1$ if two communities are identical, and $0$ otherwise. \\ \hline
$k_{i}$ & Total weight of edges incident to vertex $v_{i}$. \\ \hline
$u_{i}$ & Index of the community to which $v_{i}$ belongs. \\
\end{tabular}
\end{table}

Readers encountering this formula likely fall into two groups. Those who are familiar with $Q$ may find nothing new. We kindly encourage them to skip to the next subsubsection. Those who have never encountered $Q$ in their respective fields may experience uhhh moments. But that is fine. The qualitative definition and the belief behind it are all you need. 

$Q$ is \emph{qualitatively} defined as the global sum of all local differences between the following components \cite{b21, b24}:

\medskip

\begin{itemize}
    \item the normalized \emph{actual} sum of intra--community edge weights of a community $U_{l}$ under a \emph{candidate} partition
    \item the normalized \emph{expected} sum of intra--community edge weights of $U_{l}$ under a \emph{null} model.
\end{itemize}

\medskip

\noindent Herein, the null model is a hypothetical model if all edges in $\mathbf{G}$ were randomly rewired while retaining the degrees $k_{i}$ and $k_{j}$ as well as the community indices $u_{i}$ and $u_{j}$ for all vertex indices $i$ and $j$. That is, the belief behind $Q$ is that a candidate partition is deemed more \emph{distinct} when its intra--community edges are significantly denser than those of a null model whose edges are connected without any strategy.

Why is such a null model necessary? In the absence of a gold--standard partition, a \emph{baseline} is essential for comparison to justify why one partition is superior to another. 

\subsubsection{How Does CD Work?}\label{subsubsec:how does cd work}

\medskip

\noindent\textbf{Belief as Ingredients and Simplified Taxonomy of CD Algorithms}. According to \cite{b117}, four main methodological streams have contributed to CD over the last two decades: 

First, modularity--based algorithms are precisely those that largely formulate CD as a problem to \emph{optimize} $Q$ or its alternatives (e.g., CPM \cite{b122}). Second, spectral clustering methods perform CD by exploiting \emph{spectral properties} (i.e., eigenvalues and eigenvectors) of the graph Laplacian matrix derived from a given adjacency matrix.

\begin{figure}[h!]
    \centering
    \includegraphics[width=\linewidth]{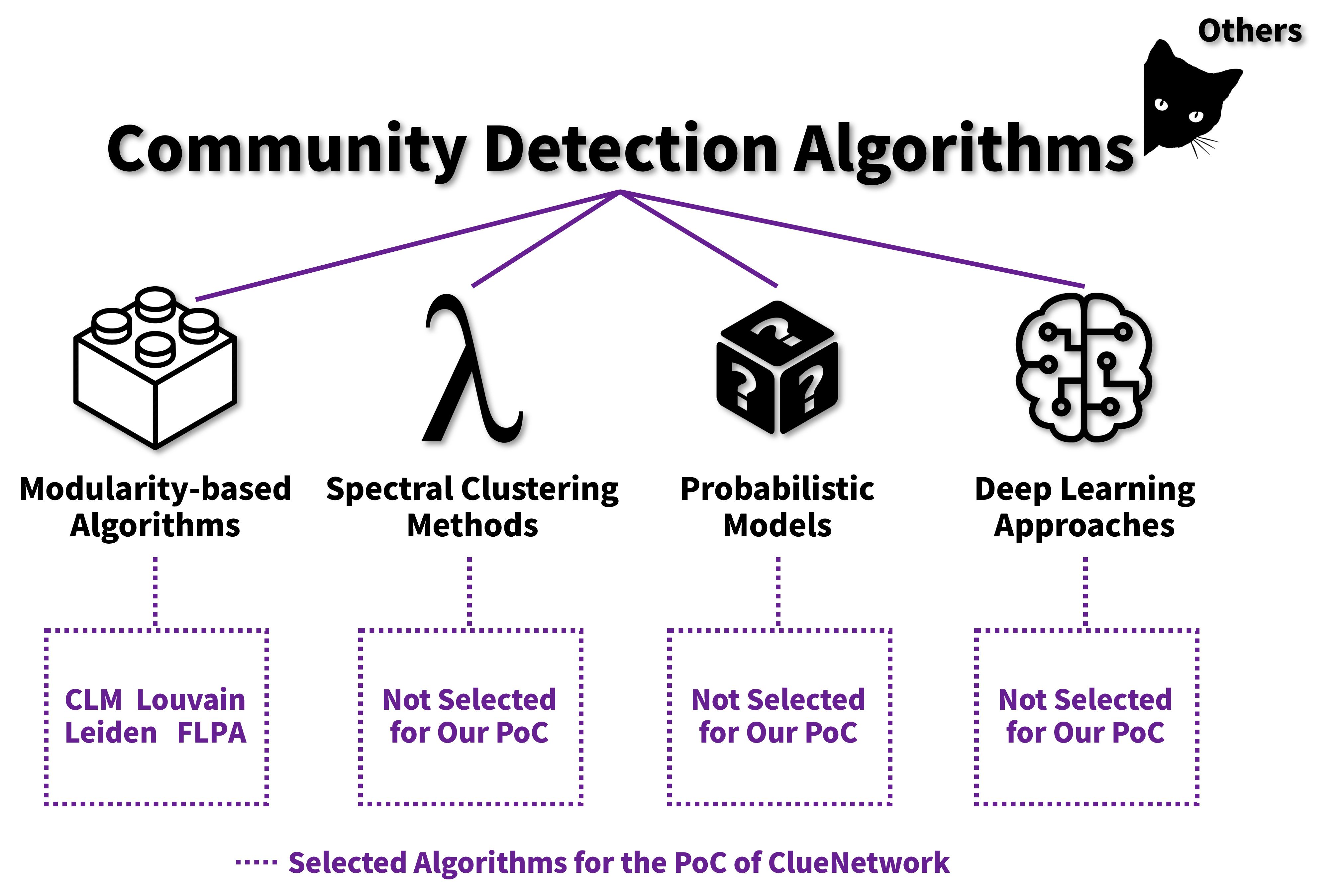}
    \caption{\textbf{Simplified Taxonomy of Community Detection Algorithms.}}
    \label{Figure 9}
\end{figure}

Third, probabilistic models fundamentally regard a network as an uncertain system. They also assume that if certain vertices belong to the same community, such vertices are more likely to be connected. Hence, they aim to uncover the underlying structure (i.e., community structure) of such uncertainty by \emph{estimating} latent variables and parameters from observed variables and their probability distributions.

Fourth, deep learning approaches leverage the capability of deep neural networks to learn patterns in complex data structures (e.g., networks) and to represent their basic elements (e.g., vertices), dependencies (e.g., edges), and larger elements (e.g., subgraphs) as informative features. Communities are then identified by exploiting such \emph{features}.

In terms of SNC, a CD algorithm $x$ as ``some putative agent'' \cite{b9} also injects beliefs as ingredients into an SN, such as ``a keyphrase $v_{i}$ belongs to a topic $U_{l}$.'' That is, $x$ judges the world to be one where $v_{i}$ belongs to $U_{l}$. Based on their respective philosophies of distinctiveness, CD algorithms hold such beliefs by focusing on $Q$, spectral properties, probabilistic patterns, or representational features. Readers interested in which algorithms belong to these streams may consult the comprehensive review paper \cite{b117}. Furthermore, \cite{b115} and \cite{b116} are excellent review papers that are specialized in deep learning--based or probabilistic CD algorithms.

\medskip

\noindent\textbf{Selected CD Algorithms}. The experiment in \cite{b117} suggests that no stream is clearly superior to the others. Peers may choose the streams with which they are most familiar. We have employed highly accessible modularity--based algorithms as \emph{shortcuts} to the PoC. As these are well--established, only their high--level summaries are provided here. We clarify that their beliefs regarding SN distinctiveness do not differ, as their shared objective function is its very interpretation. They differ only in \emph{how} they optimize such a function.

\begin{figure}[h!]
    \centering
    \includegraphics[width=\linewidth]{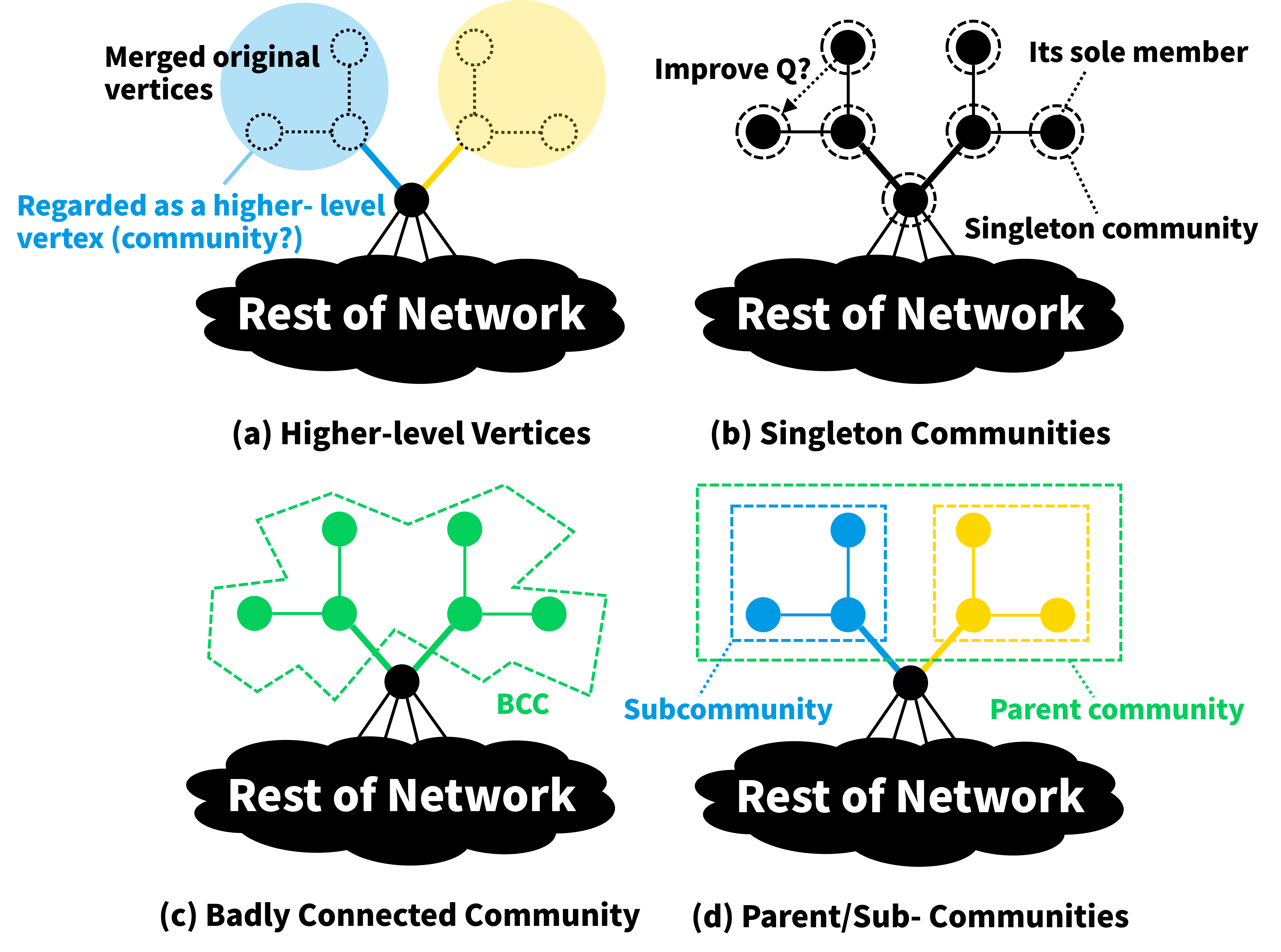}
    \caption{\textbf{Examples of Higher--level Vertices, Singleton Communities, Badly Connected Community, Parent Community, and Subcommunities.}}
    \label{Figure 10}
\end{figure}

The Clauset--Newman--Moore algorithm (CNM) \cite{b24} is an early \emph{greedy} algorithm. It selects a vertex merger (Fig. \ref{Figure 10}--a) that yields the best possible $Q$ change at each iteration, without considering long--term alternatives. Such merging is iterated until all original vertices are merged into the same community, after which the partition that yields the best $Q$ among all recorded partitions is selected as the final partition.

CNM's limitation is that original vertices are \emph{irreversibly} merged because CNM \emph{directly} regards higher--level vertices as communities (Fig. \ref{Figure 10}--a). Hence, at the initial step, the Louvain algorithm (Louvain) \cite{b25} regards each vertex as the sole member of a singleton community (Fig. \ref{Figure 10}--b) rather than the community itself. Vertices can then be moved to other communities \emph{multiple times} until no further improvement in $Q$ is possible. Only after such a phase does Louvain regard temporary communities as higher--level vertices (Fig. \ref{Figure 10}--a) to iterate the phase (Fig. \ref{Figure 10}--b), yielding the final partition.

The Leiden algorithm \cite{b118} cautions that Louvain's focus on vertex moves can generate badly connected communities whose certain members can only reach each other through paths extending outside their community (Fig. \ref{Figure 10}--c). After early vertex moves, Leiden refines a given temporary partition by allowing members of every parent community to form subcommunities (Fig. \ref{Figure 10}-d). Retaining the temporary partition, Leiden then regards subcommunities as higher--level vertices. Here, they can move to other parent communities.

Such moving and refining are iterated until no further improvement in $Q$ is possible, yielding the final partition. Leiden also uses a strategy during such refinement similar to the \emph{exploitation versus exploration} method in reinforcement learning \cite{b123}. That is, while moves yielding maximal improvements in $Q$ are preferentially exploited, those yielding mere improvements can also be randomly selected. Such randomness (pa. $\beta$) leaves room for uncovering optimal partitions that CNM or Louvain would otherwise miss.

Finally, although the Fast Label Propagation Algorithm (FLPA) \cite{b124} is categorized as a modularity--based algorithm \cite{b117}, it does not explicitly optimize any objective function. Instead, the quality of its outputs can be evaluated in terms of $Q$. FLPA operates on a highly intuitive mechanism. It iteratively reassigns community labels until the label of each vertex becomes the one \emph{most shared} among its neighbors.

\section{Local Evaluation}\label{sec:local evaluation}

\medskip

Section \ref{sec:theoretical foundation} has covered one fold of the subject matter through a review of SNs as clues and the three SNC stages. This section first defines and justifies the local evaluation criteria for ranking SNs (Subsection \ref{subsec:local evaluation criteria}), and then presents illustrative experiments based on the criteria (Subsubsections \ref{subsubsec:experimental setup for ake}--\ref{subsubsec:experimental setup for cd}). Hence, it is a significant part of the other fold for the PoC of \emph{ClueNetwork}. Here, while Subsection \ref{subsec:general experimental setup} describes the general experimental setup, Subsubsections \ref{subsubsec:experimental setup for ake}--\ref{subsubsec:experimental setup for cd} describe the stage--tailored setups. 

\subsection{Local Evaluation Criteria}\label{subsec:local evaluation criteria}

\medskip

Firstly, this subsection addresses \textbf{RQ5} (``How are the local evaluation criteria defined and justified?''). Here, $h\text{F}_{1}$ and $\text{RI}$ are defined and justified in Subsubsections \ref{subsubsec:harmonic F1 Score} and \ref{subsubsec:robustness improvement}, respectively. Given that we have already defined and justified Modularity $Q$ in Subsubsection \ref{subsubsec:why does cd matter, what is it}, readers may consult that section for details on $Q$.

\subsubsection{Harmonic F1 Score}\label{subsubsec:harmonic F1 Score}

\medskip

\noindent\textbf{Definition}. For the PoC, the selected AKE algorithms are evaluated by using the harmonic $\text{F}1$ score $h\text{F}_{1}$. The definitions\footnote{We have referred to the natural--language--based description adopted by \cite{b49}, which is more intuitive than the confusion--matrix--based description.} of $h\text{F}_{1}$ and relevant criteria are presented in Table \ref{table:Table 6}. Herein, the ``total number of gold keyphrases'' is the number of keyphrases assigned as \emph{gold} standards to a document. This number, denoted as a variable $\upsilon$, is the number of prioritization \emph{opportunities} allowed for an AKE algorithm $x$ on that document \cite{b127}. That is, the task of $x$ is to maximize, within $\upsilon$, matches between prioritized candidates and the gold keyphrases by ranking candidates; the essence of the evaluation is to assess such prioritizing capabilities of algorithms.

\begin{table}[h!]
\caption{Possible Evaluation Criteria in AKE}\label{table:Table 6}
\centering
\renewcommand{\arraystretch}{2}
\begin{tabular}{>{\arraybackslash}m{2.4cm}|>
{\arraybackslash}m{5.2cm}}
\hline
Name (Symbol) & Definition \\
\hline\hline
Exact Precision ($\text{P}$) & $=\frac{\text{Number of Exactly Matched Candidates}}{\text{Total Number of Prioritized Candidates}}$ \\ \hline
Exact Recall ($\text{R}$) & $=\frac{\text{Number of Exactly Matched Candidates}}{\text{Total Number of Gold Keyphrases}}$ \\ \hline
Exact F1 ($\text{F}_{1}$) & $=\frac{2(\text{P} \times \text{R})}{\text{P} + \text{R}}$ (Harmonic Mean of $\text{P}$ and $\text{R}$) \\ \hline
Partial Precision ($p\text{P}$) & $=\frac{\text{Number of Partially Matched Candidates}}{\text{Total Number of Prioritized Candidates}}$ \\ \hline
Partial Recall ($p\text{R}$) & $=\frac{\text{Number of Partially Matched Candidates}}{\text{Total Number of Gold Keyphrases}}$ \\ \hline
Partial F1 ($p\text{F}_{1}$) & $=\frac{2(p\text{P} \times p\text{R})}{p\text{P} + p\text{R}}$ (Harmonic Mean of $p\text{P}$ and $p\text{R}$) \\ \hline
Harmonic F1 ($h\text{F}_{1}$) & $=\frac{2(\text{F}_{1} \times p\text{F}_{1})}{\text{F}_{1} + p\text{F}_{1}}$ (Harmonic Mean of $\text{F}_{1}$ and $p\text{F}_{1}$) \\ \hline
\end{tabular}
\end{table}

Interestingly, using $\upsilon$ instead of a constant makes precision and recall \emph{equivalent}, as their denominators become identical. If a constant is used instead, while precision rewards algorithms that maximize matches \emph{within} that constant, recall rewards algorithms that retrieve as many gold keyphrases as possible, \emph{regardless of} the constant. In either case, $\text{F}_{1}$ \cite{b125} is the \emph{standard} criterion in the field (e.g., \cite{b49, b51, b58, b59, b62, b63, b71, b79, b81, b82, b85}, and \cite{b86})\hypertarget{justification of hf1}{.}

\medskip

\noindent\textbf{Justification of Selection}. Nonetheless, $p\text{F}_{1}$ serves as a compelling alternative in the field (e.g., \cite{b49}, \cite{b51}, and \cite{b81}), given that $\text{F}_{1}$ imposes strict penalties on prioritized candidates (i.e., extracted keyphrases) unless they exactly match gold keyphrases, regardless of semantic relatedness \cite{b51}.

In contrast, in partial matching, if gold keyphrases are \{(happy, cat), (meme)\} and extracted ones are \{(huh, cat), (meme)\}, these sets are decomposed into \{(happy), (cat), (meme)\} and \{(huh), (cat), (meme)\}, respectively. $\text{F}_{1}$ calculated based on such newly formed sets is $p\text{F}_{1}$. It sometimes prioritizes trivial unigrams that lack sufficient contextual relevance \cite{b51}. That is, there is a \emph{trade--off} between $\text{F}_{1}$ and $p\text{F}_{1}$.

In general, a harmonic mean is accepted as a \emph{canonical} mathematical mechanism (e.g., $\text{F}_{1}$) that strictly penalizes the final value if either of two trading--off components is excessively low. Then, is there any reason not to adopt $h\text{F}_{1}$ as a form of nested harmonic mean? If the field can accept $p\text{F}_{1}$, it can also accept the stricter $h\text{F}_{1}$, given that $h\text{F}_{1}$ does not measure an entirely different \emph{dimension}. We clarify that this paper does not argue that $h\text{F}_{1}$ must be used exclusively for SNC.

\subsubsection{Robustness Improvement ($\text{RI}$)}\label{subsubsec:robustness improvement}

\medskip

For the PoC, the selected EW measures are evaluated by using $\text{RI}$. Unlike $h\text{F}_{1}$, \emph{deduced} from $\text{F}_{1}$ and $p\text{F}_{1}$, $\text{RI}$ has been derived by introducing a touch of novelty to the well--established percolation theory, a necessity born out of scenarios where gold--standard-based evaluation remains elusive (Subsubsection \ref{subsubsec:what is ew, and why does it matter}). Hence, more thorough background, motivation, and justification are presented below.

\medskip

\noindent\textbf{Facets of Distribution}. Let us begin with a fact based on the distributional hypothesis. If an EW measure passes the minimal veracity\footnote{The term ``veracity'' occurs repeatedly throughout this paper. We clarify that our notion of it refers to the property of aligning with ground--truth.} pretest, \emph{within} the system where the given $\mathbf{DTM}$ is assumed to be true, it does not generate false edges, and its belief that the relatedness between terms $a$ and $b$ is $0.7$ reflects \emph{a certain facet} of the $\mathbf{DTM}$. That is, if CF or Cosine holds that belief, although neither determines the $0.7$ by encapsulating all possible facets of the $\mathbf{DTM}$, the $0.7$ is an \emph{inevitable} consequence derived from the $\mathbf{DTM}$, not others. Hence, the beliefs of CF and Cosine are \emph{distinct}, not \emph{correct or incorrect}, and a possible error between the $0.7$ and a possibly existing but elusive gold weight is no longer an issue.

\medskip

\begin{figure}[h!]
    \centering    \includegraphics[width=0.75\linewidth]{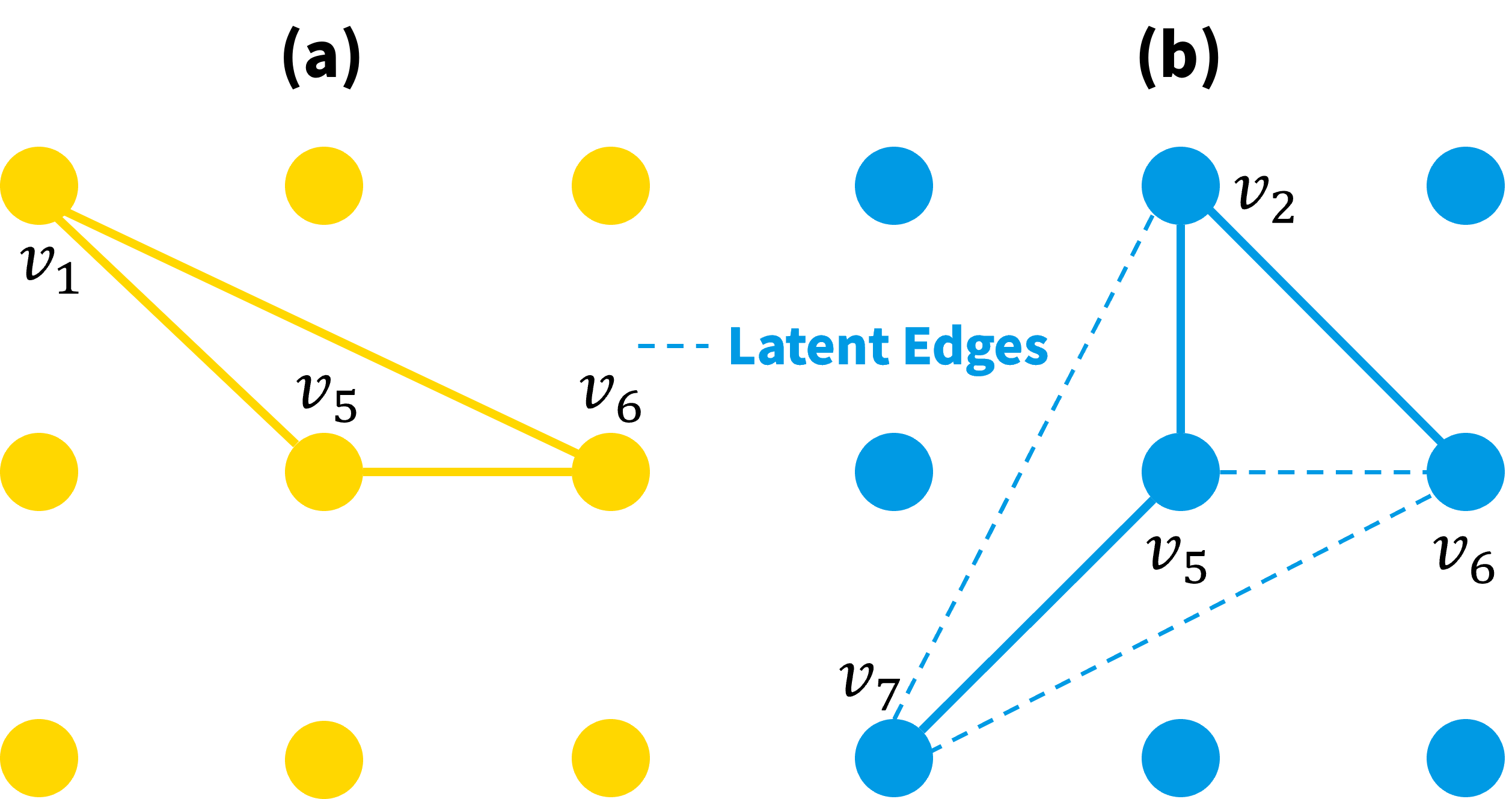}
    \caption{\textbf{Which SN Maximizes Interpretability?}}
    \label{Figure 11}
\end{figure}

\noindent\textbf{From Veracity To Interpretability}. Against this backdrop, we enter an alternative detour to facilitate ranking EW measures. Let us suppose that nine ``extracted keyphrases,'' hereafter referred to as ``keyphrases'' for brevity, form two SNs (Fig. \ref{Figure 11}), each consisting of three edges. Although it is elusive which SN is more veracious, Fig. \ref{Figure 11}--b appears more suitable for exploratory research (Subsection \ref{subsec:surrogates and clues}).

Why? Its \emph{hubs} (mediators) $v_2$ and $v_5$ manifest the latent edges among $v_2, v_6$, and $v_7$. The presence of such hubs prompts an interpreter to generate hypotheses that mediated vertices also somehow percolate their semantics through each other. In contrast, Fig. \ref{Figure 11}-a forms the \emph{closed} triangle, decreasing potential wow moments of abduction\hypertarget{definition of max interpretability}{.} Hence, we define the interpretability of an SN as follows and regard the objective of EW as the maximization of interpretability:

\medskip

\begin{itemize}
    \item the property of prompting an interpreter's \emph{hypothesis generation} whereby hubs are deployed.
\end{itemize}

\medskip

\noindent\textbf{Fair Rule is All We Need}. Here, a consideration arises. As explained above, CF and Cosine operate under different beliefs, making it natural that hub deployments they yield \emph{differ}. Therefore, directly comparing hub deployments of resulting SNs from CF and Cosine is \emph{unfair}. In physics, however, there is an established fair criterion: $\frac{\text{Output}-\text{Baseline}}{\text{Maximum Possible}-\text{Baseline}}$ \cite{b128}.

\begin{table}[h!]
\caption{Comparison of Fair Rule Settings (Ours: Top, \cite{b128}: Bottom) for Targets with Varying Baselines. $f^{(\dots)}$, $\omega$, and $\mathbf{K}$ are defined later.}\label{table:Table 7}
\centering
\begin{tabular}{>{\arraybackslash}m{2.1cm}|>
{\arraybackslash}m{2.5cm}|>
{\arraybackslash}m{2.1cm}}
Baseline & \emph{Treatment} & Output \\
\hline\hline
$f^{\text{null}}$@($\omega$, $\mathbf{K}$) & EW Measure & $f^{\text{empirical}}$@($\omega$, $\mathbf{K}$) \\ \hline
Students' Average \newline Pre--Test Score & Instructional Method & Students' Average \newline Post--Test Score \\
\end{tabular}
\end{table}

This normalization--based criterion quantifies effects of different treatments applied to different baselines; \cite{b128} used it to compare effects of different instructional methods across different physics courses. The basic concept and formula of $\text{RI}$ are isomorphic to it (Table \ref{table:Table 7}). Then, the remaining question is as follows: how such an output (i.e., pre--normalization interpretability) and its baseline are defined in terms of EW.

\medskip

\noindent\textbf{Output}. Let us first define such an output. According to percolation theory \cite{b129}, as more vertices act as hubs, they collectively form the largest connected subgraph, termed a giant component $\mathbf{G}_{\mathbf{GC}}(\mathbf{V}_{\mathbf{GC}}, \mathbf{E}_{\mathbf{GC}})$ \cite{b130}. Then, by constructing an SN as a $\mathbf{G}_{\mathbf{GC}}$ via an EW measure and removing its vertices one--by--one at random, can we identify its critical threshold $f_{c}$ at which $\mathbf{G}_{\mathbf{GC}}$ fragments? Yes, as it is the very approach established by physicists \cite{b131}--\cite{b133}.

Herein, $f_{c}$ is defined as the fraction of removed vertices $(|\mathbf{V}_{\mathbf{GC}}(f)|/|\mathbf{V}(0)|)/(|\mathbf{V}_{\mathbf{GC}}(0)|/|\mathbf{V}(0)|) \rightarrow 0$, where $|\mathbf{V}_{\mathbf{GC}}(0)|/|\mathbf{V}(0)|$ and $|\mathbf{V}_{\mathbf{GC}}(f)|/|\mathbf{V}(0)|$ denote each probability that a randomly chosen vertex belongs to $\mathbf{G}_{\mathbf{GC}}$ before and after removing a fraction $f$ of vertices from $\mathbf{G}$, respectively \cite{b132, b133}. For finite networks of $|\mathbf{V}| < \infty$, such fragmentation can be regarded as the moment when the size of the \emph{second} largest connected component\footnote{Why the second largest? Analogous to chopping a tree, the moment it splits in half is far more critical than when minor woodchips fly off.} reaches its maximum size during such an \emph{empirical} simulation \cite{b134}.

In short, a higher $f_{c}^{\text{empirical}}$ of an SN ($\mathbf{G}$) signifies that more hubs are deployed in it, thereby exhibiting greater robustness against random vertex failure simulation. Hence, we define the output as $f_{c}^{\text{empirical}}$. For reference, $f_{c}$ can also be derived through \emph{analytical} approximation methods. We clarify that $f_{c}$ in this paper is confined to $f_{c}^{\text{empirical}}$, given challenges in applying such methods to empirical networks (Appendix \ref{Appendix E}). 

\medskip

\noindent\textbf{Key Properties}. Unlike students' average pre--test score, a baseline $f_{c}^{\text{null}}$ against $f_{c}^{\text{empirical}}$ of $\mathbf{G}$ is not pre--given. Hence, $f_{c}^{\text{null}}$ of a \emph{null} model $\mathbf{G}_{\text{null}}$ generated from $\mathbf{G}$ is necessary as an artificial baseline (cf. \cite{b21} and \cite{b24}). When generating $\mathbf{G}_{\text{null}}$, preserving all properties of $\mathbf{G}$ makes comparison meaningless. $\mathbf{G}_{\text{null}}$ should reflect properties \emph{to be preserved} as preserved, and properties \emph{to be varied} as varied \cite{b141}. The former can be $\omega$ and $\mathbf{K}$, while the latter $L$ and $T_{\triangle}$.

First, $\omega$ is a specified number of edges to be represented. Together with $l$ (a specified number of vertices to be represented), it constitutes the desired resolution parameters for SN sparsification. As allowing $100$ edges for one EW measure while restricting another to $50$ edges is unfair, $\omega$ must be fixed. Second, $\mathbf{K}$ is a vertex degree sequence\footnote{For instance, if $\mathbf{V}=\{v_1,v_2,v_3\}$ and the degrees of $v_1,v_2, \text{ and } v_3$ are $1,1 \text{ and } 2$, respectively, then $\mathbf{K}=\{1,1,2\}$.}, where the degree of a vertex is the number of edges it has to other vertices \cite{b10}. $\mathbf{K}$ must be fixed for \emph{each} SN to ensure valid comparison between $f_{c}^{\text{empirical}}$ and $f_{c}^{\text{null}}$. Why? If an EW measure determines $\mathbf{K}$ of $\mathbf{G}$ one way, generating $\tilde{\mathbf{G}}_{\text{null}}$ with another $\tilde{\mathbf{K}}$ would be akin to comparing against $\tilde{f}_{c}^{\text{null}}$ grounded in another $\tilde{\mathbf{G}}$.

\medskip

Third, the average shortest path length $L$ \cite{b10} is defined as

\begin{equation} \label{Eq 5}
\begin{aligned}
L=\frac{1}{|\mathbf{V}|(|\mathbf{V}|-1)}\sum_{v_{i} \neq v_{j}}d(v_{i},v_{j}),
\end{aligned}
\end{equation}

\medskip

\noindent where $d(v_i, v_j)$ is the smallest number of edges to be traversed from $v_{i}$ to $v_{j}$. For an SN, $L$ certainly signifies the average semantic percolation cost between keyphrases (e.g., \emph{a}bc\emph{d} versus \emph{a}b\emph{d}). That is, $L$ is an \emph{SN interpreter's} visual scanning effort. Therefore, to evaluate the effect of an EW measure, $L(\mathbf{G}_{\text{null}})$ should be greater than or at least equal to $L(\mathbf{G})$.

Fourth, transitivity $T_{\triangle}$ \cite{b142} is defined as \eqref{Eq 6}, and, for an SN, $T_{\triangle}(\mathbf{G}_{\text{null}})$ should also be greater than or at least equal to $T_{\triangle}(\mathbf{G})$. Why? A high proportion of triangles can delay an SN interpreter's visual scanning due to their circular connectivity. Indeed, \cite{b139} suggests that a higher $T_{\triangle}$ can hinder smooth interactions (i.e., percolation) among vertices.

\begin{equation} \label{Eq 6}
\begin{aligned}
T_{\triangle}=\frac{6N_{\triangle}}{2N_{\text{3}}},
\end{aligned}
\end{equation}

\medskip

\noindent where $N_{3}$ and $N_{\triangle}$ are the numbers of connected triples (e.g., not only $\triangle$ but also a--b--c) and triangles (closed triples), respectively; each triangle manifests in six different permutations depending on a starting vertex and a direction (e.g., \emph{a}bca, \emph{b}cab, \emph{c}abc, $...$, and c\emph{b}ac) of semantic cycling (thus, $N_{\triangle}$ is multiplied by $6$); for a connected triple, only two directions (e.g., \emph{a}bc and \emph{c}ba) through which semantics percolate need to be considered (thus, $N_{3}$ is multiplied by $2$).

\medskip

\noindent\textbf{Baseline}. To adhere to these requirements for $\omega$, $\mathbf{K}$, $L$, and $T_{\triangle}$, we have slightly modified the established ClustRNet algorithm (null model generator) \cite{b141}. ClustRNet generates $\mathbf{G}_{\text{null}}$ by introducing randomness while adhering to the requirements for $\omega$, $\mathbf{K}$, and $T_{\triangle}$, but it does not account for $L$. Our modified version (Algorithm \ref{Algorithm 1}) accounts for $L$.

As ClustRNet also employs a configuration model \cite{b135} as the initial candidate null model $\mathbf{G}'_{(0)}$, ClustRNet is prone to enter an infinite loop if any candidate null model $\mathbf{G}'$ fails to remain connected. Algorithm \ref{Algorithm 1} avoids such a loop by setting $\mathbf{G}'_{(0)}$ as a copy of a \emph{connected} $\mathbf{G}$. Edge rewiring is then performed to \emph{monotonically increase} both $T_{\triangle}(\mathbf{G}')$ and $L(\mathbf{G}')$. Adhering to $\mathbf{K}$, Algorithm \ref{Algorithm 1} terminates when $T_{\triangle}(\mathbf{G}')$ and $L(\mathbf{G}')$ approach $(1+z)T_{\triangle}(\mathbf{G})$ and $(1+z)L(\mathbf{G})$, respectively, or when a maximum $h$ iterations are reached.

How, then, is such a connected $\mathbf{G}$ obtained? First, for a given $\mathbf{TTM} \in \mathbb{R}^{l \times l}$, established Kruskal's algorithm \cite{b143} extracts a Maximum Spanning Tree (MST) that serves as the backbone of $\mathbf{G}$ by prioritizing higher--weighted $\mathbf{TTM}$ entries. Next, additional higher--weighted $\mathbf{TTM}$ entries are sequentially added to the MST until the total number of edges reaches $\omega$, thereby constructing $\mathbf{G}$. Ultimately, the critical threshold of  $\mathbf{G}_{\text{null}}$ generated for $\mathbf{G}$ is the very baseline $f_{c}^{\text{null}}$.

\medskip

\noindent\textbf{Definition}. Therefore, $0 \leq \text{RI} \leq 1$ is \emph{mathematically} defined as \eqref{Eq 7}, where $\mathbf{G}$'s $f_c^{\text{empirical}}$ and $\mathbf{G}_{\text{null}}$'s $f_{c}^{\text{null}}$ are obtained through respective vertex random failure simulations:

\begin{equation} \label{Eq 7}
\begin{aligned}
\text{RI}=\begin{cases}
    \frac{f_{c}^{\text{empirical}} - f_{c}^{\text{null}}}{1-f_{c}^{\text{null}}}, \text{ if } f_{c}^{\text{empirical}} > f_{c}^{\text{null}}\text{;} \\ \\
    0, \text{ otherwise.}
\end{cases}
\end{aligned}
\end{equation}

\medskip

\noindent \eqref{Eq 7} can also be \emph{qualitatively} regarded as the proportion of additional SN interpretability gained relative to the total possible SN interpretability improvement. A higher $\text{RI}$ signifies a greater effect of an EW measure on hub deployment activating multiple alternative routes for semantic percolation. Here, the well--established $\frac{\text{Output}-\text{Baseline}}{\text{Maximum Possible}-\text{Baseline}}$ \cite{b128}, percolation theory \cite{b129}, $\omega$, $\mathbf{K}$ \cite{b10}, $L$ \cite{b10}, $T_{\triangle}$ \cite{b142}, and ClustRNet \cite{b141} are not our creations. Nonetheless, the idea of integrating them to \emph{facilitate} ranking raw SNs (ultimately, their EW measures) humbly constitutes a touch of novelty\hypertarget{justification of ri}{.}

\medskip

\noindent\textbf{Justification}. Here, one can legitimately demand, ``Validate whether $\text{RI}$ actually harmonizes with the common sense of interpretability improvement.'' Reasonable, and we will do so. To this end, we first operationally define the ``Interpretability imProvement (\text{IP})'' as two aspects:

\begin{table*}[b!]
\caption{Overview of Benchmark Datasets}\label{table:Table 8}
\centering
\begin{tabular}{>{\arraybackslash}m{3.3cm}|>{\arraybackslash}m{2.7cm}|>
{\arraybackslash}m{3cm}|>{\arraybackslash}m{0.8cm}|>{\arraybackslash}m{1.5cm}|>
{\arraybackslash}m{1.6cm}|>{\arraybackslash}m{1.6cm}}
\hline
Document Type & Dataset & Annotator Type & \#Case & Avg \#Tokens & Max \#Tokens & Avg \#Goldens \\
\hline\hline
\multirow{2}{3.3cm}{Scientific (full--text papers)} & SemEval--2010 \cite{b144} & Authors and Students & $244$ & $8821.97$ & $17347$ & $15.00$ \\
 & NUS \cite{b145} & Authors and Students & $211$ & $9773.89$ & $18375$ & $11.07$ \\
\hline
\multirow{3}{3.3cm}{Scientific (paper abstracts)} & Inspec \cite{b146} & Professional Indexers & $2000$ & $159.75$ & $672$ & $9.65$ \\
 & KDD \cite{b147} & Authors & $754$ & $216.79$ & $446$ & $4.07$ \\
 & WWW \cite{b147} & Authors & $1330$ & $185.62$ & $664$
 & $4.80$ \\
\hline
Scientific (paper paragraphs) & SemEval--2017 \cite{b148} & An Expert and Students & $500$ & $237.55$ & $499$ &  $17.29$ \\
\hline
\multirow{2}{3.3cm}{Common (news articles)} & DUC--2001 \cite{b63} & Students & $308$ & $964.20$ & $6375$ & $8.06$ \\
 & 500N--KP--Crowd \cite{b149} & Crowdsourced Annotators & $500$ & $591.27$ & $7947$ & $48.93$ \\
\hline
\end{tabular}
\end{table*}

\begin{equation} \label{Eq 8}
\begin{aligned}
\text{IP}=\begin{cases}
\text{IP}_{1}=\frac{L(\mathbf{G}_{\text{null}})-L(\mathbf{G})}{L(\mathbf{G}_{\text{null}})} \;\ \big(L(\mathbf{G}_{\text{null}}) \geq L(\mathbf{G})\big) \\ \\
\text{IP}_{2}=\frac{T_{\triangle}(\mathbf{G}_{\text{null}})-T_{\triangle}(\mathbf{G})}{T_{\triangle}(\mathbf{G}_{\text{null}})} \;\ \big(T_{\triangle}(\mathbf{G}_{\text{null}}) \geq T_{\triangle}(\mathbf{G})\big)
\end{cases}
\end{aligned}
\end{equation}

\medskip

\noindent Based on the established $L$ and $T_{\triangle}$, \eqref{Eq 8} quantifies the relative reduction (achieved by using an EW measure) in an SN interpreter's visual scanning effort for identifying relatedness between two keyphrases. Ultimately, hypothesis generation is \emph{subject to} the interpreter’s \emph{expertise}. Nonetheless, it is an \emph{objective} fact that larger $\text{IP}_{1}$ and $\text{IP}_{2}$ at least foster a more interpretable \emph{condition}. Hence, in Subsubsection \ref{subsubsec:results of ew}, we validate correlations between $\text{RI}$ and these two aspects.

\begin{algorithm}[t!]\label{Algorithm 1}
\caption{Null Model Generator}
\SetKw{KwEnd}{end}
\SetKw{KwDo}{do}
\SetKw{KwEmpty}{}
\SetKw{KwMain}{Main Steps}
\KwIn{Connected $\mathbf{G}(\mathbf{V}, \mathbf{E}, \mathbf{K})$, $h$, and $z$}
\KwOut{Connected $\mathbf{G}_{\text{null}}(\mathbf{V}, \mathbf{E}_{\text{null}}, \mathbf{K})$}
$\triangleright$ \KwMain \\
Generate the initial $\mathbf{G}'_{(0)}$ by copying $\mathbf{G}$ \\
\For{$i$ in $[1,\;\ h]$}
{
$\mathbf{G}'_{(i)} \leftarrow \mathbf{G}'_{(i-1)}$ \\
Select $v_{a}$ ($k_{a}>1$) $\in \mathbf{V}'_{(i)}$ uniformly at random \\ 
Select two neighbors, $v_{b}$ and $v_{c}$, of $v_{a}$ uniformly at random such that $k_{b}>1$, $k_{c}>1$, and $v_{b} \neq v_{c}$ \\
Select a neighbor, $v_{d}$ of $v_{b}$, and a neighbor, $v_{e}$ of $v_{c}$, uniformly at random such that $v_{d} \neq v_{a}$, $v_{e} \neq v_{a}$, and $v_{d} \neq v_{e}$ \\ 
\If{$(v_{b},v_{c})$ and $(v_{d},v_{e})$ $\notin \mathbf{E}'_{(i)}$}{Remove $(v_{b},v_{d})$ and $(v_{c},v_{e})$ from $\mathbf{E}'_{(i)}$ \\
Add $(v_{b},v_{c})$ and $(v_{d},v_{e})$ to $\mathbf{E}'_{(i)}$ \\
$\text{VALID}:=\text{FALSE}$ \\
\If{$T_{\triangle}(\mathbf{G}'_{(i)})>T_{\triangle}(\mathbf{G}'_{(i-1)})$, $L(\mathbf{G}'_{(i)})>L(\mathbf{G}'_{(i-1)})$, $T_{\triangle}(\mathbf{G}'_{(i)}) \leq (1+z)T_{\triangle}(\mathbf{G})$, $L(\mathbf{G}'_{(i)}) \leq (1+z)L(\mathbf{G})$, and $\mathbf{G}'_{(i)}$ is connected}{$\text{VALID}:=\text{TRUE}$ \\
\If{$T_{\triangle}(\mathbf{G}'_{(i)}) \geq 0.95(1+z)T_{\triangle}(\mathbf{G})$ and $L(\mathbf{G}'_{(i)}) \geq 0.95(1+z)L(\mathbf{G})$}{$\mathbf{G}_{\text{null}}:=\mathbf{G}'_{(i)}$ \\ \textbf{return} $\mathbf{G}_{\text{null}}$}
\textbf{end}}
\If{$\text{VALID}$ is $\text{FALSE}$}{Add $(v_{b},v_{d})$ and $(v_{c},v_{e})$ to $\mathbf{E}'_{(i)}$ \\
Remove $(v_{b},v_{c})$ and $(v_{d},v_{e})$ from $\mathbf{E}'_{(i)}$ \\ \textbf{end}}
\textbf{end}}
\KwEnd} 
\Return $\mathbf{G}_{\text{null}}$
\end{algorithm}

\subsection{General Experimental Setup}\label{subsec:general experimental setup}

\medskip

We now explain the general experimental setup prepared to answer \textbf{RQ6} (``Which stage--specific selected methods yield high performance?''). Later, Subsubsections \ref{subsubsec:results of ake}--\ref{subsubsec:results of cd} themselves serve as the very answers.

\medskip

\noindent\textbf{Datasets}. Eight benchmark datasets (Table \ref{table:Table 8}), widely employed in the field of AKE (e.g., \cite{b49, b51}, \cite{b58, b59}, \cite{b61}--\cite{b63}, \cite{b71, b79, b81, b82, b85}, and \cite{b86}), were used in the illustrative experiments. Every incomplete case (missing ID, document, or gold keyphrases) had been excluded from all datasets. As one (\href{https://huggingface.co/midas}{link 1}) of the dataset distributors (\href{https://huggingface.co/midas}{link 1})(\href{https://huggingface.co/taln-ls2n}{link 2}) provided some datasets in tokenized forms, we restored them to original or near--original forms. Regarding the latter, the algorithms remain under fair conditions and should be adaptive. Given that BERT variants typically have an input limit of 512 tokens \cite{b88}, KeyBERT and MDERank were only evaluated on KDD and SemEval--2017. Although MPNet has a shorter limit of 384 tokens \cite{b90}, average pooling allows LMRank to overcome that constraint \cite{b81}.

\medskip

\noindent\textbf{General Process} of local evaluation in \emph{ClueNetwork} is as follows. In Stage 1 (Subsection \ref{subsec:experiment 1}), average $h\text{F}_{1}$ scores of AKE algorithms on each dataset are evaluated. In Stage 2 (Subsection \ref{subsec:experiment 2}), top--$l$ keyphrases of each AKE algorithm are first identified, and then corresponding $\mathbf{TTM}$s $\in \mathbb{R}^{l \times l}$ are built by using EW measures. Based on $\omega$, raw SNs are extracted from those $\mathbf{TTM}$s, and then their $\text{RI}$ scores are evaluated. In Stage 3 (Subsection \ref{subsec:experiment 3}), CD algorithms are applied to the SNs and corresponding $Q$ scores are evaluated.

\medskip

\noindent\textbf{For Reproducibility}. All relevant code, random seeds, parameters, preprocessing, intermediate or final outputs, and the versions of employed Python packages have been disclosed at \href{https://github.com/potentialreviewer/Ha-Kim-2026a}{here (link)}. As our laptops are digital fossils, all relevant experiments have been conducted in a cloud environment:

\medskip

\begin{itemize}
    \item Cloud Environment: Google Colab (Pro)
    \item System Type: x86\_64 with $64$--bit architecture
    \item CPU: Intel(R) Xeon(R) CPU @ 2.20 GHz (Family 6, Model 85, Stepping 7), $6$ cores with $12$ threads
    \item GPU: NVIDIA L4 ($22.5$ GB VRAM, Driver v550.54.15)
    \item RAM: $54$ GB (Details are not provided by Google)
    \item Storage: $250$ GB NVMe SSD
    \item OS: Ubuntu 22.04.4 LTS (Kernel 6.6.105+)
    \item Programming Language: Python 3.12.12
\end{itemize}

\begin{table*}[t!]
\caption{Experimental Settings for AKE Algorithms (algos.)}\label{table:Table 9}
\centering
\begin{tabular}{>{\arraybackslash}m{3.9cm}|>{\arraybackslash}m{2.5cm}|>
{\arraybackslash}m{7.5cm}|>{\arraybackslash}m{1.6cm}}
\hline
Algo. (Existing Implementation) & Target Dataset & Employed Parameters & Preprocessing \\
\hline\hline
TF (We implemented) & \multirow{7}{2.5cm}{All the datasets} & $n=3$ & \multirow{4}{1.6cm}{Type $1$} \\ \cline{1-1} \cline{3-3}
TF--IDF (\href{https://github.com/boudinfl/pke/blob/master/pke/unsupervised/statistical/tfidf.py}{Link}) &  & $n=3$ & \\ \cline{1-1} \cline{3-3}
KP--Miner (\href{https://github.com/boudinfl/pke/blob/master/pke/unsupervised/statistical/kpminer.py}{Link}) &  & $n=5$, $lasf$ $=$ $3$, $cutoff$ $=$ $400$, $\sigma=3.0$, $\alpha=2.3$, and $P_{f}=1$ & \\ \cline{1-1} \cline{3-3}
YAKE! (\href{https://github.com/boudinfl/pke/blob/master/pke/unsupervised/statistical/yake.py}{Link}) &  & $n=3$, $\text{(window)}=2$, and $use\_stems=True$ & \\ \cline{1-1} \cline{3-4}
TextRank (TR) (\href{https://github.com/boudinfl/pke/blob/master/pke/unsupervised/graph_based/textrank.py}{Link}) &  & $\text{(window)}=2$, $T=33\%$, and $\delta=0.85$ & \multirow{3}{1.6cm}{Type $2$} \\ \cline{1-1} \cline{3-3}
SingleRank (SR) (\href{https://github.com/boudinfl/pke/blob/master/pke/unsupervised/graph_based/singlerank.py}{Link}) &  & $\text{(window)}=10$ and $\delta=0.85$ & \\ \cline{1-1} \cline{3-3}
PositionRank (PR) (\href{https://github.com/boudinfl/pke/blob/master/pke/unsupervised/graph_based/positionrank.py}{Link}) &  & $\text{(window)}=10$ and $\delta=0.85$ & \\ \hline
KeyBERT (\href{https://github.com/MaartenGr/KeyBERT}{Link}) & KDD, SemEval--2017 & (language model) $=$ "all--distilroberta--v1" & Type $3$ \\
\hline
MDERank (\href{https://github.com/LinhanZ/mderank}{Link}) & KDD, SemEval--2017 & (language model) $=$ "bert--base--uncased" & Type $4$ \\
\hline
\multirow{4}{3.9cm}{LMRank (\href{https://github.com/NC0DER/LMRank}{Link})} & All except Inspec and 500N--KPC & (language model) $=$ "all--mpnet--base--v2", $deduplicate$ $=$ $False$, $keeps\_noun\_adjs$ $=$ $True$, $positional\_feature$ $=$ $True$, and $\mu=1.0$ & \multirow{4}{1.6cm}{Type $5$} \\ \cline{2-3} & Inspec, 500N--KPC & (language model) $=$ "all--mpnet--base--v2", $deduplicate$ $=$ $False$, $keeps\_noun\_adjs$ $=$ $True$, $positional\_feature$ $=$ $False$, and $\mu=1.0$ & \\
\hline
\end{tabular}
\end{table*}

\subsection{Experiment 1: Automatic Keyphrase Extraction}\label{subsec:experiment 1}

\medskip

We now report the AKE--tailored experimental setup (Subsubsection \ref{subsubsec:experimental setup for ake}) and which of the selected AKE algorithms maximized SN keyness (Subsubsection \ref{subsubsec:results of ake}).

\subsubsection{Experimental Setup for AKE}\label{subsubsec:experimental setup for ake}

\medskip

\noindent \textbf{Implementations}. We implemented TF through PositionRank (Table \ref{table:Table 9}) based on the implementations of the Python Keyphrase Extraction toolkit (PKE) \cite{b150}. Given that the NN--based algorithms are not yet supported by PKE, we tailored their existing implementations for our work.

\medskip

\noindent \textbf{Parameters}. Most AKE algorithms incorporate one or more user--specifiable parameters, namely hyperparameters. To focus on the PoC of \emph{ClueNetwork}, we employed the \emph{default} parameters whenever appropriate (Table \ref{table:Table 9}). These defaults are also grounded in empirical evidence from the original papers or their subsequent implementations. Readers may opt for alternatives for future applications.

\medskip

\noindent \textbf{Preprocessing}, the algorithm itself, and its data adaptability are crucial factors \emph{jointly} determining the performance of an AKE algorithm. Most AKE algorithms employ their own preprocessing strategies to extract keyphrases exclusively from given phrases. One can specify an $n$--gram parameter (Table \ref{table:Table 9}) to exclude $(n+1)$--grams and higher, while another can consider only Noun Phrases (NPs) as candidates. As preprocessing manifests in such diverse types, to paradoxically avoid verbosity, we summarize the types employed by the selected algorithms and their details in Table \ref{table:Table 9} and Appendix \ref{Appendix F}, respectively. In short, we have adhered to the strategies of the original papers or implementations. To emphasize preprocessing, we intervened exceptionally in KeyBERT, and details are discussed in Subsubsection \ref{subsubsec:results of ake}. 

\subsubsection{Results of AKE}\label{subsubsec:results of ake}

\medskip

\noindent \textbf{Statistical Analysis}. While most AKE algorithms achieve moderate ($\text{F}_{1}$, $p\text{F}_{1}$, or $h\text{F}_{1}$) scores for many documents and rarely reach near--perfect scores, they also record zero scores for many other documents. Their score distributions often follow \emph{zero--inflated} distributions. Indeed, across the selected algorithms and datasets, Table \ref{table:Table 10} shows the low (average) $h\text{F}_{1}$ scores and high variances. For brevity, $95\%$ Bootstrap Confidence Intervals (BCIs) \cite{b154} for these averages are separately disclosed at \href{https://github.com/potentialreviewer/Ha-Kim-2026a/blob/main/notebooks/AKE_Statistical_Analysis.ipynb}{our GitHub (link)}, where $10^{4}$ resamples were drawn with replacement from each distribution.

\begin{figure}[h!]
    \centering    \includegraphics[width=0.75\linewidth]{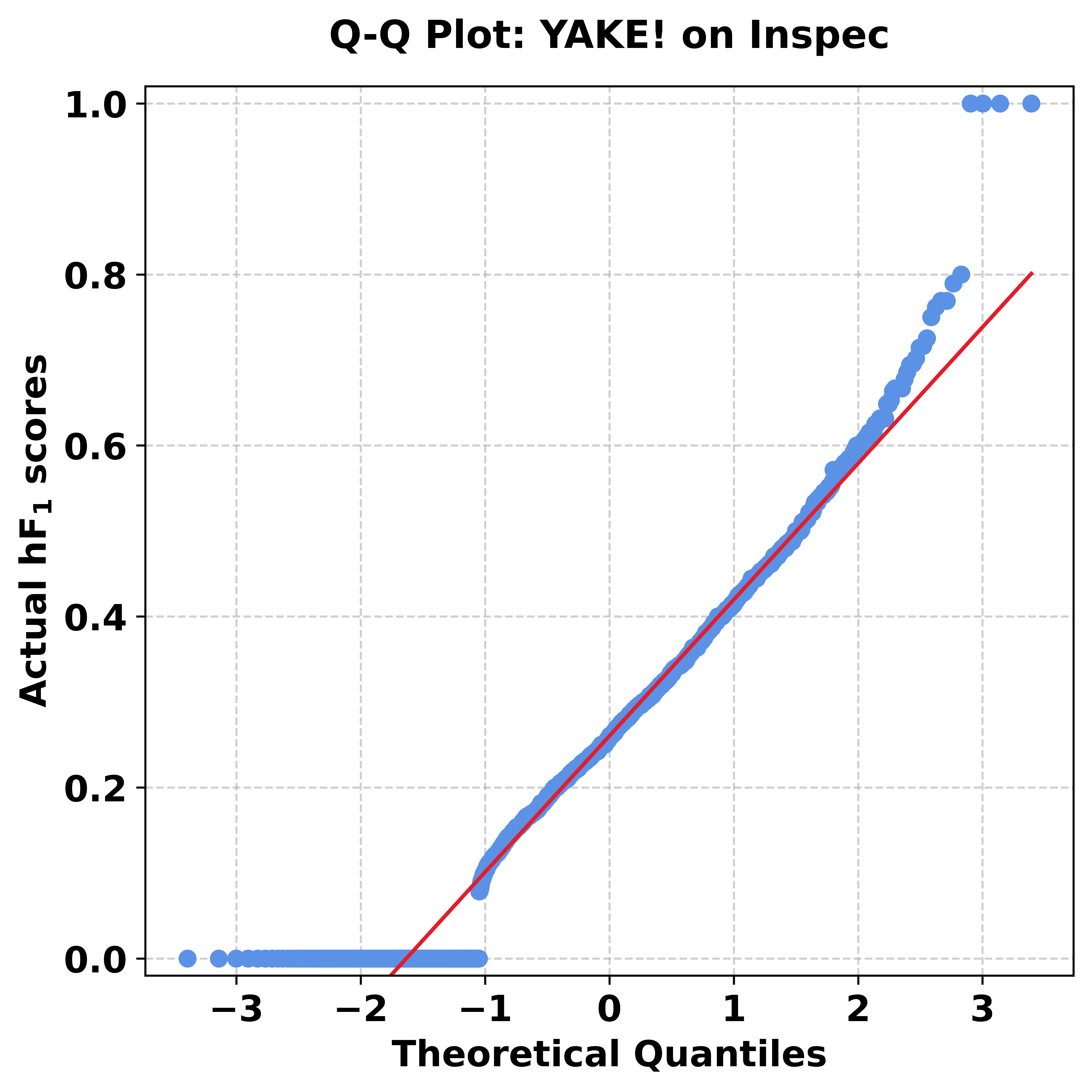}
    \caption{\textbf{Q-Q Plot of YAKE! on Inspec.}}
    \label{Figure 12}
\end{figure}

Consequently, both the Q--Q plots and the Kolmogorov--Smirnov (K--S) test \cite{b155} --- whose null hypothesis that a given distribution follows the normal distribution is rejected when $p < 0.05$ --- indicated that the normality was violated for $61$ of the $68$ $h\text{F}_{1}$ distributions. An example is presented in Fig. \ref{Figure 12}. The remaining $67$ pairs of Q--Q plots and K--S test $p$--values are also disclosed separately \href{https://github.com/potentialreviewer/Ha-Kim-2026a/blob/main/notebooks/AKE_Statistical_Analysis.ipynb}{(link)}.

\begin{table*}[t!]
\caption{$h\text{F}_{1}$--based Evaluation Results ($68$ Pairs of Averages and Standard Deviations)}\label{table:Table 10}
\centering
\begin{tabular}{>{\arraybackslash}m{1.8cm}>
{\arraybackslash}m{1cm}|>{\arraybackslash}m{0.7cm}>{\arraybackslash}m{1cm}>{\arraybackslash}m{1.2cm}>
{\arraybackslash}m{0.9cm}|>
{\arraybackslash}m{0.7cm}>
{\arraybackslash}m{0.7cm}>
{\arraybackslash}m{0.7cm}|>
{\arraybackslash}m{1.0cm}>
{\arraybackslash}m{1.0cm}>{\arraybackslash}m{1.0cm}}
\hline
\multirow{2}{1.8cm}{Benchmark} & \multirow{2}{1cm}{$h\text{F}_{1}\text{@}\upsilon$} & \multicolumn{4}{c|}{Statistics--based Algorithms} & \multicolumn{3}{c|}{Graph--based Algorithms} & \multicolumn{3}{c}{NN--based Algorithms} \\ 
 & & TF & TF--IDF & KP--Miner & YAKE! & TR & SR & PR & KeyBERT & MDERank & LMRank \\
\hline\hline
\multirow{2}{1.8cm}{SemEval--2010} & Avg & $0.134$ & \boldsymbol{$0.207$} & \underline{\boldsymbol{$0.287$}} & $0.190$ & $0.035$ & $0.053$ & $0.112$ & - & - & $0.164$ \\
& Std & $0.109$ & $0.115$ & $0.119$ & $0.119$ & $0.059$ & $0.073$ & $0.092$ & - & - & $0.099$ \\
\hline
\multirow{2}{1.8cm}{NUS} & Avg & $0.144$ & \boldsymbol{$0.258$} & \underline{\boldsymbol{$0.341$}} & $0.233$ & $0.021$ & $0.039$ & $0.113$ & - & - & $0.145$ \\
& Std & $0.144$ & $0.167$ & $0.171$ & $0.150$ & $0.054$ & $0.074$ & $0.118$ & - & - & $0.136$ \\
\hline
\multirow{2}{1.8cm}{Inspec} & Avg & $0.125$ & $0.234$ & $0.092$ & $0.260$ & $0.149$ & \boldsymbol{$0.287$} & $0.287$ & - & - & \underline{\boldsymbol{$0.364$}} \\
& Std & $0.133$ & $0.163$ & $0.133$ & $0.162$ & $0.157$ & $0.188$ & $0.180$ & - & - & $0.200$ \\
\hline
\multirow{2}{1.8cm}{KDD} & Avg & $0.092$ & \underline{\boldsymbol{$0.175$}} & \boldsymbol{$0.174$} & $0.119$ & $0.084$ & $0.089$ & $0.148$ & $0.146$ & $0.138$ & $0.164$ \\
& Std & $0.159$ & $0.200$ & $0.211$ & $0.179$ & $0.155$ & $0.154$ & $0.187$ & $0.191$ & $0.189$ & $0.199$ \\
\hline
\multirow{2}{1.8cm}{WWW} & Avg & $0.123$ & \underline{\boldsymbol{$0.186$}} & \boldsymbol{$0.175$} & $0.118$ & $0.084$ & $0.074$ & $0.122$ & - & - & $0.142$ \\
& Std & $0.165$ & $0.192$ & $0.200$ & $0.160$ & $0.153$ & $0.149$ & $0.182$ & - & - & $0.182$ \\
\hline
\multirow{2}{1.8cm}{SemEval--2017} & Avg & $0.217$ & $0.275$ & $0.140$ & $0.284$ & $0.216$ & $0.426$ & $0.399$ & \underline{\boldsymbol{$0.485$}} & \boldsymbol{$0.463$} & $0.290$ \\
& Std & $0.111$ & $0.119$ & $0.111$ & $0.127$ & $0.125$ & $0.155$ & $0.149$ & $0.154$ & $0.148$ & $0.153$ \\
\hline
\multirow{2}{1.8cm}{DUC--2001} & Avg & $0.106$ & $0.160$ & $0.180$ & $0.178$ & $0.124$ & \boldsymbol{$0.247$} & \underline{\boldsymbol{$0.281$}} & - & - & $0.240$ \\
& Std & $0.125$ & $0.149$ & $0.154$ & $0.143$ & $0.134$ & $0.167$ & $0.184$ & - & - & $0.161$ \\
\hline
\multirow{2}{1.8cm}{500N--KPC} & Avg & \underline{\boldsymbol{$0.363$}} & \boldsymbol{$0.337$} & $0.193$ & $0.309$ & $0.194$ & $0.297$ & $0.315$ & - & - & $0.166$ \\
& Std & $0.107$ & $0.095$ & $0.103$ & $0.106$ & $0.082$ & $0.153$ & $0.138$ & - & - & $0.088$ \\
\hline
\end{tabular}
\end{table*}

Given such substantial non--normality, the original averages are not reliable representative statistics. It is safer to use new averages derived from \emph{rank--transformed} distributions and to conduct a non--parametric statistical analysis that relies on such transformation. Hence, to determine whether significant differences exist among the averages (Table \ref{table:Table 10}), we performed the Friedman test \cite{b156} followed by the Nemenyi post--hoc test \cite{b157} at the dataset level. For our matched--samples design, where a document, an AKE algorithm, and a resulting $h\text{F}_{1}$ score \emph{distribution} served as a block, a treatment, and a sample (i.e., \emph{set} of observations), respectively, this non--parametric approach was suitable.

For all datasets, Friedman indicated significant differences in the average $h\text{F}_{1}$ ranks among at least one pair of the AKE algorithms. That is, the null hypothesis (``No such difference exists'') was rejected, as all $p$--values were less than $0.05$. Nemenyi then identified which pairs exhibited significance at the dataset level. Its null hypothesis (``No such difference exists'') is also rejected when $p < 0.05$.

\begin{figure*}[t!]
    \centering    \includegraphics[width=0.9\textwidth]{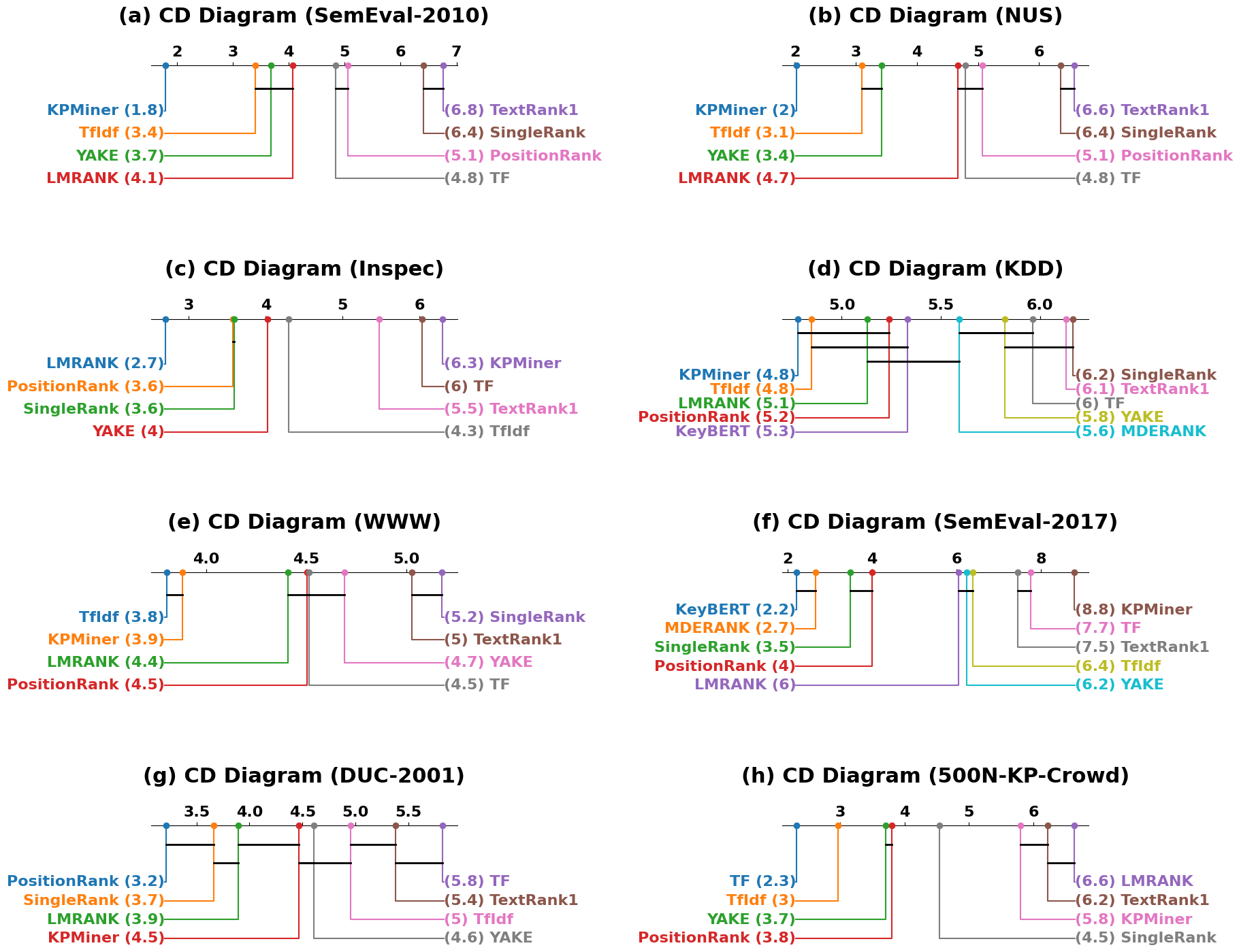}
    \caption{\textbf{Critical Difference Diagrams of AKE Algorithms across Datasets. Each value in parentheses is an average of $h\text{F}_{1}$ ranks across documents.}}
    \label{Figure 13}
\end{figure*}

The $p$--values are disclosed separately \href{https://github.com/potentialreviewer/Ha-Kim-2026a/blob/main/notebooks/AKE_Statistical_Analysis.ipynb}{(link)}, while the results are summarized in the Critical Difference (CD) diagrams \cite{b158}. The numbers in parentheses denote the average $h\text{F}_{1}$ ranks. No significant difference exists between any two algorithms directly tied by a string (Fig. \ref{Figure 13}).

\medskip

\noindent \textbf{Qualitative Analysis}. Based on the significant differences ($p < 0.05$) presented in Fig. \ref{Figure 13} and the statistics in Table \ref{table:Table 8}, we provide the following post--hoc interpretation:

First, TF achieved mid--to--low average $h\text{F}_{1}$ ranks across most datasets. An exception was 500N--KP--Crowd, characterized by an exceptionally high number of gold keyphrases per document ($48.9$ per document). Meanwhile, KP--Miner exhibited low performance on 500N--KP--Crowd, as its strict cutoff and least allowable seen frequency constrained full exploitation of these abundant prioritization opportunities.

Second, these trends were reversed for the full--text paper datasets (SemEval--2010 and NUS) and the abstract datasets (KDD and WWW). As the former tend to present contributions and novelties early, KP--Miner benefited from its cutoff. Given the lack of prioritization opportunities ($4.8$ per document) for the latter, algorithms must prioritize candidates based on salient features. In this regard, the heuristics of KP--Miner and its baseline (TF--IDF) were effective.

Third, the graph--based algorithms achieved mid--to--low average $h\text{F}_{1}$ ranks for SemEval--2010, NUS, KDD, and WWW. As words that contribute to topics but do not serve as components of gold keyphrases occur frequently in full--text papers \cite{b51}, they hinder graph--based algorithms that focus on word associations. Given the limited prioritization opportunities for KDD and WWW ($4$ to $4.8$ per document), word associations appeared less effective than salient features.

Fourth, nonetheless, PositionRank and SingleRank outperformed the statistics--based algorithms on Inspec, SemEval--2017, and DUC--2001, which feature at least twice as many prioritization opportunities ($8$ to $17.3$ per document). As word associations become noteworthy with increasing prioritization opportunities, these algorithms benefited accordingly. Notably, PositionRank achieved its highest performance on the news article dataset (DUC--2001), which typically presents topics in opening sentences.

Fifth, when sufficient prioritization opportunities were available ($17.3$ per document), KeyBERT and MDERank achieved the highest performance on SemEval--2017. While KeyBERT, which regards $n$--grams as candidates by default, had achieved low performance in \cite{b81} due to overlapping $n$--grams reducing diversity of extracted keyphrases, even KeyBERT paired with the less performant (\href{https://www.sbert.net/docs/sentence_transformer/pretrained_models.html}{link}) all--distilroberta--v1 outperformed LMRank paired with all--mpnet--base--v2 by regarding NPs as candidates (Table \ref{table:Table 9} with Appendix \ref{Appendix F}). This underscores the significance of preprocessing.

Sixth, while KeyBERT and MDERank faced the input limit of $512$ tokens, average pooling did not lead LMRank to a definitive breakthrough. Although we expected LMRank to deliver superior performance on the full--text paper datasets, it instead peaked on the abstract dataset (Inspec), where prioritization opportunities per document are $9.7$. When such opportunities exceeded or fell below that level, LMRank struggled with prioritization (e.g., SemEval--2017 and WWW).

\subsection{Experiment 2: Edge Weighting}\label{subsec:experiment 2}

\medskip

We now report the EW--tailored experimental setup (Subsubsection \ref{subsubsec:experimental setup for ew}) and which of the selected EW measures maximized SN interpretability (Subsubsection \ref{subsubsec:results of ew}).

\subsubsection{Experimental Setup for EW}\label{subsubsec:experimental setup for ew}

\medskip

\noindent \textbf{Implementations}. DMD was implemented based on \cite{b159}, while the remaining measures were implemented by us.

\medskip

\noindent \textbf{Parameters}. Given that complex SNs overload human cognitive capacity, setting resolution parameters is essential. Accordingly, $l$ and $\omega$ were set to the author--fit levels of $50$ vertices and $100$ edges, respectively. For the null model generator (Algorithm \ref{Algorithm 1}), $h$ and $z$ were set to $10^{4}$ and $1.0$, respectively, to secure the discriminative power of null models.

\medskip

\noindent \textbf{Vertex Random Failure Simulations}. We evaluated each row SN based on its average $\text{RI}$ obtained from $10^{3}$ simulation runs, where the random seed is fixed to $2026$. A single run may yield unreliable $f_{c}^{\text{empirical}}$ and $f_{c}^{\text{null}}$ due to the risk of premature hub removals. In contrast, $10^{3}$ substantially exceeds the empirical threshold of $30$ required for the Central Limit Theorem to hold while avoiding computational overhead.

\subsubsection{Results of EW}\label{subsubsec:results of ew}

\medskip

\noindent \textbf{Veracity Pretest}. As explained in Subsubsection \ref{subsubsec:separate penguin from orca}, we conducted the minimal veracity pretest on the EW measures. CF, Dice, Jaccard, and Cosine passed the test, as they recorded $\text{MCC}=1$ for all $\mathbf{DTM}$s across every dataset (Table \ref{table:Table 11}).

\begin{table}[h!]
\caption{Veracity Pretest Results of EW Measures (P: PASS, F: FAIL)}\label{table:Table 11}
\centering
\begin{tabular}{>{\arraybackslash}m{0.8cm}|>
{\arraybackslash}m{0.35cm}|>{\arraybackslash}m{0.55cm}|>{\arraybackslash}m{0.85cm}|>{\arraybackslash}m{1.15cm}|>
{\arraybackslash}m{0.8cm}|>
{\arraybackslash}m{0.95cm}}
Data & CF & Dice & Jaccard & Euclidean & Cosine & DMD \\
\hline\hline
Sem10 & P & P & P & F: $0/8$ & P & F: $0/8$ \\ \hline
NUS & P & P & P & F: $0/8$ & P & F: $1/8$ \\ \hline
Inspec & P & P & P & F: $0/8$ & P & F: $0/8$ \\ \hline
KDD & P & P & P & F: $0/10$ & P & F: $0/10$ \\ \hline
WWW & P & P & P & F: $0/8$ & P & F: $0/8$ \\ \hline
Sem17 & P & P & P & F: $0/10$ & P & F: $0/10$ \\ \hline
DUC & P & P & P & F: $0/8$ & P & F: $0/8$ \\ \hline
500N & P & P & P & F: $0/8$ & P & F: $0/8$ \\
\end{tabular}
\end{table}

In contrast, Euclidean and DMD failed the test, consistently recording $\text{MCC}<1$. DMD recorded $\text{MCC}=1$ for only a single $\mathbf{DTM}$ generated from NUS, but that instance is trivial. While the other measures exhibit their anti--false edge abilities in SNC, Euclidean and DMD almost always generate at least one false edge. Therefore, these measures were excluded from the subsequent experiments.

\medskip

\noindent \textbf{Outliers}. To obtain true $f_{c}$ by removing the topological \emph{shielding} effect \cite{b141}, Algorithm \ref{Algorithm 1} requires its input $\mathbf{G}$ to be connected. In other words, $f_{c}$ can be overestimated when isolated vertices are removed earlier, even when the actual size of $\mathbf{G}_{\text{GC}}$ is small. Hence, only four of the $272$ raw SNs were excluded. They were all SNs extracted from the $\mathbf{DTM}$ built by using LMRank for SemEval--2017. While the use of MST backbones with $l=50$ and $\omega=100$ is sufficient to ensure connected SNs, such corner cases seldom arise when multigrams with sparse co--occurrences are extracted as keyphrases. However, these four outliers were not entirely excluded from the PoC, as further discussed in Section \ref{sec:global optimization}.

\medskip

\noindent \textbf{Statistical Analysis}. If Algorithm \ref{Algorithm 1} cannot decrease $L(\mathbf{G})$ and $T_{\triangle}(\mathbf{G})$ within $h$ iterations, it returns a mere copy of the SN ($\mathbf{G}$) as its null model $\mathbf{G}_{\text{null}}$. As there is no difference between $f_{c}^{\text{empirical}}$ and $f_{c}^{\text{null}}$, \eqref{Eq 7} assigns a zero--$\text{RI}$ to $\mathbf{G}$. Thus, $\text{RI}$ scores of EW measures sometimes follow zero--inflated distributions at the dataset level. Interestingly, certain SNC pipelines are prone to yielding such zero--$\text{RI}$ SNs, whose details are further discussed in the qualitative analysis.

Indeed, Table \ref{table:Table 12} shows the low average $\text{RI}$ scores and high variances. To examine normality, the K--S test and the Shapiro--Wilk (S--W) test were carefully considered but found to be inappropriate (Appendix \ref{Appendix G}). Accordingly, we focused on the Q--Q plots to visually inspect the normality. The Q--Q plots \href{https://github.com/potentialreviewer/Ha-Kim-2026a/blob/main/notebooks/Edge_Weighting.ipynb}{(link)} indicated that the normality was clearly violated for $23$ of the $32$ $\text{RI}$ distributions. Every 95\% BCI (with $10^{4}$ resamples) of the average $\text{RI}$ scores was also wide \href{https://github.com/potentialreviewer/Ha-Kim-2026a/blob/main/notebooks/Edge_Weighting.ipynb}{(link)}, supporting the non--parametric approach based on rank transformation per distribution (Subsubsection \ref{subsubsec:results of ake}).

\begin{table}[h!]
\caption{$\text{RI}$--based Evaluation Results ($32$ Pairs of Avgs and Stds)}\label{table:Table 12}
\centering
\begin{tabular}{>{\arraybackslash}m{0.8cm}>
{\arraybackslash}m{0.5cm}|>{\arraybackslash}m{0.8cm}|>{\arraybackslash}m{0.8cm}|>
{\arraybackslash}m{1.0cm}|>
{\arraybackslash}m{0.9cm}}
Data & $\text{RI}$ & CF & Dice & Jaccard & Cosine \\
\hline\hline
\multirow{2}{0.8cm}{Sem10} & Avg & $0.015$ & \boldsymbol{$0.018$} & \boldsymbol{$0.018$} & \underline{\boldsymbol{$0.042$}} \\
& Std & $0.043$ & $0.023$ & $0.023$ & $0.054$ \\ \hline
\multirow{2}{0.8cm}{NUS} & Avg & $0.000$ & \boldsymbol{$0.034$} & \boldsymbol{$0.034$} & \underline{\boldsymbol{$0.056$}} \\
& Std & $0.000$ & $0.058$ & $0.058$ & $0.060$ \\ \hline
\multirow{2}{0.8cm}{Inspec} & Avg & $0.049$ & \boldsymbol{$0.146$} & \boldsymbol{$0.146$} & \underline{\boldsymbol{$0.168$}} \\
& Std & $0.086$ & $0.128$ & $0.128$ & $0.076$ \\ \hline
\multirow{2}{0.8cm}{KDD} & Avg & $0.030$ & \boldsymbol{$0.085$} & \boldsymbol{$0.085$} & \underline{\boldsymbol{$0.162$}} \\
& Std & $0.064$ & $0.125$ & $0.125$ & $0.116$ \\ \hline
\multirow{2}{0.8cm}{WWW} & Avg & $0.032$ & \boldsymbol{$0.114$} & \boldsymbol{$0.114$} & \underline{\boldsymbol{$0.216$}} \\
& Std & $0.057$ & $0.142$ & $0.142$ & $0.126$ \\ \hline
\multirow{2}{0.8cm}{Sem17} & Avg & $0.058$ & \boldsymbol{$0.176$} & \boldsymbol{$0.176$} & \underline{\boldsymbol{$0.264$}} \\
& Std & $0.056$ & $0.076$ & $0.076$ & $0.071$ \\ \hline
\multirow{2}{0.8cm}{DUC21} & Avg & $0.051$ & \boldsymbol{$0.110$} & \boldsymbol{$0.110$} & \underline{\boldsymbol{$0.125$}} \\
& Std & $0.079$ & $0.057$ & $0.057$ & $0.060$ \\ \hline
\multirow{2}{0.8cm}{500N} & Avg & $0.006$ & \boldsymbol{$0.046$} & \boldsymbol{$0.046$} & \underline{\boldsymbol{$0.068$}} \\
& Std & $0.016$ & $0.060$ & $0.060$ & $0.076$ \\
\end{tabular}
\end{table}

Subsequently, to determine whether significant differences exist among the averages (Table \ref{table:Table 12}), Friedman followed by Nemenyi was performed at the dataset level. Here, a $\mathbf{DTM}$, an EW measure, and a resulting $\text{RI}$ score distribution served as a block, a treatment, and a sample, respectively. The number of blocks (i.e., the number of AKE algorithms used) per dataset was small, meaning that each distribution contains only $8$ to $10$ observations. As Friedman is prone to Type II errors in such cases \cite{b161}, we incorporated the Iman--Davenport correction \cite{b161} to obtain more reliable $p$--values.

\begin{figure*}[t!]
    \centering    \includegraphics[width=0.9\textwidth]{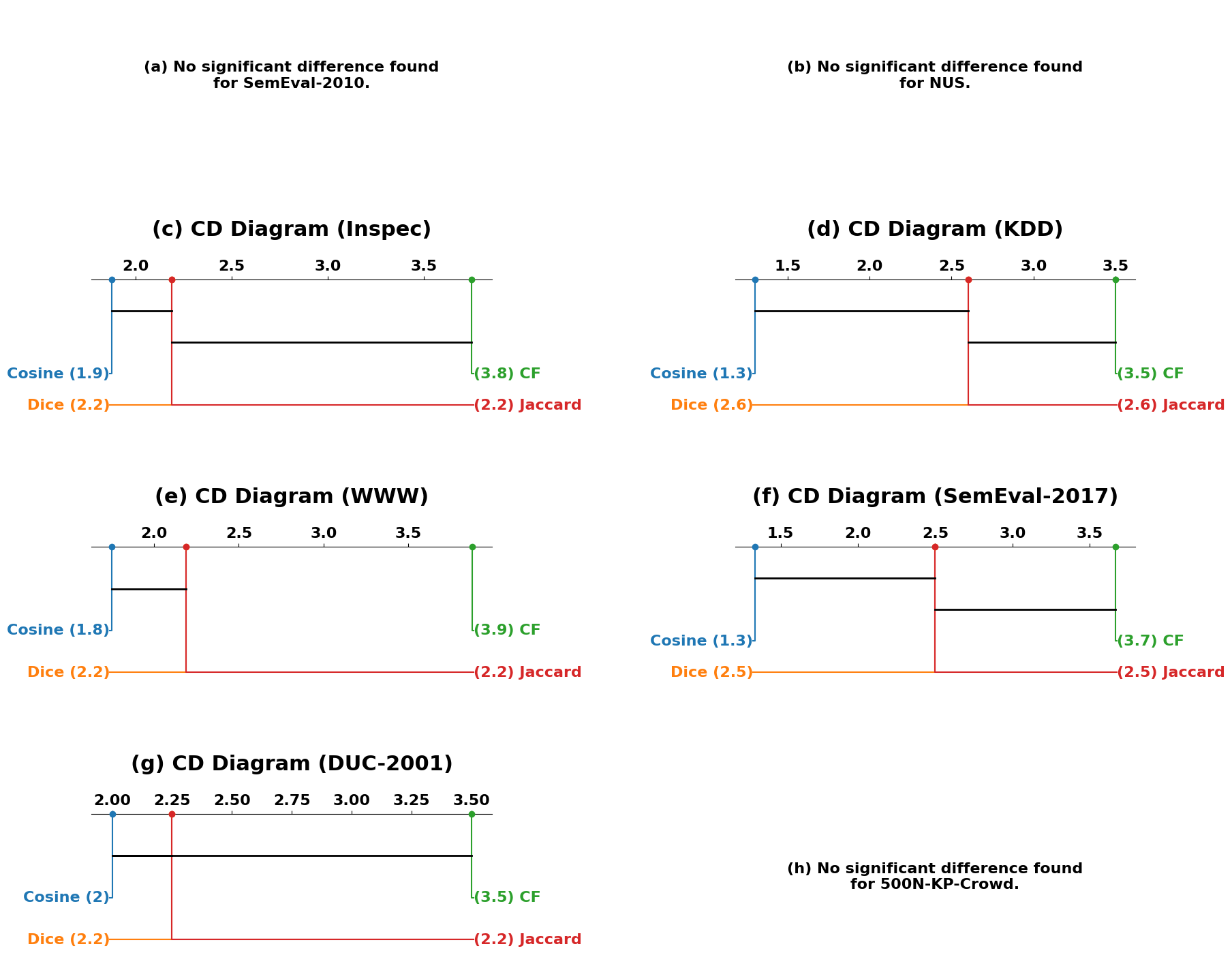}
    \caption{\textbf{Critical Difference Diagrams of EW Measures across Datasets. Each value in parentheses is an average of RI ranks across $\mathbf{DTM}$s.}}
    \label{Figure 14}
\end{figure*}

Interestingly, clear significant differences in the average $\text{RI}$ ranks ($p < 0.05$) were observed only for the short document datasets (Inspec, KDD, WWW, and SemEval--2017). While Friedman with Iman--Davenport activated the CD diagram for DUC--2001, Nemenyi conservatively tied all measures together by a string (Fig. \ref{Figure 14}--g). All $p$--values are provided separately \href{https://github.com/potentialreviewer/Ha-Kim-2026a/blob/main/notebooks/Edge_Weighting.ipynb}{(link)} for those seeking further verification.

We clarify that SNC ultimately depends on method selection pipelines. Although the presence or absence of statistically significant differences facilitates a better understanding of EW measures, even a marginal numerical difference in $\text{RI}$ can be \emph{factored into} a global SNC pipeline (Subsection \ref{subsec:ablation study}). Maintaining this premise, we provide a post--hoc interpretation. In particular, why did Cosine significantly outperform CF across half of the datasets (Figs. \ref{Figure 14}--c, d, e, and f)?

\medskip

\noindent \textbf{Qualitative Analysis}. Our tentative answer is that while certain SNC pipelines \emph{flexibly} incorporate contextual information, others \emph{rigidly} adhere to Co--occurrence Frequency Information (CFI), resulting in differences in $\text{RI}$ scores.

\begin{figure*}[t!]
    \centering    \includegraphics[width=0.9\linewidth]{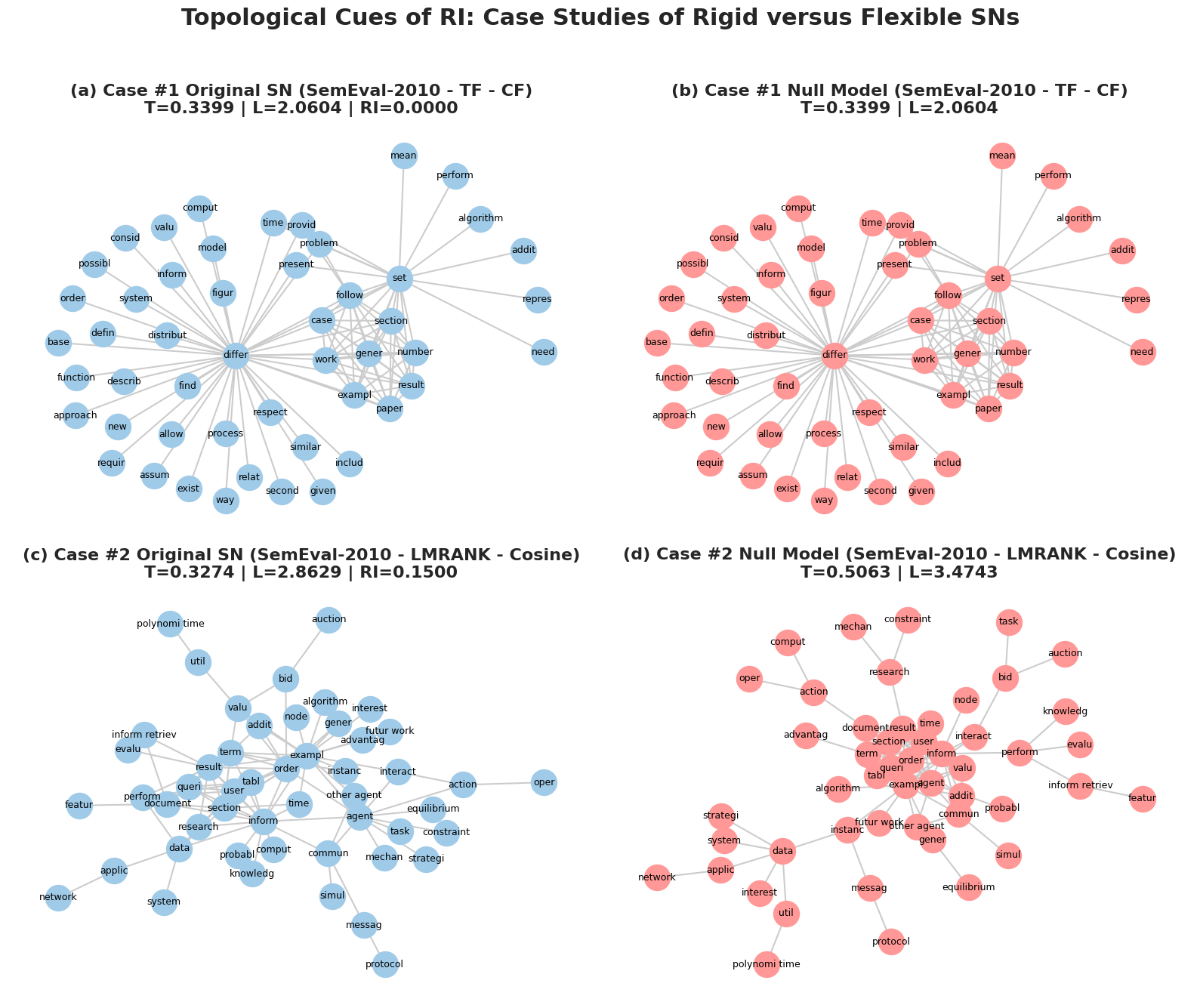}
    \caption{\textbf{Cases of Rigid SN (Top) and Flexible SN (Bottom). T and L represent transitivity and the average shortest path length, respectively.}}
    \label{Figure 15}
\end{figure*}

More specifically, when CF paired with TF was applied to SemEval--2010, a \emph{partially star--topological} SN, hereafter referred to as a \emph{rigid} SN, was formed by concentrating edges to a \emph{few} dominantly co--occurring keyphrases (Fig. \ref{Figure 15}--a). Most $k_{i} \in \mathbf{K}$ were also pinned to $1$, thereby making edge--rewiring--based Algorithm \ref{Algorithm 1} return a mere copy of the SN (Fig. \ref{Figure 15}--b). So, zero--$\text{RI}$ was yielded (Fig. \ref{Figure 15}-a). Algorithm \ref{Algorithm 1} precisely prevents such pipelines from obtaining high $\text{RI}$.

Why should we \emph{prevent} this? Even if we do not demand that EW measures go beyond the level of minimal facet veracity, we can still consider \emph{how many} facet \emph{types} each measure can consider. It is also self--evident that any measure that affords a researcher broader room for diverse interpretations is more desirable in terms of abduction. So, we give edge--prioritization opportunities of fair $\omega$ to every measure. If certain measures nonetheless adhere to CFI and exhaust $\omega$ for dominant keyphrases, thereby diminishing such room, it is karma\footnote{Indeed, when an output remains at its baseline, $\frac{\text{Output}-\text{Baseline}}{\text{Maximum Possible}-\text{Baseline}}$ \cite{b128} regards the treatment effect as \emph{zero} (Table \ref{table:Table 7}).} that the measures are penalized.

\begin{figure*}[t!]
    \centering    \includegraphics[width=\linewidth]{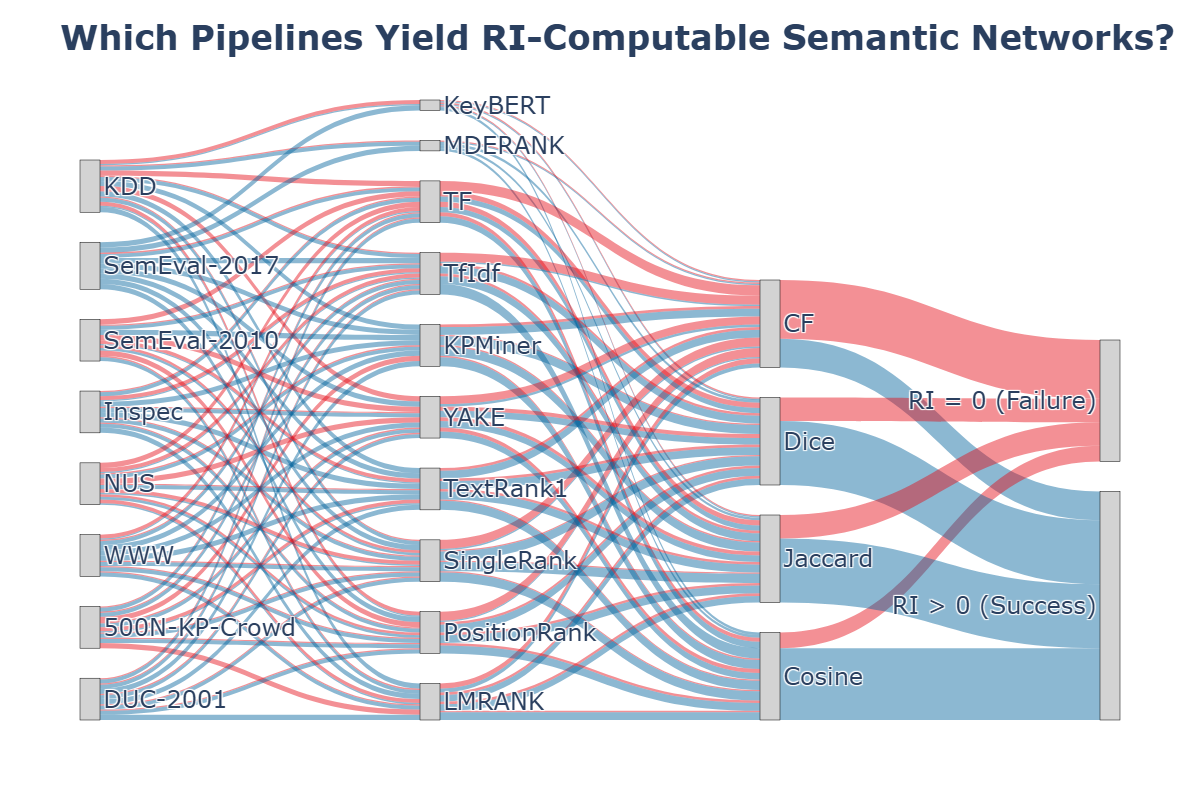}
    \caption{\textbf{Sankey Diagram of SNC Pipelines. Pipelines with the same outcome are merged into a superpipeline. The red and blue pipelines indicate failures ($\text{RI}=0$) and successes ($\text{RI}>0$), respectively. To accommodate colorblind readers, an interactive version of the diagram can be rendered by downloading and executing our code and data \href{https://github.com/potentialreviewer/Ha-Kim-2026a/blob/main/notebooks/Edge_Weighting.ipynb}{(link)}. In that version, the flow of each pipeline (e.g., TF $\rightarrow$ CF) can be examined. For any starting point (dataset, AKE algorithm, or EW measure), the collective magnitude of subpipelines that participate in the flow from the point can also be examined.}}
    \label{Figure 16}
\end{figure*}

Can we, then, observe an actual pattern where CFI--biased EW measures induce zero--$\text{RI}$ inflation? Yes. The Sankey diagram \cite{b162} in Fig. \ref{Figure 16} visualizes which pipelines succeeded or failed in constructing raw SNs with $\text{RI} > 0$ upon completing the EW stage. CF, Dice--Jaccard, and Cosine generated zero--$\text{RI}$ SNs in this order. This is a natural outcome, given that CF considers solely CFI, Dice--Jaccard merely adds normalization to CFI (Table \ref{table:Table 4}), and for Cosine, target vectors can be populated with entries that can signify diverse semantics depending on which AKE algorithms it is partnered with.

As may be noticed, not only EW measures but also AKE algorithms can serve as relevant factors. For instance, since it was ultimately its partner that passed contextual information to Cosine, when LMRank paired with Cosine was applied to SemEval--2010, a \emph{partially mesh--topological} SN with $\text{RI}>0$, hereafter referred to as a \emph{flexible} SN, was successfully formed. Its many vertices certainly appear to satisfy $k_{i}>1$ (Fig. \ref{Figure 15}--c). According to Fig. \ref{Figure 16}, pipelines based on KP--Miner (which uses a strict \emph{cutoff}), TextRank (which exploits a word co--occurrence \emph{binary} graph), or NN--based algorithms generated fewer zero--$\text{RI}$ SNs. In contrast, pipelines based on SingleRank or PositionRank (which exploit word co--occurrence \emph{frequency} graphs), TF, TF--IDF, or YAKE! generated more zero--$\text{RI}$ SNs.

Here, an interesting observation is that long document datasets can \emph{structurally} inject CFI into SNC pipelines. The long document datasets (SemEval--2010 and NUS) contain a large number of tokens, i.e., rich CFI (Table \ref{table:Table 8}). Indeed, for these datasets, significant differences among the EW measures completely vanished (Figs. \ref{Figure 14}--a and b). This aligns with Fig. \ref{Figure 16}, which says that only $34.4$\% ($11/32$) of pipelines for long--document NUS yielded flexible SNs, whereas short--document SemEval--2017 yielded the most at $97.2$\% ($35/36$).

\begin{table*}[t!]
\caption{Results of Spearman's $\rho$ Correlation Analysis between RI and IP}\label{table:Table 13}
\centering
\begin{tabular}{>{\arraybackslash}m{2.3cm}|>{\arraybackslash}m{0.4cm}|>
{\arraybackslash}m{0.4cm}|>
{\arraybackslash}m{0.4cm}|>
{\arraybackslash}m{0.7cm}|>
{\arraybackslash}m{1.8cm}|>
{\arraybackslash}m{2.5cm}|>
{\arraybackslash}m{0.7cm}}
\hline
Type & $X$ & $Y$ & $Z$ & $\rho$ & $95$\% BCI with \newline $10^{4}$ resamples & $p$--value & Power \\
\hline\hline
General $\rho(X\text{, }Y)$ & $\text{RI}$ & $\text{IP}_{1}$ & - & $0.914$ & $(0.892,0.929)$ & $(3.961 \times 10^{-141})^{***}$ & $1.000$ 
\\ \hline
General $\rho(X\text{, }Y)$ & $\text{RI}$ & $\text{IP}_{2}$ & - & $0.874$ & $(0.841,0.899)$ & $(2.762 \times 10^{-107})^{***}$ & $1.000$ 
\\ \hline
Partial $\rho(X.Z\text{, }Y.Z)$ & $\text{RI}$ & $\text{IP}_{1}$ & $\text{IP}_{2}$ & $0.597$ & $(0.496,0.682)$ & $(4.337 \times 10^{-29})^{***}$ & $1.000$ 
\\ \hline
Partial $\rho(X.Z\text{, }Y.Z)$ & $\text{RI}$ & $\text{IP}_{2}$ & $\text{IP}_{1}$ & $0.276$ & $(0.138,0.407)$ & $(4.094 \times 10^{-6})^{***}$ & $0.996$ 
\\ \hline
\end{tabular}
\end{table*}

Nonetheless, document lengths do not solely determine CFI. KDD is a short--document dataset comparable to SemEval--2017, but only $57.5$\% ($23/40$) of pipelines for KDD yielded flexible SNs. Why? KDD \cite{b147} consists of certain abstracts from proceedings of the ACM Conference on Knowledge Discovery and Data Mining (KDD). They share \emph{domain--specific} terms, i.e., rich CFI.

\medskip

\noindent \textbf{RI Validation}. As explained in Subsubsection \ref{subsubsec:robustness improvement}, we conducted $\text{RI}$ validation. We first conducted the normality test on $\text{IP}_{1}$, $\text{IP}_{2}$, and $\text{RI}$ for all $268$ raw SNs across the datasets. According to the results of the K--S test ($p<0.05$) and visual inspection of the Q--Q plots \href{https://github.com/potentialreviewer/Ha-Kim-2026a/blob/main/notebooks/Edge_Weighting.ipynb}{(link)}, normality was not confirmed for any of the variables. Accordingly (cf. Subsubsection \ref{subsubsec:results of ake}), we conducted Spearman's $\rho$ analysis \cite{b98}, with the results presented in Table \ref{table:Table 13}. Herein, $X.Z$ [$Y.Z$] denotes variable $X$ [$Y$] after controlling for the effect of $Z$ in $X$ [$Y$]; $p^{***}$ indicates that $p \leq 0.001$, confirming that the null hypothesis of no correlation was clearly rejected.

First, the general correlations $\rho(\text{RI}\text{, }\text{IP}_{1})$ and $\rho(\text{RI}\text{, }\text{IP}_{2})$ were high at $0.914$ and $0.874$, respectively, strongly supporting the premise that higher $\text{RI}$ aligns with higher $\text{IP}$. Second, the partial correlation $\rho(\text{RI}.\text{IP}_{2}, \text{IP}_{1}.\text{IP}_{2})$ remained moderate at $0.597$, whereas $\rho(\text{RI}.\text{IP}_{1}, \text{IP}_{2}.\text{IP}_{1})$ was weak at $0.276$. That is, the strong general correlation between $\text{RI}$ and $\text{IP}$ was driven primarily by $\rho(\text{RI}.\text{IP}_{2}, \text{IP}_{1}.\text{IP}_{2})$ and supplemented by $\rho(\text{RI}.\text{IP}_{1}, \text{IP}_{2}.\text{IP}_{1})$. Nonetheless, this observed weak $\rho(\text{RI}.\text{IP}_{1}, \text{IP}_{2}.\text{IP}_{1})$ does not diminish the validity of $\text{RI}$:

Although $\text{RI}$ rewards smooth semantic percolation, simultaneously maximizing both $\text{IP}_{1}$ and $\text{IP}_{2}$ involves an inherent structural trade--off. When hubs cluster, $L(\mathbf{G})$ tends to decrease, but because triangles also tend to increase, $T_{\triangle}(\mathbf{G})$ tends to increase (e.g., small--world networks). Despite this constraint, $\text{RI}$ is assigned to SNs satisfying $L(\mathbf{G}_{\text{null}}) \geq L(\mathbf{G})$ and $T_{\triangle}(\mathbf{G}_{\text{null}}) \geq T_{\triangle}(\mathbf{G})$ (Algorithm \ref{Algorithm 1}). Therefore, the observed weak $\rho(\text{RI}.\text{IP}_{1}, \text{IP}_{2}.\text{IP}_{1})$ suggests that certain nontrivial SNs achieved relatively high $\text{IP}_{2}$ despite this constraint, rather than invalidating the relationship between $\text{RI}$ and $\text{IP}_{2}$.

\subsection{Experiment 3: Community Detection}\label{subsec:experiment 3}

\medskip

We now report the CD--tailored experimental setup (Subsubsection \ref{subsubsec:experimental setup for cd}) and which of the selected EW measures maximized SN distinctiveness (Subsubsection \ref{subsubsec:results of cd}).

\subsubsection{Experimental Setup for CD}\label{subsubsec:experimental setup for cd}

\medskip

\noindent \textbf{Implementations}. We implemented Leiden and the other selected CD algorithms based on Python igraph \cite{b163} and NetworkX \cite{b164}, respectively.

\medskip

\noindent \textbf{Parameters}. First, while not detailed earlier, $Q$ has a resolution parameter $\gamma>0$, where higher or lower $\gamma$ is likely to create more or fewer communities, respectively \cite{b118}:

\begin{equation} \label{Eq 9}
\begin{aligned}
Q=\frac{1}{2m}\sum_{ij}\bigg[\mathbf{A}_{ij}-\gamma\frac{k_{i}k_{j}}{2m}\bigg]\delta(c_{i},c_{j})
\end{aligned}
\end{equation}

\medskip

\noindent If $\gamma$ is individually tuned for each SN, the SNs would correspond to different objective functions, rendering a fair comparison impossible. $\gamma$ is conventionally fixed at $1$ to neutralize its influence, a practice followed in this paper. Readers may opt for their own $\gamma$ in future applications.

Second, unlike the original Leiden paper \cite{b118}, the implementation provided by igraph requires users to specify the number of algorithm iterations. Our informal check based on Optuna \cite{b165}, a Python library for automated hyperparameter search, indicated that, for most SNs in this PoC, fewer than $200$ iterations were sufficient to reach their maximum Leiden $Q$. We fixed it at $200$ to save time. Third, the randomness parameter $\beta$ of Leiden was tuned for each SN by using Optuna. Exploring an extensive range of $\beta$ with numerous trials per SN is time--consuming. We found that most SNs in this PoC reached maximum Leiden $Q$ within $50$ trials in $0.1 \leq \beta \leq 10$. We selected these time--manageable compromises.

\subsubsection{Results of CD}\label{subsubsec:results of cd}

\medskip

\noindent \textbf{Statistical Analysis}. CD was conducted on the 272 raw SNs, including the four outliers that had been excluded at the EW stage. $95$\% BCIs of these average $Q$ scores (Table \ref{table:Table 14}) are provided separately \href{https://github.com/potentialreviewer/Ha-Kim-2026a/blob/main/notebooks/Community_Detection.ipynb}{(link)}. For these $32$ $Q$ distributions ($8$ datasets times $4$ algorithms), while the K--S test indicated that only the FLPA distribution for NUS violated normality, the Q--Q plots of FLPA violated normality across all datasets \href{https://github.com/potentialreviewer/Ha-Kim-2026a/blob/main/notebooks/Community_Detection.ipynb}{(link)}. Accordingly, we decided to maintain the non--parametric approach to ensure consistency in our analysis.

\begin{table}[h!]
\caption{Summary of $32$ Modularity $Q$ Distributions}\label{table:Table 14}
\centering
\begin{tabular}{>{\arraybackslash}m{0.8cm}>
{\arraybackslash}m{0.5cm}|>{\arraybackslash}m{0.8cm}|>{\arraybackslash}m{1.0cm}|>
{\arraybackslash}m{1.0cm}|>
{\arraybackslash}m{0.9cm}}
Data & $Q$ & CNM & Louvain & Leiden & FLPA \\
\hline\hline
\multirow{2}{0.8cm}{Sem10} & Avg & $0.312$ & \boldsymbol{$0.317$} & \underline{\boldsymbol{$0.321$}} & $0.117$ \\
& Std & $0.075$ & $0.076$ & $0.075$ & $0.118$ \\ \hline
\multirow{2}{0.8cm}{NUS} & Avg & $0.348$ & \boldsymbol{$0.351$} & \underline{\boldsymbol{$0.354$}} & $0.162$ \\
& Std & $0.094$ & $0.094$ & $0.094$ & $0.187$ \\ \hline
\multirow{2}{0.8cm}{Inspec} & Avg & \boldsymbol{$0.402$} & $0.398$ & \underline{\boldsymbol{$0.408$}} & $0.228$ \\
& Std & $0.124$ & $0.123$ & $0.123$ & $0.196$ \\ \hline
\multirow{2}{0.8cm}{KDD} & Avg & $0.346$ & \boldsymbol{$0.349$} & \underline{\boldsymbol{$0.353$}} & $0.198$ \\
& Std & $0.135$ & $0.136$ & $0.134$ & $0.183$ \\ \hline
\multirow{2}{0.8cm}{WWW} & Avg & \boldsymbol{$0.391$} & $0.388$ & \underline{\boldsymbol{$0.394$}} & $0.245$ \\
& Std & $0.130$ & $0.130$ & $0.130$ & $0.207$ \\ \hline
\multirow{2}{0.8cm}{Sem17} & Avg & $0.429$ & \boldsymbol{$0.431$} & \underline{\boldsymbol{$0.438$}} & $0.258$ \\
& Std & $0.105$ & $0.105$ & $0.104$ & $0.198$ \\ \hline
\multirow{2}{0.8cm}{DUC21} & Avg & $0.429$ & \boldsymbol{$0.430$} & \underline{\boldsymbol{$0.435$}} & $0.294$ \\
& Std & $0.178$ & $0.179$ & $0.176$ & $0.274$ \\ \hline
\multirow{2}{0.8cm}{500N} & Avg & \boldsymbol{$0.406$} & $0.406$ & \underline{\boldsymbol{$0.410$}} & $0.303$ \\
& Std & $0.127$ & $0.131$ & $0.129$ & $0.193$ \\
\end{tabular}
\end{table}

\begin{figure*}[t!]
    \centering    \includegraphics[width=0.9\textwidth]{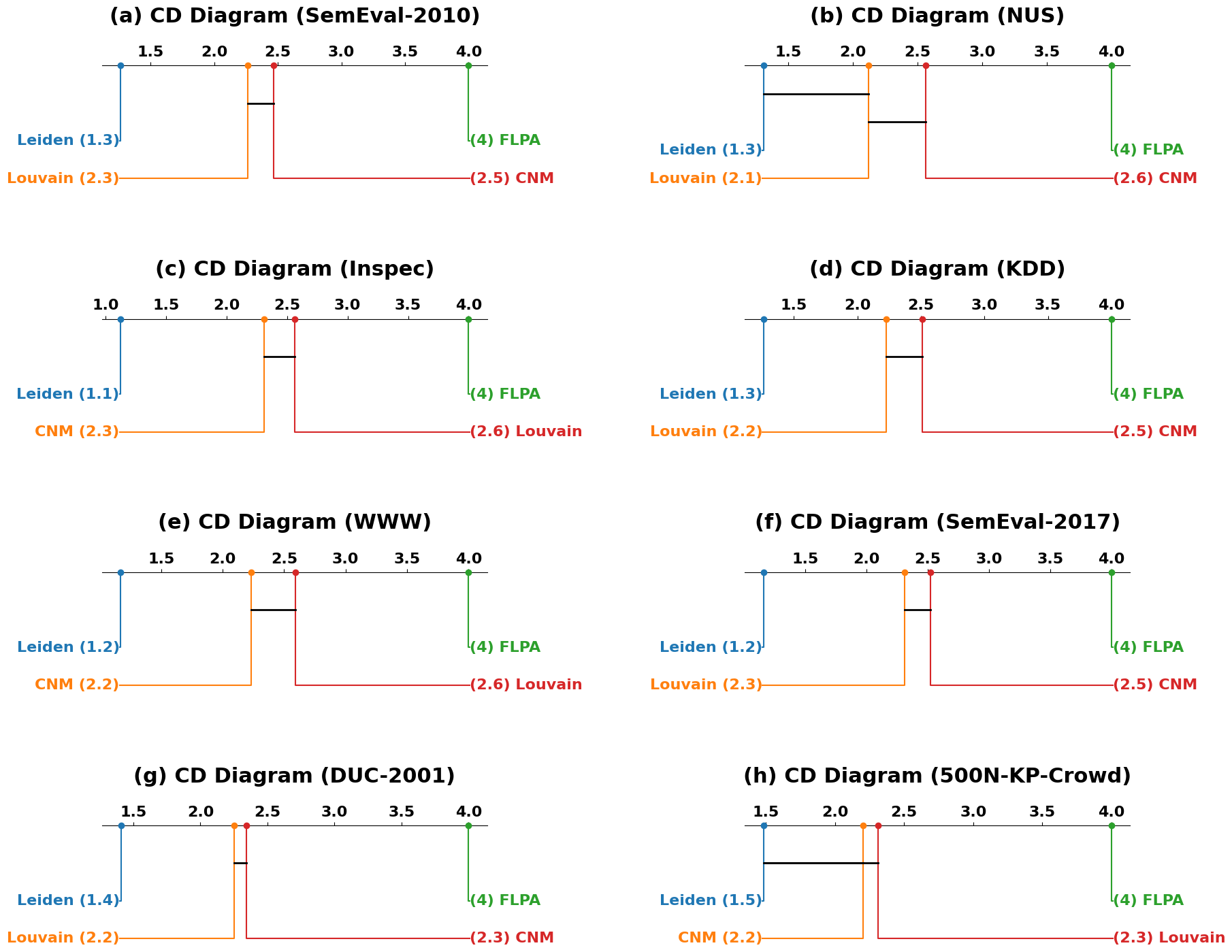}
    \caption{\textbf{Critical Difference Diagrams of CD Algorithms across Benchmark Datasets. Each value in parentheses is an average of $Q$ ranks across SNs.}}
    \label{Figure 17}
\end{figure*}

To determine whether significant differences exist among the average $Q$ ranks, Friedman followed by Nemenyi was performed at the dataset level. A raw SN, a CD algorithm, and a resulting $Q$ score distribution corresponded to a block, a treatment, and a sample, respectively. For all datasets, Friedman indicated that significant differences in the average $Q$ ranks existed among at least one pair of CD algorithms ($p<0.05$). Nemenyi then identified which pairs exhibited significance at the dataset level (Fig. \ref{Figure 17}). All $p$--values have been disclosed separately \href{https://github.com/potentialreviewer/Ha-Kim-2026a/blob/main/notebooks/Community_Detection.ipynb}{(link)} for brevity.

\medskip

\noindent \textbf{Qualitative Analysis}. Based on the significant differences ($p<0.05$) shown in Fig. \ref{Figure 17} and the statistics in Table \ref{table:Table 14}, we provide the following post--hoc interpretation.

First, the average $Q$ scores followed the descending order of Leiden, Louvain, CNM, and FLPA across most datasets (Table \ref{table:Table 14}). These results were as expected, given that the heuristic--based FLPA did not explicitly optimize $Q$, whereas the other three algorithms followed the incremental advancement path of modularity--based algorithms.

Second, however, no significant differences in average $Q$ ranks were observed between Louvain and CNM, the middle--tier algorithms (Fig. \ref{Figure 17}). Even for a few datasets (Inspec, WWW, and 500N--KP--Crowd), CNM's greedy optimization surpassed Louvain's approach in terms of average $Q$ scores, though the differences were marginal (Table \ref{table:Table 14}). This lack of distinction likely stemmed from the sparsity of vertex adjacency information and edge weight information, as the size of each SN was small ($l=50$ and $\omega=100$). Algorithms capable of maximizing an objective function even under such information scarcity can be prioritized for future applications.

\section{Global Optimization}\label{sec:global optimization}

\medskip

In this section, Subsection \ref{subsec:problem reformulation} first answers \textbf{RQ7} (``How are local evaluation results across the SNC stages integrated to identify the optimal SN?''). To answer upfront, the results are integrated by reformulating SNC as a Process Optimization Problem (POP) and factoring the results into a global objective function $J$. Here, an SNC process that maximizes $J$ is regarded as the optimal policy (or pipeline) that generates the optimal SN for a textual dataset. However, providing $J$ in isolation may pose challenges for its reuse.

Thus, Subsection \ref{subsec:cluenetwork} presents the comprehensive framework \emph{ClueNetwork}, which incorporates all the key components (\eqref{Eq 1}, \eqref{Eq 2}, $l$, $\omega$, Algorithm \ref{Algorithm 1}, $h\text{F}_{1}$, $\text{RI}$, a CSF, the veracity pretest, and $J$). As such, \emph{ClueNetwork} facilitates \emph{ranking} candidate policies (ultimately their resulting SNs) for a dataset. Of course, Subsection \ref{subsec:justification of j} answers \textbf{RQ8} (``How is $J$ defined and justified?''), thereby establishing the framework's core, $J$, not as a mere heuristic, but as a scientifically realistic criterion. Only after that does Subsection \ref{subsec:ablation study} finalize the PoC of \emph{ClueNetwork} by demonstrating that it is an actually working framework through an ablation study.

\subsection{Problem Reformulation}\label{subsec:problem reformulation}

\medskip

\noindent\textbf{Semantic Network Construction (SNC) is a Process Optimization Problem (POP)}. A POP can be regarded as a problem divided into multiple stages, each requiring a decision \cite{b166, b167}. Specifically, the current $i$\textsuperscript{th} stage takes one of its possible states $\big(s_{i} \in \mathbf{S}_{i}=\{s_{i,1}, s_{i,2},\dots,s_{i,|
\mathbf{S}_{i}|}\}\big)$ associated with a decision ($x_{i-1}$) at the previous $(i-1)$\textsuperscript{th} stage. Then, a decision at the current stage $\big(x_{i} \in \mathbf{X}_{i}=\{x_{i,1}, x_{i,2},\dots,x_{i,|
\mathbf{X}_{i}|}\}\big)$ transitions the current state ($s_{i}$) into a state ($s_{i+1}$) at the subsequent $(i+1)$\textsuperscript{th} stage.

\begin{figure}[h!]
    \centering   \includegraphics[width=\linewidth]{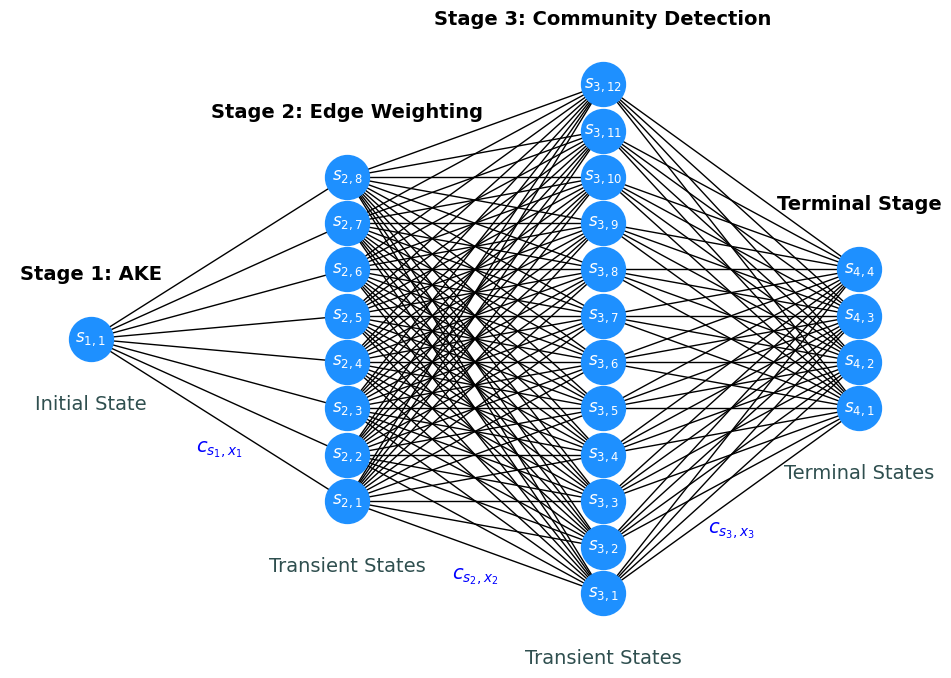}
    \caption{\textbf{SNC for a Given Dataset.}}
    \label{Figure 18}
\end{figure}

SNC precisely follows this structure. That is, the selections of an AKE algorithm ($x_{1}$), EW measure ($x_{2}$), and CD algorithm ($x_{3}$) at the first, second, and third stages determine a $\mathbf{DTM}$ ($s_{2}$), raw SN ($s_{3}$), and final SN ($s_{4}$). Such selections yield function values $h\text{F}_{1}(x_{1})$, $\text{RI}(x_{2})$, and $Q(x_{3})$. Accordingly, components of SNC can be reformulated as follows:

\medskip

\begin{itemize}
    \item $\mathbf{S}_{1}$: Set of textual datasets in the world.
    \item $s_{1}$: Given dataset, where $s_{1} \in \mathbf{S}_{1}$.
    \item $\mathbf{X}_{1}$: Set of given AKE algorithms.
    \item $x_{1}$: Selected algorithm, where $x_{1} \in \mathbf{X}_{1}$.
    \item $c_{s_{1},x_{1}}=(average)\text{ }h\text{F}_{1}(x_{1})$: Confidence for SN keyness obtained by applying $x_{1}$ to $s_{1}$.
    \item $\mathbf{S}_{2}$: Set of $\mathbf{DTM}$s $\big(|\mathbf{S}_{2}|=|\mathbf{X}_{1}|\big)$.
    \item $s_{2}$: Given $\mathbf{DTM}$, where $s_{2} \in \mathbf{S}_{2}$.
    \item $\mathbf{X}_{2}$: Set of given \emph{veracious} EW measures.
    \item $x_{2}$: Selected measure, where $x_{2} \in \mathbf{X}_{2}$.
\end{itemize}

\medskip

At this point, we clarify that zero--$\text{RI}$ SNs and unconnected SNs (e.g., the four outliers mentioned in Subsubsection \ref{subsubsec:results of ew}) should not be excluded from SNC. Although they exhibit no observable effects (i.e., $\text{RI}$ scores) in terms of SN interpretability, they are also built by using EW measures that pass the (minimal) veracity pretest. Hence, it is fair to assign them the same baseline score of $1$ for SN veracity by separating the dimension of veracity from that of interpretability. That is, $\text{RI}$ scores should be treated as additional points.

\medskip

\begin{itemize}
    \item $c_{s_{2},x_{2}}=\big(1+\text{RI}(x_{2})\big)/2$: Confidence for the overall quality (veracity and interpretability) of interactions between keyphrases in an SN obtained by applying $x_{2}$ to $s_{2}$.
\end{itemize}

\medskip

\noindent Herein, $c_{s_{2},x_{2}}$ is calculated as the average of the baseline score (i.e., $1$) and the actual $\text{RI}$ score ($0 \leq \text{RI} \leq 1$). Then,

\medskip

\begin{itemize}
    \item $\mathbf{S}_{3}$: Set of raw SNs $\big(|\mathbf{S}_{3}|=|\mathbf{X}_{1}|\times|\mathbf{X}_{2}|\big)$.
    \item $s_{3}$: Given raw SN, where $s_{3} \in \mathbf{S}_{3}$.
    \item $\mathbf{X}_{3}$: Set of given CD algorithms.
    \item $x_{3}$: Selected algorithm, where $x_{3} \in \mathbf{X}_{3}$.
    \item $c_{s_{3},x_{3}}=Q(x_{3})$: Confidence for SN distinctiveness obtained by applying $x_{3}$ to $s_{3}$.
    \item $\mathbf{S}_{4}$: Set of final SNs $\big(|\mathbf{S}_{4}|=|\mathbf{X}_{1}|\times|\mathbf{X}_{2}|\times|\mathbf{X}_{3}|\big)$.
    \item $s_{4}$: Constructed final SN, representing a terminal state.
    \item $s_i \xrightarrow{x_i} s_{i+1}$: \emph{Deterministic} move to a subsequent state $s_{i+1}$ resulting from taking $x_{i}$ at the current state $s_{i}$.
\end{itemize}

\medskip

In the fields of operations research \cite{b167} and reinforcement learning \cite{b168}, $x_{i}$, $c_{s_{i},x_{i}}$, and $s_i \xrightarrow{x_i} s_{i+1}$ are termed an ``action (or decision),'' a ``reward,'' and a ``transition,'' respectively. In terms of SNC, the goal is to identify the optimal policy (the sequence of selected actions that \emph{jointly} maximizes a \emph{global confidence} in the optimality of a final SN for a given dataset).

\medskip

\begin{itemize}
    \item $\mathbf{\Pi}$: Set of possible policies, where $|\mathbf{\Pi}|=|\mathbf{S}_{4}|$.
    \item $\pi=(x_{1}, x_{2}, x_{3})$: Policy (process or pipeline), which is a specific combination of an AKE algorithm ($x_{1}$), EW measure ($x_{2}$), and CD algorithm ($x_{3}$), where $\pi \in \mathbf{\Pi}$.
    \item $\pi^{*}=(x_{1}^{*},x_{2}^{*},x_{3}^{*})$: Optimal policy, which is the sequence of selected actions that jointly maximizes a global confidence for SN optimality, where $\pi^{*} \in \mathbf{\Pi}$.
\end{itemize}

\medskip

\noindent Here, to identify $\pi^{*}$ for a given dataset, information regarding possible rewards at each stage is required. Hence,

\medskip

\begin{itemize}
    \item $\mathbf{C}_{i}$: Set of rewards at the $i$\textsuperscript{th} stage, where $c_{s_i, x_i} \in \mathbf{C}_{i}$.
\end{itemize}

\medskip

\noindent\textbf{Global Objective Function}. Based on these components, we can reformulate a joint confidence for SN optimality as a global objective function:

\begin{equation} \label{Eq 10}
\begin{aligned}
& \text{Maximize} \quad J(\pi) = \prod_{i=1}^{3} c_{s_{i},x_{i}} = c_{s_{1},x_{1}} \times c_{s_{2},x_{2}} \times c_{s_{3},x_{3}}, \\ \\
& \text{where} \quad 
\begin{cases} 
c_{s_{1},x_{1}} = h\text{F}_{1}(x_{1}) \\ 
c_{s_{2},x_{2}} = \frac{1+\text{RI}(x_{2})}{2} \\ 
c_{s_{3},x_{3}} = Q(x_{3}) 
\end{cases}, \\ \\
& \text{subject to} \quad 0 \leq c_{s_{i},x_{i}} \leq 1, \;\ \forall \;\ i \in \{1, 2, 3\}.
\end{aligned}
\end{equation}

\medskip

\noindent This multiplicative expression links the local confidences like a chain. \emph{ClueNetwork} regards some $\pi$ maximizing this expression as the optimal\hypertarget{one can demand}{} policy $\pi^{*}$ that yields the optimal SN. Here, one can demand,

\medskip

\begin{itemize}
    \item ``Why do we multiply $c_{s_{1},x_{1}}$, $c_{s_{2},x_{2}}$, and $c_{s_{3},x_{3}}$? Provide \emph{criteria} for the optimal SN and \emph{evidence} that the current expression of $J$ \emph{harmonizes with} them, thereby demonstrating that a higher $J$ guarantees a better SN.''
\end{itemize}

\medskip

\noindent Legitimate, and we will do so in Subsection \ref{subsec:justification of j}. Nonetheless, let us first present \emph{ClueNetwork} under the assumption that $J$ is somehow justified.

\medskip

\subsection{ClueNetwork}\label{subsec:cluenetwork}

\medskip

\subsubsection{Framework}\label{subsubsec:framework}

\medskip

\begin{figure*}[p]
    \centering   \includegraphics[width=0.88\linewidth]{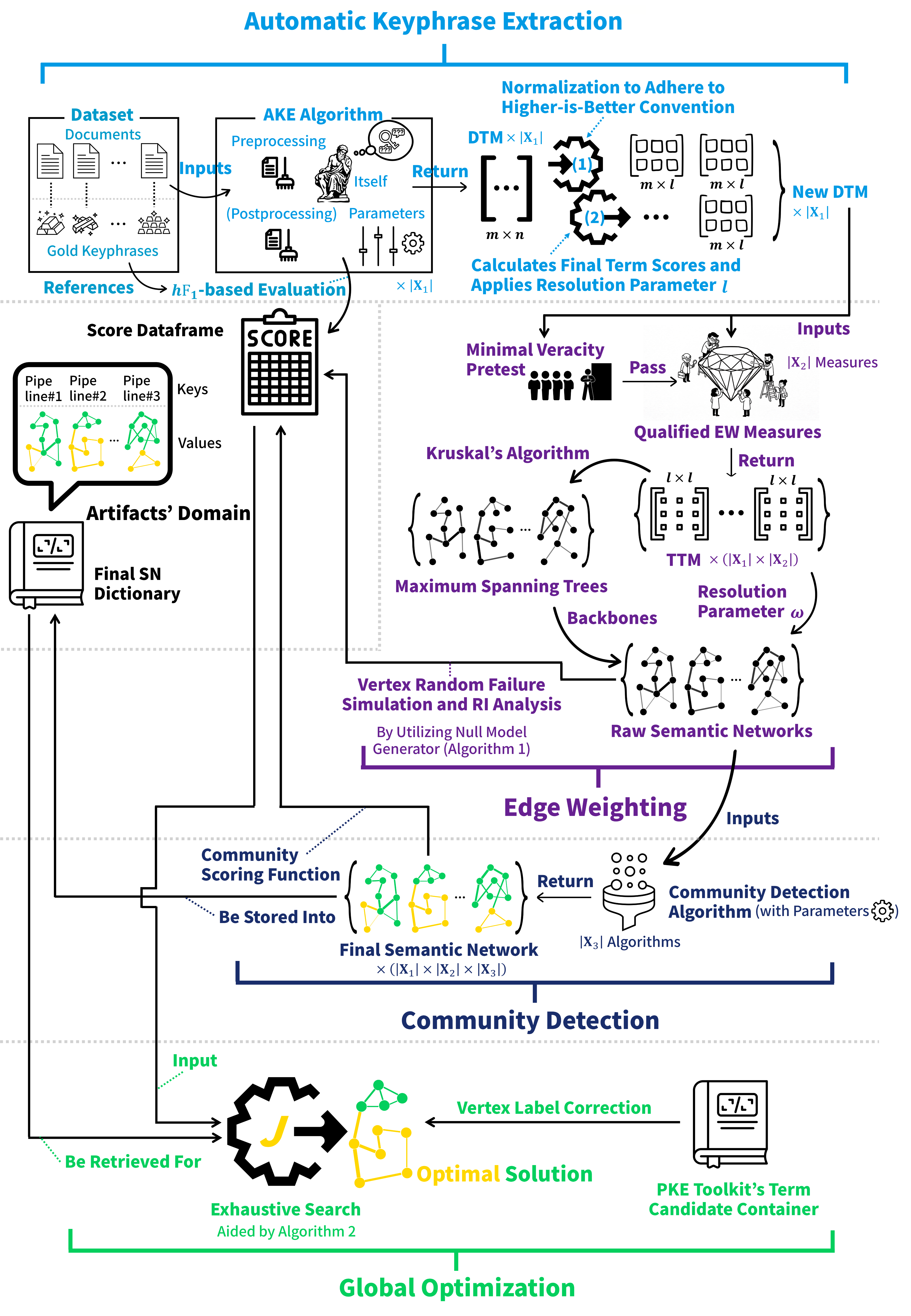}    \caption{\textbf{Workflow of ClueNetwork. For detailed descriptions, please refer to the main text.}}
    \label{Figure 19}
\end{figure*}

\emph{ClueNetwork} is presented in Fig. \ref{Figure 19}. Due to the figure's unavoidable large size and placement, we respectfully encourage readers to view this PDF side--by--side by displaying the figure in one window and this main text in the other.

\medskip

\noindent\textbf{Stage 1 (AKE)}. First, as previously emphasized, an AKE algorithm is an \emph{ensemble} of its preprocessing strategy, itself, and its parameters. Algorithms given independently perform AKE on a given dataset depending on their philosophies of keyness. Here, certain algorithms require \emph{postprocessing}. Why? The point is that both gold keyphrases and scored terms should be \emph{stemmed} (e.g., ``definition(s)'' $\xrightarrow{\text{stemming}}$ ``defin'' ) to ensure a consistent $h\text{F}_{1}$--based evaluation.

For instance, NN--based algorithms cannot extract appropriate contextual embeddings from stemmed forms (i.e., lexical forms). Hence, they first obtain embeddings and keyness scores for \emph{surface} forms, and then perform poststemming. Thereafter, if multiple surface forms share the same stemmed form, only the stemmed form becomes a candidate, and its keyness score is determined as that of the surface form with the highest keyness score (or earliest offset).

Second, each algorithm's average $h\text{F}_{1}$ across all documents is recorded into a score dataframe (please refer to the ``Artifacts' Domain'' in Fig. \ref{Figure 19}), and each returns a $\mathbf{DTM}\in\mathbb{R}^{m \times n}$. Third, every row of each $\mathbf{DTM}$ is normalized by using \eqref{Eq 1}. Fourth, each dimension--reduced new $\mathbf{DTM}$ $\in \mathbb{R}^{m \times l}$ is obtained by calculating final candidate scores based on \eqref{Eq 2} and applying the resolution parameter $l$.

\medskip

\noindent\textbf{Stage 2 (EW)}. The obtained $\mathbf{DTM}$s are utilized in two ways. First, they serve as references for the minimal veracity pretest on given EW measures. Second, only measures that have passed the test are allowed to focus on different facets of every $\mathbf{DTM}$ to return $\mathbf{TTM}$s $\in \mathbb{R}^{l \times l}$. Given that users might conceive flawed hypotheses without the pretest as a safeguard, the provision of the pretest constitutes \emph{ClueNetwork}'s touch of novelty. Third, MSTs are built by applying established Kruskal's algorithm \cite{b143} to the $\mathbf{TTM}$s.

Fourth, the MSTs serve as backbones of raw SNs --- each raw SN is built by sequentially adding higher--weighted $\mathbf{TTM}$ entries to the corresponding MST until the total number of edges reaches $\omega$. Fifth, raw SNs are evaluated via vertex random failure simulation and $\text{RI}$ analysis, and then resulting $\text{RI}$ scores are restored into the score dataframe. Given that users would otherwise have to prepare elusive gold edge weights without this evaluation mechanism, it also constitutes \emph{ClueNetwork}'s touch of novelty.

\medskip

\noindent\textbf{Stage 3 (CD)}. First, selected CD algorithms are applied to the raw SNs, thereby obtaining final SNs. The final SNs are stored in a dictionary, where the keys are identifiers representing configurations of SNC pipelines (e.g., \texttt{(LMRank, Cosine, Leiden)}) and the values are corresponding final SNs. Second, the final SNs are evaluated by a CSR, and then their resulting scores are stored in the score dataframe.

\medskip

\noindent\textbf{Global Optimization}. After all local evaluations have been completed, the framework integrates them into an exhaustive search mechanism for the optimal SN. First, if the local evaluation results have appropriately accumulated into the score dataframe $\mathcal{D}$, it resembles Table \ref{table:Table 15}.

\begin{table}[h!]
\caption{Conceptual Illustration of Score Dataframe $\mathcal{D}$.}\label{table:Table 15}
\centering
\begin{tabular}{>{\arraybackslash}m{1.0cm}|>
{\arraybackslash}m{0.8cm}|>{\arraybackslash}m{1.0cm}|>{\arraybackslash}m{0.7cm}|>{\arraybackslash}m{0.7cm}|>
{\arraybackslash}m{0.7cm}}
$x_{1}$ & $x_{2}$ & $x_{3}$ & $c_{s_{1},x_{1}}$ & $c_{s_{2},x_{2}}$ & $c_{s_{3},x_{3}}$ \\
\hline\hline
TF & CF & CNM & (...) & (...) & (...) \\ \hline
TF & CF & Louvain & (...) & (...) & (...) \\ \hline
TF & CF & Leiden & (...) & (...) & (...) \\ \hline
TF & CF & FLPA & (...) & (...) & (...) \\ \hline
TF--IDF & CF & CNM & (...) & (...) & (...) \\ \hline
$\vdots$ & $\vdots$ & $\vdots$ & $\vdots$ & $\vdots$ & $\vdots$ \\ \hline
LMRank & Cosine & Leiden & (...) & (...) & (...) \\ \hline
LMRank & Cosine & FLPA & (...) & (...) & (...) \\
\end{tabular}
\end{table}

\noindent Second, such $c_{s_{1},x_{1}}$, $c_{s_{2},x_{2}}$, and $c_{s_{3},x_{3}}$ columns in Table \ref{table:Table 15} serve as $\mathbf{C}_{1}$, $\mathbf{C}_{2}$, and $\mathbf{C}_{3}$. The element--wise product of $\mathbf{C}_{1}$, $\mathbf{C}_{2}$, and $\mathbf{C}_{3}$, produces a column $\mathbf{J}$ containing all $J(\pi)$ scores. Third, by appending $\mathbf{J}$ to $\mathcal{D}$ and then sorting $\mathcal{D}$ in descending order, $\pi^{*}$ and $J(\pi^{*})$ from the top row are identified. The auxiliary Algorithm \ref{Algorithm 2} summarizes these steps. Finally, the identified $\pi^{*}$ is used as the key to retrieve the corresponding optimal SN from the final SN dictionary.

\begin{algorithm}[h!]\label{Algorithm 2}
\caption{Auxiliary Algorithm}
\SetKw{KwEnd}{end}
\SetKw{KwDo}{do}
\SetKw{KwEmpty}{}
\SetKw{KwMain}{Main Steps}
\KwIn{$\mathcal{D}$}
\KwOut{$\pi^{*}$ and $J(\pi^{*})$} $\triangleright$ \KwMain \\
$\mathbf{J}:= \mathbf{C}_{1}\odot\mathbf{C}_{2}\odot\mathbf{C}_{3}$, where $\mathcal{D}$ includes $\mathbf{C}_{i\in\{1,2,3\}}$ \\
Update $\mathcal{D}$ by appending $\mathbf{J}$ \\
Sort $\mathcal{D}$ in descending order based on $\mathbf{J}$ \\
Identify the top row in $\mathcal{D}$ \\
Regard $(x_{1},x_{2},x_{3})$ from the top row as $\pi^{*}$ \\
Regard $J(\pi)$ from the top row as $J(\pi^{*})$ \\
\Return $\pi^{*}$ and $J(\pi^{*})$
\end{algorithm}

\noindent\textbf{Vertex Label Correction}. The steps described can be readily implemented by anyone possessing standard programming skills. Consequently, \emph{ClueNetwork} can be readily reused. Nonetheless, there is a noteworthy consideration. As previously explained, candidate terms in the AKE stage are compared with gold keyphrases in their \emph{stemmed} forms. Since such stemmed forms are preserved throughout the subsequent stages (cf. Fig. \ref{Figure 15}), vertex labels of an identified optimal SN should be corrected from stemmed forms to corresponding \emph{surface} forms to ensure interpretability.

How? During preprocessing in AKE, a container, such as the PKE toolkit's \texttt{Candidate} container\footnote{The conceptualization of such a container may be aided by Lines 208--235 in \href{https://github.com/boudinfl/pke/blob/master/pke/base.py}{this script (link)} and Lines 30--49 in \href{https://github.com/boudinfl/pke/blob/master/pke/data_structures.py}{this script (link)}.}, that maps each stemmed form (i.e., lexical form) to its corresponding list of surface forms can be built. Thereafter, the optimal SN's stemmed--form vertex labels are used as keys to retrieve corresponding surface forms from the container, thereby correcting the labels. Since AKE algorithms independently perform preprocessing, such containers should be created and managed separately for each algorithm.

\subsubsection{Handling Lack of Gold Keyphrases}\label{subsubsec:handling lack of gold keyphrases}

\medskip

Here, one might ask, ``Okay, you have presented a framework. But how can I obtain $h\text{F}_{1}$ scores for any AKE algorithm when my dataset lacks gold keyphrases?'' A timely question. Most textual datasets, unless they consist of academic papers with author keywords, can lack gold keyphrases. However, there exist at least two practical solutions, although we defer them to future applications (given the scope and length constraints):

\medskip

\noindent\textbf{If You Can Be Experts, Then the Experts Can Be Gold Standards}. The notion that persons possessing expertise on certain targets can annotate them has been accepted in the field of KR \cite{b35}. So, a research team can also first sample a subset of documents from a dataset. The team can \emph{read} each document, simultaneously \emph{becoming} its experts, and then \emph{annotate} it with \emph{multi--annotator validation}. While an ideal sample size can vary depending on the data, it should significantly exceed $30$ (the minimum size required for the Central Limit Theorem to hold), given that an average $h\text{F}_{1}$ obtained from such a sample is an algorithm's $\widehat{c_{s_{1},x_{1}}}$. 

\medskip

\noindent\textbf{Silver Keyphrases}. For a researcher without a team who finds manual annotation burdensome, leveraging a SOTA LLM with appropriate prompts can be considered to automatically assign silver keyphrases to documents \cite{b169}. This solution is also powerful because such an LLM facilitates annotation for an entire dataset beyond a small sample. However, the solution requires \emph{meta--evaluation} of whether silver keyphrases can truly represent reality. Furthermore, if the backbone of a participating NN--based algorithm and the annotator LLM are contextually close, the competition can be biased. Notwithstanding these considerations, it remains a promising solution given recent advancements in NLP.

\subsection{Justification of J}\label{subsec:justification of j}

\medskip

Let us now respond to the previously deferred demand, ``Provide \emph{criteria} for the optimal SN and \emph{evidence} that the current expression of $J$ \emph{harmonizes with} them.'' Here, Subsubsection \ref{subsubsec:theoretical justification} provides a theoretical response, ``In fact, the current $J$ \emph{itself} is the very criterion, and $J$ already not only harmonizes with but also \emph{coincides with} the criterion. This is not circular reasoning.'' Nonetheless, some peers might not accept this response. So, Subsubsection \ref{subsubsec:empirical justification} empirically demonstrates through perturbation analysis that the current $J$ harmonizes with reality more than additive expressions, such as weighted sum or simple average. Before these responses, Subsubsection \ref{subsubsec:unsuitable justification approaches} rules out irrelevant expectations by clarifying which approaches are unsuitable for this justification.

\subsubsection{Unsuitable Justification Approaches}\label{subsubsec:unsuitable justification approaches}

\medskip

\noindent\textbf{Gold--SN--Based Approaches}. Throughout this paper, we have already shown that gold SNs remain elusive --- specifically, \hyperlink{Yi Sang}{here (link)} and \hyperlink{many edges}{here (link)}. Just as planets with countless diamonds far more valuable than gold might exist yet remain inaccessible to us, we no longer consider gold SNs.

\medskip

\noindent\textbf{Human Judgments}. Likewise, we do not recruit experts or undergraduate students and ask them to judge qualities of SNs extracted by using $J$. It is highly labor--intensive and makes our already thin wallets even thinner.

\medskip

\noindent\textbf{Hypothesis--Testing--Based Approaches}. Given that this paper is not a social science study, we do not derive hypotheses from SNs and test them. We have already clarified that, unlike KGs, SNs as clues are not collections of propositions \hyperlink{unlike KGs}{(link)}. Therefore, the veracity of hypotheses derived from SNs must be decoupled from qualities of the SNs themselves.

\medskip

\noindent\textbf{Extrinsic Approaches}. Sometimes, we also serve as peer reviewers and encounter papers that justify their frameworks using extrinsic approaches. Here, while we say, ``Congratulations. Your frameworks happen to align well with extrinsic tasks'' to authors whose frameworks perform well on such tasks, we do not simply say, ``Your papers should be rejected'' to authors whose frameworks perform poorly on them.

Why? While the Platonic Representation Hypothesis \cite{b170} argues that, as AIs advance, their representations appear to converge toward the Forms (?!) of vector representations, the recently proposed Aristotelian\footnote{Understanding why Aristotle appears here requires an understanding of his metaphysics \cite{b172}, certainly beyond the scope. It suffices to know that, whereas Plato regarded \emph{fixity} as the condition of being, Aristotle did not (i.e., the banana mentioned \hyperlink{banana}{here (link)} also \emph{existed}) \cite{b172}.} view \cite{b171} suggests that AIs' representations are only converging toward shared local neighborhood relationships rather than a globally shared geometry. Likewise, justifying $J$ based on performance obtained by sharing SN representations to extrinsic representation--based systems seems to involve both luck and mismatch, given the possible absence of a globally shared geometry. Therefore, extrinsic task performance may not directly measure qualities of $J$--driven SNs themselves.

\subsubsection{Theoretical Justification: Bayesian Perspective}\label{subsubsec:theoretical justification}

\medskip

\noindent\textbf{Decision--making Under Uncertainty}. Let us begin with the fact that this paper's notion of optimal SN has \emph{consistently} been confined to an SN that achieves the \emph{highest joint confidence} for the local objectives \hyperlink{joint confidence}{(link)}. Also, let us acknowledge the fact that, for any textual dataset, any expression of $J$ operates under ``uncertainty'' about which SN actually represents it, yet $J$ must decide on the most ``desirable'' (in our terminology, ``optimal'') SN, while whether that decision actually corresponds to reality remains ``uncontrollable'' \cite{b173}.

\medskip

\noindent\textbf{Subjective Probabilities}. Hence, what any $J$ ultimately incorporates into its expression is a set of subjective probabilities concerning an SN. And de Finetti, one of the proponents of Bayesian statistics, defined subjective probability as \cite{b174}:

\medskip

\begin{itemize}
    \item ``the \emph{degree of belief}, attributed by a given person at a given instant and with a given set of information, in the occurrence of an event.''
\end{itemize}

\medskip

\noindent Therefore, if ``instant,'' ``person,'' ``set of information,'' and ``event'' herein can be defined in terms of SNC, we can also define subjective probabilities for the optimal SN:

\medskip

\begin{itemize}
    \item Instant: $s_{i} \in \mathbf{S}_{i}$, where $i \in \{2, 3, 4\}$.
    \item Person: a ``putative'' \cite{b9} \emph{rational} peer reviewer that evaluates whether a given SN has achieved [keyness, interpretability, or distinctiveness].
    \item Set of Information: $c_{s_{i},x_{i}}$ $\in \{\chi|\chi$ is known to \emph{the} person$\}$. 
    \item Event: $\mathcal{E}_i$ and $\mathcal{E}_i^{c}$, the events that the SN has achieved or not achieved [keyness, interpretability, or distinctiveness], respectively, where $\mathcal{E}_i,\mathcal{E}_i^{c}$ $\subsetneq$ $\Omega_{i}$, the sample space, and $\Omega_{i}=\{\texttt{achieved, not achieved}\}$.
\end{itemize}

\medskip

Then, would the \emph{rational} reviewer evaluate the subjective probabilities based on irrelevant information, such as average prices of the reviewer's breakfast, lunch, and dinner? Certainly not. The reviewer would evaluate them based on the obtained relevant scores ($c_{s_{1},x_{1}}$, $c_{s_{2},x_{2}}$, and $c_{s_{3},x_{3}})$ of the local objectives (SN keyness, interpretability, and distinctiveness). That is, the subjective probabilities can be formulated as:

\begin{equation} \label{Eq 11}
\begin{gathered}
c_{s_{1},x_{1}},\; c_{s_{2},x_{2}},\; \text{and }\;c_{s_{3},x_{3}}
\\[2pt]
\overset{\text{can be treated as}}{\longrightarrow}
\\[2pt]
P(\mathcal{E}_{2}),
P(\mathcal{E}_{3}\mid\mathcal{E}_{2}),
\text{ and }
P(\mathcal{E}_{4}\mid\mathcal{E}_{2}\cap\mathcal{E}_{3}), \text{ respectively}.
\end{gathered}
\end{equation}

\medskip

\noindent\textbf{Subjective Probabilities Are All $J$ Can Say}. As de Finetti clarifies \cite{b175}, such subjective probabilities acknowledge that uncertain things are uncertain, rather than directly making assertions about reality. Indeed, there is little else that any $J$ can represent concerning SNs beyond $P(\mathcal{E}_{2})$, $P(\mathcal{E}_{3}\mid\mathcal{E}_{2})$, and $P(\mathcal{E}_{4}\mid\mathcal{E}_{2}\cap\mathcal{E}_{3})$. Nevertheless, these are ``the criteria for the optimal SN.'' Why? Because, insofar as a $J$ systematically incorporates $c_{s_{1},x_{1}}$, $c_{s_{2},x_{2}}$, and $c_{s_{3},x_{3}}$ to represent \emph{relevant} evaluation results, an SN selected based on $J$ should be regarded as the \emph{rational}, and thereby \emph{optimal}, SN.

\medskip

\noindent\textbf{Then, Why Do We Multiply Them?} $P(\mathcal{E}_{2})$, $P(\mathcal{E}_{3}\mid\mathcal{E}_{2})$, and $P(\mathcal{E}_{4}\mid\mathcal{E}_{2}\cap\mathcal{E}_{3})$ are \emph{local} confidences (degrees of belief) in different aspects of an SN. What we ultimately need is a \emph{joint} confidence in their joint artifact. Then, the ``putative'' \cite{b9} rational reviewer would treat this global confidence as given by \eqref{Eq 12}--b concerning the joint sample space defined in \eqref{Eq 12}--a, in which case $J$ can also be re--expressed as \eqref{Eq 12}--c:

\begin{equation} \label{Eq 12}
\begin{aligned}
\Omega =\Omega_{2} \times \Omega_{3} \times \Omega_{4} \cdots (a) \\[6pt]
J= c_{s_{1},x_{1}} \times c_{s_{2},x_{2}} \times c_{s_{3},x_{3}}
\overset{\text{can also be treated as}}{\longrightarrow}
\\[2pt]
P(\mathcal{E}_{2} \cap \mathcal{E}_{3} \cap \mathcal{E}_{4}) = P(\mathcal{E}_{2}) P(\mathcal{E}_{3}|\mathcal{E}_{2}) P(\mathcal{E}_{4}|\mathcal{E}_{2} \cap \mathcal{E}_{3}) \cdots (b)
\\[6pt]
\underset{\pi}{\max} \quad J(\pi) = \underset{\pi}{\max}  \quad P(\mathcal{E}_{2} \cap \mathcal{E}_{3} \cap \mathcal{E}_{4}\mid\pi) \cdots (c)
\end{aligned}
\end{equation}

\medskip

\noindent In short, the reason we multiply $P(\mathcal{E}_{2})$, $P(\mathcal{E}_{3}\mid\mathcal{E}_{2})$, and $P(\mathcal{E}_{4}\mid\mathcal{E}_{2}\cap\mathcal{E}_{3})$ is that doing so logically follows from the established \emph{chain rule of probabilities}.

\medskip

\noindent\textbf{Conditions of Coherence}. One may point out that the fact that $c_{s_{1},x_{1}}$, $c_{s_{2},x_{2}}$, $c_{s_{3},x_{3}}$, and $J$ are \emph{relevant} to SNC is \emph{insufficient}, by itself, to make it \emph{rational} to treat them as $P(\mathcal{E}_{2})$, $P(\mathcal{E}_{3}\mid\mathcal{E}_{2})$, $P(\mathcal{E}_{4}\mid\mathcal{E}_{2}\cap\mathcal{E}_{3})$, and $P(\mathcal{E}_{2} \cap \mathcal{E}_{3} \cap \mathcal{E}_{4})$, respectively. Indeed, de Finetti clarifies that, for anything to qualify as a subjective probability, it must also satisfy \emph{coherence}, formalized by the following conditions \cite{b175}:

\medskip

\begin{itemize}
    \item Boundedness: $0 \leq P(\mathcal{E}_{i}) \leq 1$.
    \item Normalization: $P(\Omega_{i})=1$.
    \item Finite Additivity: $P(\bigcup_{j=1}^{n}\mathcal{E}_{i,j})    =\sum_{j=1}^{n}P(\mathcal{E}_{i,j})$, where $\mathcal{E}_{i,j} \cap \mathcal{E}_{i,k}=\emptyset \text{ } (j \neq k)$ and $n=|\{\text{possible }\mathcal{E}_{i,j}\}|<\infty$.
\end{itemize}

\medskip

\noindent Here, these conditions express the requirement that probabilities evaluated by a person for events in an event set must not contradict one another; de Finetti explains that, if any person wishes to avoid ``decisions whose consequences are manifestly undesirable (leading to certain loss)'' \cite{b175}, these conditions must be satisfied. Readers interested in cases of such \emph{irrational} decisions that de Finetti demonstrates as resulting from violations of these conditions may consult \cite{b176}.

De Finetti opposes constraints beyond these conditions because doing so is prone to excluding valid evaluations of probabilities as invalid \cite{b174}. Accordingly, for $c_{s_{1},x_{1}}$, $c_{s_{2},x_{2}}$, $c_{s_{3},x_{3}}$, and $J$, which constitute the \emph{criteria} for SNC, to function as \emph{starting points} for supporting exploratory research, satisfying these conditions alone is \emph{theoretically} sufficient. Then, do these criteria satisfy these conditions?

\medskip

\noindent\textbf{Demonstration}. Yes. First, $c_{s_{1},x_{1}}$ (cf. Table \ref{table:Table 6}) and $c_{s_{2},x_{2}}$ (cf. \eqref{Eq 7}) range from $0$ to $1$ and thus automatically satisfy boundedness. Although Modularity $Q$ originally ranges from $-1$ to $1$ \cite{b21, b24}, under the nonnegativity constraint in \eqref{Eq 10}, negative $Q$ values are truncated at zero. That is, $c_{s_{3},x_{3}}$ is given by $\max\{Q(s_{4}),0\}$ and thus satisfies boundedness. Second, 
for any $c_{s_{i},x_{i}}$, $\Omega_{i}$ is $\{\texttt{achieved, not achieved}\}$ and values ($c_{s_{i},x_{i}}$ and $1-c_{s_{i},x_{i}}$) assigned to these two mutually exclusive events always sum to $1$. Therefore, any $c_{s_{i},x_{i}}$ satisfies both normalization and finite additivity.

Third, therefore, $c_{s_{1},x_{1}}$, $c_{s_{2},x_{2}}$, and $c_{s_{3},x_{3}}$ can be treated as $P(\mathcal{E}_{2})$, $P(\mathcal{E}_{3}\mid\mathcal{E}_{2})$, and $P(\mathcal{E}_{4}\mid\mathcal{E}_{2}\cap\mathcal{E}_{3})$, respectively, and thus $J$ necessarily not only satisfies boundedness, normalization, and finite additivity but, from the chain rule of probabilities, can also be treated as $P(\mathcal{E}_{2} \cap \mathcal{E}_{3} \cap \mathcal{E}_{4})$. Finally, therefore, our theoretical response to the demand to ``Provide \emph{criteria} for the optimal SN and \emph{evidence} that the current $J$ \emph{harmonizes with} them'' is as follows: ``$c_{s_{1},x_{1}}$, $c_{s_{2},x_{2}}$, $c_{s_{3},x_{3}}$, and the current $J$ \emph{themselves} are the \emph{very} criteria, and $J$ already not only harmonizes with but also \emph{coincides with} the criteria. This is not circular reasoning. Rather, it is a logically necessary consequence from the Bayesian perspective.''

\medskip

\noindent\textbf{De Finetti's Counterarguments to Calibrationists}. Here, one might ask, ``Okay, the current $J$ is theoretically coherent. But would it be \emph{unscientific} if \emph{actual} evidence that $J$ harmonizes with reality still has not been provided?'' De Finetti would probably say no. Why? The claim that the scientific legitimacy of subjective probabilities should be grounded in empirical evidence presupposes that objective probabilities are \emph{inherent in} reality. De Finetti points out that there is no evidence for such a presupposition, as what we actually observe is events and their frequencies, whereas probability is an \emph{additional} concept\footnote{This is why de Finetti consistently speaks not of \emph{discovering} probabilities, but of \emph{evaluating} probabilities \cite{b173}--\cite{b176}.} introduced by human beings \cite{b173}.

One might nevertheless press the point, ``Apart from \emph{petitio principii} (begging the question) involved in the presupposition, could some \emph{actual} reviewers not construct $J$ as other expressions, such as a weighted sum? If so, values of the current $J$ would differ from empirically obtained $J$ values. In that case, the current $J$ would need to be \emph{calibrated}.'' That is, one would then have us submit our 1,088 SNs to this \emph{IEEE Access} journal and ask Reviewers \#1 through \#32640\footnote{Why 32,640? Reviewers for each SN should significantly exceed $30$, as a conventional rule of thumb for invoking the Central Limit Theorem.} to evaluate their qualities on a scale from $0$ to $1$.

Apart from its impracticality, de Finetti would probably point out that such calibration is not constitutive of $J$ itself. Why? As de Finetti explains \cite{b176}: a subjective probability is a \emph{prior} (yet rational) evaluation of an event based on information available under uncertainty, but it is not a prediction about the future outcome. Once the outcome becomes known (once uncertainty is removed), information is \emph{updated}, and so a \emph{posterior} evaluation can arise. It is therefore an \emph{error categoriae} (category mistake) to judge the prior evaluation based on the posterior evaluation.

In short, $J$ does not assert any proposition, yet represents a rational belief. In other words, calibration is useful, yet $J$ concerns \emph{starting points}, not \emph{endpoints}, in science.

\subsubsection{Empirical Justification: Perturbation Analysis}\label{subsubsec:empirical justification}

\medskip

\noindent\textbf{To Avoid Re--Re--Revision}. Nonetheless, might some reviewers still not accept the Bayesian perspective, or that of de Finetti? If so, this paper could be rejected again, which is certainly not what we want. Fortunately, although it is difficult to show that ``the current $J$ aligns with truth,'' showing that ``it conflicts with falsehood'' is easy. Furthermore, under a well--designed perturbation analysis, these two are \emph{logically equivalent}. Please consider the following.

\medskip

\noindent\textbf{Comparative Expressions of $J$}. First, we decided to confirm whether $J$ should indeed take its current multiplicative form. We conducted a perturbation analysis on the current form as well as alternative forms in which $J$ was modified into general weighted sum forms. Here, the simple average (arithmetic mean), Inverse--Variance Weighting (IVW) method \cite{b177}, and CRITIC (CRiteria Importance Through Intercriteria Correlation) method \cite{b178} were employed as weighting methods. Although these methods differ in detail (Table \ref{table:Table 16}), in terms of SNC, they fundamentally share an optimistic perspective that a loss in one stage can be compensated to some extent by another. This sets them apart from our $J$, which penalizes global confidence in SN optimality if a local objective is not sufficiently met in even a single stage.

\begin{table}[h!]
\caption{Weight Formulas Tailored to Our Perturbation Analysis}\label{table:Table 16}
\centering
\renewcommand{\arraystretch}{2}
\begin{tabular}{>{\arraybackslash}m{1.5cm}|>
{\arraybackslash}m{2.9cm}|>
{\arraybackslash}m{2.8cm}}
Method & Weight $w_{i}$ of $c_{s_{i},x_{i}}$ & Philosophy \\
\hline\hline
Arithmetic \newline Mean & $1/3$ & The three criteria are equally important. \\ \hline
IVW & $\frac{1/\sigma^{2}_{i}}{\sum_{j=1}^{3}1/\sigma^{2}_{j}}$ & More stable criteria are given greater weight. \\ \hline
CRITIC & $\frac{\sigma_{i}\sum_{j=1}^{3}(1-\rho_{ij})}{\sum_{k=1}^{3}\big[ \sigma_{k} \sum_{j=1}^{3} (1-\rho_{kj})\big]}$ & More discriminative and less redundant criteria are given greater weight. \\ \hline
\multicolumn{3}{m{7.2cm}}{For a given textual dataset, \newline $\sigma_{i}^{2}$ is the variance of applied SNC pipelines' $c_{s_{i},x_{i}}$ values; \newline $\rho_{ij}$ is Spearman's $\rho$ between distributions of $c_{s_{i},x_{i}}$ and $c_{s_{j},x_{j}}$.} \\
\end{tabular}
\end{table}

\noindent\textbf{Perturbation Analysis Procedure} was as follows. First, the SNC pipelines (KP--Miner, Cosine, Leiden) for NUS and (KeyBERT, Cosine, Leiden) for SemEval--2017 served as targets. For brevity while maintaining generalizability, these optimal pipelines (cf. Table \ref{table:Table 17}) for the long--document dataset (NUS) and short--document dataset (SemEval-2017) were chosen. Second, a unit of noise consisted of injections across respective stages. Specifically, for the SN corresponding to a target pipeline, its one vertex label (and its occurrences in the sets of prioritized keyphrases) were replaced with a dummy label Frog\#X (i.e., Frogs\#1, \#2, ..., and \#50); a (possibly distinct) vertex was removed; and (possibly) another vertex was returned to a single community.

\begin{figure}[b!]
    \centering    \includegraphics[width=0.94\linewidth]{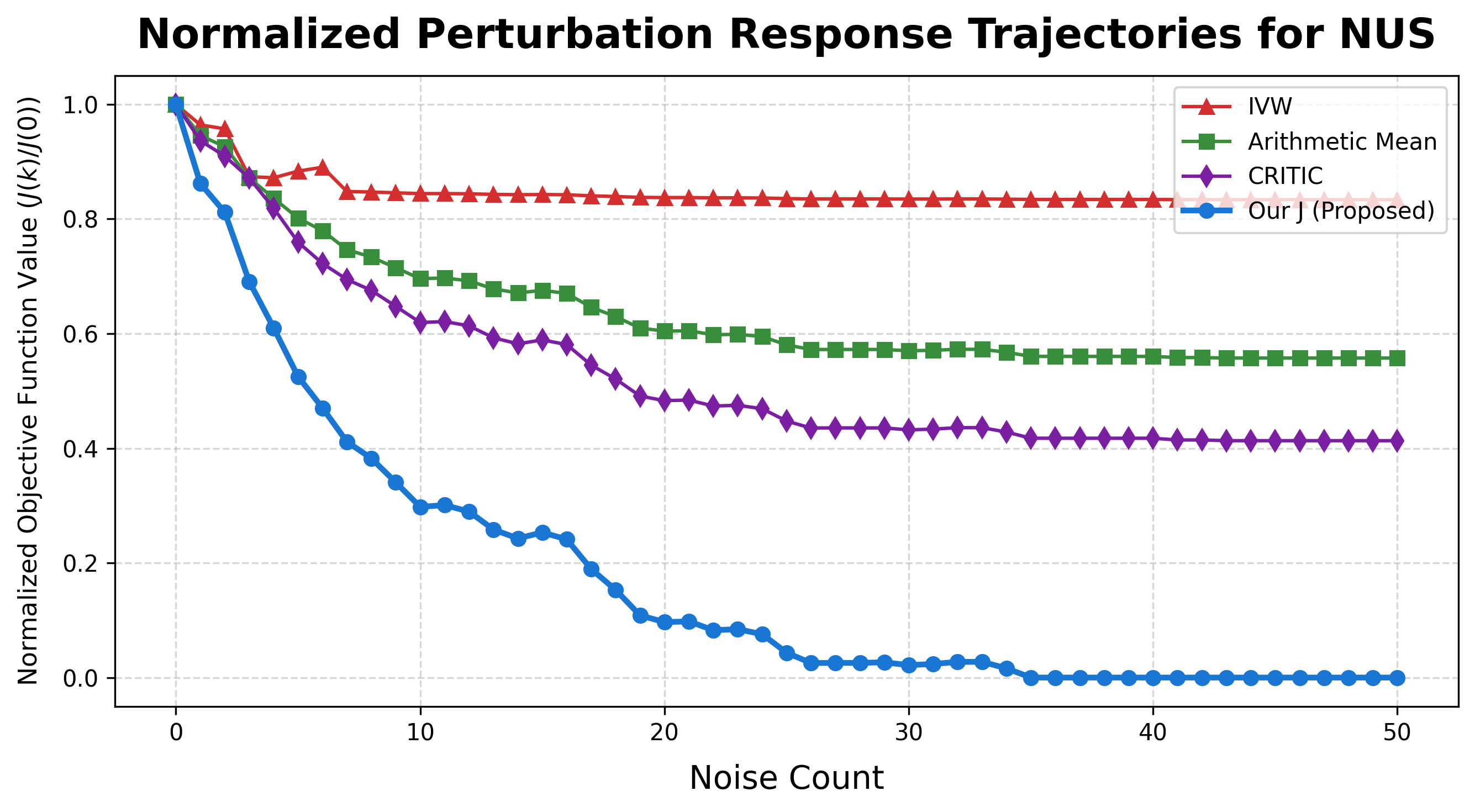}
    \caption{\textbf{Global Perturbation Analysis Results for NUS.}}
    \label{Figure 20}
\end{figure}

\begin{figure}[b!]
    \centering    \includegraphics[width=\linewidth]{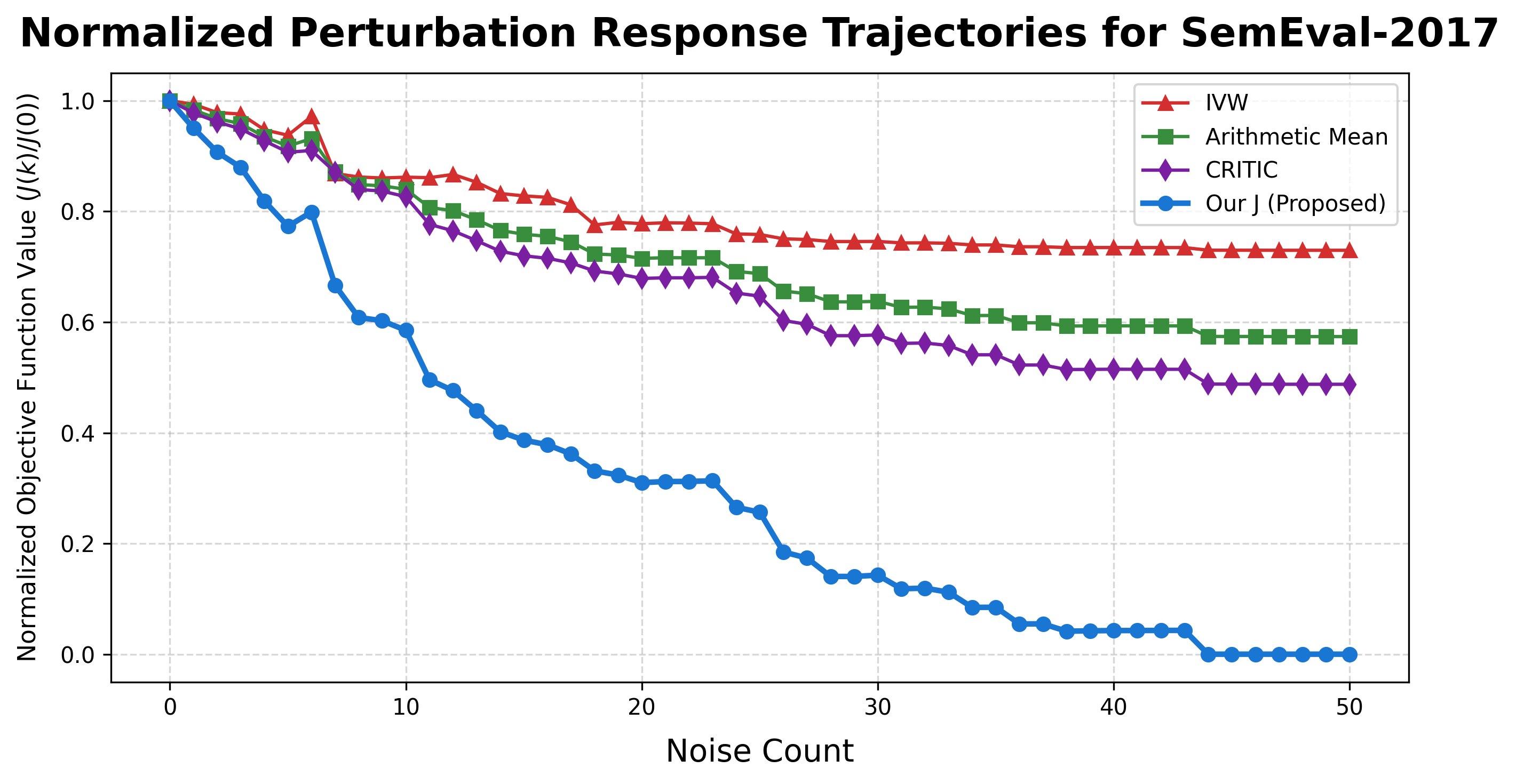}
    \caption{\textbf{Global Perturbation Analysis Results for SemEval--2017.}}
    \label{Figure 21}
\end{figure}

Third, under these conditions, $c_{s_{1},x_{1}}$, $c_{s_{2},x_{2}}$, $c_{s_{3},x_{3}}$, and $J$ were measured across all competing forms of $J$. Fourth, by repeating these steps for all $l=50$ vertices, we obtained the normalized perturbation response trajectories (Figs. \ref{Figure 20} and \ref{Figure 21}). The implementations have been disclosed separately (\href{https://github.com/potentialreviewer/Ha-Kim-2026a/blob/main/notebooks/Perturbation_Analysis_(NUS).ipynb}{link 1}, \href{https://github.com/potentialreviewer/Ha-Kim-2026a/blob/main/notebooks/Perturbation_Analysis_(SemEval_2017).ipynb}{link 2}). Herein, the $y$--axis denotes values of $J(k)/J(0)$, where $J(k)$ and $J(0)$ represent $J$ when $k$ units of noise were injected and the baseline $J$ without noise, respectively.

\medskip

\noindent\textbf{Our $J$ Reacted More Strongly Against False Information}. Such noise is certainly false. It replaces keyphrases with out--of--conext frogs, deceives an SN interpreter by rendering available hubs unusable, and forcibly isolates vertices that belong to certain communities. Therefore, there is no room for a third property or intermediate state analogous to ``gray between black and white'' or ``$0.5$ between $0$ and $1$.'' Only mutually contradictory truth and falsehood exist as options.

At this crossroads, if there is a form of $J$ that rejects falsehood more strongly, the only remaining logical possibility is to believe that it aligns with reality better than the others. Our $J$ was precisely this form, as shown in Figs. \ref{Figure 20} and \ref{Figure 21}.

\medskip

\noindent\textbf{Local Implications}. For the sake of \emph{academic honesty} and for \emph{potential users} of \emph{ClueNetwork}, we also share findings (Table \ref{table:Table 17}) that are unwelcome to us, even if they are not central to the subject matter. Herein, a relative drop rate is defined as $\frac{y(0)-y(50)}{y(0)} \times 100\%$, where $y(0)$ and $y(50)$ are values of a metric $y$ without noise and with maximum noise, respectively:

\begin{table}[h!]
\caption{Brief Ablation Study via Relative Drop Rate Analysis}\label{table:Table 17}
\centering
\begin{tabular}{>{\arraybackslash}m{4.0cm}|>
{\arraybackslash}m{2.8cm}}
Configuration (Metric) & Relative Drop Rate (\%) \\
\hline\hline
AKE ($c_{s_{1},x_{1}}$) & $3.04\%$ \\ \hline
EW ($c_{s_{2},x_{2}}$) & $14.67\%$ \\ \hline
CD ($c_{s_{3},x_{3}}$) & $100.00\%$ \\ \hline
Only with out AKE ($c_{s_{2},x_{2}} \times c_{s_{3},x_{3}}$) & $100.00\%$ \\ \hline
Only with out EW ($c_{s_{1},x_{1}} \times c_{s_{3},x_{3}}$) & $100.00\%$ \\ \hline
Only with out CD ($c_{s_{1},x_{1}} \times c_{s_{2},x_{2}}$) & $17.27\%$ \\ \hline
Full ($c_{s_{1},x_{1}} \times c_{s_{2},x_{2}} \times c_{s_{3},x_{3}}$) & $100.00\%$ \\
\end{tabular}
\end{table}

\medskip

\noindent As shown in Table \ref{table:Table 17}, our $J$'s impressive anti--false performance was primarily driven by CD, whereas EW and AKE contributed minimally. This is because $Q$ is a metric that \emph{directly} measures the partition quality of a given SN, whereas $h\text{F}_{1}$ measures the \emph{general} prioritizing ability of an applied AKE method, and $\text{RI}$ does not function independently, but rather forms $c_{s_{2},x_{2}}$ \emph{together} with the baseline score.

These local criteria were certainly chosen under limited resources. However, this result leaves the open question of which local criteria should serve as proxies for our belief in SN optimality. Nevertheless, we clarify that this result reflects the inherent difficulty of selecting local criteria, rather than a structural flaw of our $J$ itself.

\subsection{Ablation Study}\label{subsec:ablation study}

\medskip

The culmination of a PoC for any framework lies in demonstrating that it actually works. Hence, we conducted an illustrative ablation study \href{https://github.com/potentialreviewer/Ha-Kim-2026a/blob/main/notebooks/Optimization.ipynb}{(link)} using \emph{ClueNetwork}. The conventional SNC pipeline (TF -- CF -- Louvain) was first selected as the baseline. Subsequently, for each dataset, we evaluated the extent of global confidence improvement in SN optimality by comparing the objective function value of the baseline with that of the optimal SNC process. The results are summarized in Table \ref{table:Table 18}, where ``$\Rightarrow$'' denotes a method substitution. We provide the following post--hoc interpretation for Table \ref{table:Table 18}.

First, the optimal processes outperformed the baseline across all datasets. That is, \emph{ClueNetwork} actually worked. The overall low $J$ values merely reflect the intrinsic difficulty of SNC. Second, for most datasets except for 500N--KP--Crowd, simply replacing TF with the AKE algorithms of the optimal processes resulted in relative $J$ increases over the baseline, ranging from $61.78\%$ to $527.08\%$. Given that well--extracted keyphrases enhance confidence for SN optimality, researchers should strive to start on the right foot.

\begin{table*}[b!]
\caption{Results of Ablation Study Across Benchmark Datasets. The optimal solutions were identified by \emph{ClueNetwork}.}\label{table:Table 18}
\centering
\begin{tabular}{>{\arraybackslash}m{2.1cm}>
{\arraybackslash}m{5.0cm}|>{\arraybackslash}m{0.7cm}>{\arraybackslash}m{0.7cm}>{\arraybackslash}m{0.7cm}|>
{\arraybackslash}m{0.8cm}>
{\arraybackslash}m{1.1cm}>
{\arraybackslash}m{1.7cm}}
\hline
Dataset & SNC Process (also termed Pipeline or Policy) & $c_{s_{1},x_{1}}$ & $c_{s_{2},x_{2}}$ & $c_{s_{3},x_{3}}$ & $J$ & $\Delta J$ & $(\Delta J/J_{\text{baseline}})$ \newline $\times 100 \%$
\\ \hline
\multirow{4}{2.1cm}{SemEval--2010} & \textbf{TF -- CF -- Louvain (Baseline)} & $0.134$ & $0.500$ & $0.366$ & \boldsymbol{$0.025$} & - & - \\
& $+$ TF $\Rightarrow$ \textbf{KP--Miner} & $0.287$ & $0.560$ & $0.300$ & $0.048$ & $+0.024$ & $+96.70\%$ \\
& $+$ CF $\Rightarrow$ \textbf{Cosine} & $0.287$ & $0.530$ & $0.471$ & $0.072$ & $+0.047$ & $+191.31\%$ \\
& $+$ Louvain $\Rightarrow$ \textbf{Leiden (Optimal)} & $0.287$ & $0.530$ & $0.473$ & \boldsymbol{$0.072$} & \boldsymbol{$+0.047$} & \boldsymbol{$+192.87\%$} \\ \hline
\multirow{4}{2.1cm}{NUS} & \textbf{TF -- CF -- Louvain (Baseline)} & $0.144$ & $0.500$ & $0.319$ & \boldsymbol{$0.023$} & - & - \\
& $+$ TF $\Rightarrow$ \textbf{KP--Miner} & $0.341$ & $0.500$ & $0.218$ & $0.037$ & $+0.014$ & $+61.78\%$ \\
& $+$ CF $\Rightarrow$ \textbf{Cosine} & $0.341$ & $0.586$ & $0.564$ & $0.113$ & $+0.090$ & $+390.01\%$ \\
& $+$ Louvain $\Rightarrow$ \textbf{Leiden (Optimal)} & $0.341$ & $0.586$ & $0.564$ & \boldsymbol{$0.113$} & \boldsymbol{$+0.090$} & \boldsymbol{$+390.01\%$} \\ \hline
\multirow{4}{2.1cm}{Inspec} & \textbf{TF -- CF -- Louvain (Baseline)} & $0.125$ & $0.500$ & $0.250$ & \boldsymbol{$0.016$} & - & - \\
& $+$ TF $\Rightarrow$ \textbf{LMRank} & $0.364$ & $0.622$ & $0.434$ & $0.098$ & $+0.082$ & $+527.08\%$ \\
& $+$ CF $\Rightarrow$ \textbf{Jaccard} & $0.364$ & $0.646$ & $0.516$ & $0.121$ & $+0.106$ & $+674.98\%$ \\
& $+$ Louvain $\Rightarrow$ \textbf{Leiden (Optimal)} & $0.364$ & $0.646$ & $0.536$ & \boldsymbol{$0.126$} & \boldsymbol{$+0.110$} & \boldsymbol{$+704.85\%$} \\ \hline
\multirow{4}{2.1cm}{KDD} & \textbf{TF -- CF -- Louvain (Baseline)} & $0.092$ & $0.500$ & $0.248$ & \boldsymbol{$0.011$} & - & - \\
& $+$ TF $\Rightarrow$ \textbf{LMRank} & $0.174$ & $0.601$ & $0.430$ & $0.045$ & $+0.033$ & $+291.46\%$ \\
& $+$ CF $\Rightarrow$ \textbf{Jaccard} & $0.174$ & $0.637$ & $0.674$ & $0.075$ & $+0.063$ & $+550.30\%$ \\
& $+$ Louvain $\Rightarrow$ \textbf{Louvain (Optimal)} & $0.174$ & $0.637$ & $0.674$ & \boldsymbol{$0.075$} & \boldsymbol{$+0.063$} & \boldsymbol{$+550.30\%$} \\ \hline
\multirow{4}{2.1cm}{WWW} & \textbf{TF -- CF -- Louvain (Baseline)} & $0.123$ & $0.500$ & $0.195$ & \boldsymbol{$0.012$} & - & - \\
& $+$ TF $\Rightarrow$ \textbf{KP--Miner} & $0.175$ & $0.563$ & $0.379$ & $0.037$ & $+0.025$ & $+209.95\%$ \\
& $+$ CF $\Rightarrow$ \textbf{Cosine} & $0.175$ & $0.681$ & $0.571$ & $0.068$ & $+0.056$ & $+464.60\%$ \\
& $+$ Louvain $\Rightarrow$ \textbf{Leiden (Optimal)} & $0.175$ & $0.681$ & $0.575$ & \boldsymbol{$0.068$} & \boldsymbol{$+0.056$} & \boldsymbol{$+468.53\%$} \\ \hline
\multirow{4}{2.1cm}{SemEval--2017} & \textbf{TF -- CF -- Louvain (Baseline)} & $0.217$ & $0.500$ & $0.211$ & \boldsymbol{$0.023$} & - & - \\
& $+$ TF $\Rightarrow$ \textbf{KeyBERT} & $0.485$ & $0.533$ & $0.237$ & $0.061$ & $+0.038$ & $+167.37\%$ \\
& $+$ CF $\Rightarrow$ \textbf{Cosine} & $0.485$ & $0.636$ & $0.501$ & $0.154$ & $+0.132$ & $+574.90\%$ \\
& $+$ Louvain $\Rightarrow$ \textbf{Leiden (Optimal)} & $0.485$ & $0.636$ & $0.507$ & \boldsymbol{$0.156$} & \boldsymbol{$+0.134$} & \boldsymbol{$+583.33\%$} \\ \hline
\multirow{4}{2.1cm}{DUC--2001} & \textbf{TF -- CF -- Louvain (Baseline)} & $0.106$ & $0.500$ & $0.266$ & \boldsymbol{$0.014$} & - & - \\
& $+$ TF $\Rightarrow$ \textbf{LMRank} & $0.240$ & $0.575$ & $0.360$ & $0.050$ & $+0.035$ & $+250.52\%$ \\
& $+$ CF $\Rightarrow$ \textbf{Cosine} & $0.240$ & $0.558$ & $0.623$ & $0.084$ & $+0.069$ & $+490.39\%$ \\
& $+$ Louvain $\Rightarrow$ \textbf{Louvain (Optimal)} & $0.240$ & $0.558$ & $0.623$ & \boldsymbol{$0.084$} & \boldsymbol{$+0.069$} & \boldsymbol{$+490.39\%$} \\ \hline
\multirow{4}{2.1cm}{500N--KP--Crowd} & \textbf{TF -- CF -- Louvain (Baseline)} & $0.363$ & $0.500$ & $0.186$ & \boldsymbol{$0.034$} & - & - \\
& $+$ TF $\Rightarrow$ \textbf{TF--IDF} & $0.337$ & $0.500$ & $0.153$ & $0.026$ & $-0.008$ & $-24.01\%$ \\
& $+$ CF $\Rightarrow$ \textbf{Cosine} & $0.337$ & $0.553$ & $0.570$ & $0.106$ & $+0.072$ & $+213.41\%$ \\
& $+$ Louvain $\Rightarrow$ \textbf{Leiden (Optimal)} & $0.337$ & $0.553$ & $0.570$ & \boldsymbol{$0.106$} & \boldsymbol{$+0.072$} & \boldsymbol{$+213.41\%$} \\ \hline
\end{tabular}
\end{table*}

Third, as anticipated in Subsection \ref{subsubsec:results of ew}, even a marginal difference in $\text{RI}$ was factored into the results. Substituting CF with Jaccard only increased $c_{s_{2},x_{2}}$ from $0.622$ to $0.646$ for Inspec. Jaccard nonetheless altered the SN topology and edge weight information, thereby influencing the CD stage. Consequently, the relative improvement in $J$ escalated from $527.08\%$ to $674.98\%$. Fourth, however, the overall influences of the selected CD algorithms on the relative improvements in $J$ were marginal. When we replaced Louvain with Leiden for NUS, the difference in the approximate $J$ values was negligible. These results were likely attributable to the sparsity of vertex adjacency and edge weight information (Subsubsection \ref{subsubsec:results of cd}). Future applications leveraging SOTA CD algorithms are expected to enhance SN optimality.

Finally, local optimization did not guarantee global optimization. For 500N--KP--Crowd, $c_{s_{1},x_{1}}$ of TF was higher than that of TF--IDF. However, $J$ of the optimal process incorporating TF--IDF ultimately exceeded that of the baseline. This result underscores the cascading nature of the SNC stages.

\section{Discussion}\label{sec:discussion}

\medskip

The previous section has wrapped up the PoC of \emph{ClueNetwork} by demonstrating that it is a scientifically realistic and practically working framework. The proposed framework and this paper nonetheless come with both contributions and limitations. We now discuss them. Limitations of a framework or paper sometimes suggest potential directions for future work. Hence, certain limitations will be discussed alongside future work. Subsequently, the contributions will be discussed.

\subsection{Limitations and Future Work}\label{subsec:limitations and future work}

\medskip

\textbf{Extension to Dynamic SNC}. Documents are created at different points in time. Although \emph{ClueNetwork} currently does not account for it, real--time updates of documents can render static SNs \emph{outdated}. It is necessary to establish an Incremental Document Update Strategy (IDUS), track impacts of dynamic environments on each SNC stage, and compare the performance of SNC policies before and after updates. Nonetheless, we defer ensuring SN timeliness to future work.

There are two reasons. First, that extension requires independent work. To compute every document's impact within a reasonable timeframe, the IDUS should be supported by well--designed data pipelines and algorithms. We have already devoted $34$ pages of this paper solely to ensuring SN optimality. Given the extension's significance, it would, paradoxically, be an underservice to address the extension within a mere subsection. Second, \emph{ClueNetwork} can already cover a substantial range of SNs. When we explored multiple articles applying SNC \cite{b19}, we found that most research teams performed SNC for datasets from narrow time windows.

\medskip

\noindent\textbf{Development of Software Library}. \emph{ClueNetwork} is a framework, yet not a \emph{full--fledged} software library. Although \emph{ClueNetwork} in its current state ensures standard reusability required for topical review articles, it does not yet achieve the highest reusability. Developing such a library requires many considerations (the extension to dynamic SNC, coverage of local methods, time and space complexity, among others). Hence, we defer it to future independent work.

\medskip

\noindent\textbf{Inspiration}. We humbly acknowledge that \emph{ClueNetwork} is largely \emph{inspired by} perspectives of network science established by peers, many of whom are physicists or biologists. For instance, the basic idea that phases or stages of network analysis can be systematically integrated has already been shared by biologists (e.g., \cite{b11}) who integrate CD and alignment of protein--protein interaction networks (e.g., \cite{b179}--\cite{b181}). ClustRNet \cite{b141}, the origin of Algorithm \ref{Algorithm 1}, was also developed by biologists. Percolation theory, on which \emph{ClueNetwork} heavily relies, has also been established by physicists (e.g., \cite{b129}, \cite{b131}--\cite{b134}).

\subsection{Contributions}\label{subsec:contributions}

\medskip

Notwithstanding the limitations, this paper and \emph{ClueNetwork} contribute to the body of knowledge in the following ways.

\medskip

\noindent\textbf{Revisiting Semantic Networks as Clues}. Nowadays, numerous frameworks leveraging KGs for AI--assisted systems (e.g., \cite{b182}) are being updated day by day. SNs participate in this trend rarely and thus are placed in a position where they could be \emph{misunderstood} as inferior approximations of KGs. Accordingly, Subsection \ref{subsec:surrogates and clues} has devoted substantial pages to elucidate that SNs can be legitimate symbolic systems representing \emph{non--propositional} knowledge in terms of abduction. By \emph{decoupling} SNs and hypotheses derived from them, Subsection \ref{subsec:surrogates and clues} has also clarified that SNs cannot be evaluated through fully gold--standard--based approaches.

\medskip

\noindent\textbf{Substantial Methodological Refinement}. Against this backdrop, \emph{ClueNetwork} reformulates the notion of an optimal SN for a textual dataset not as an elusive gold SN, but as some SN that achieves the \emph{highest joint confidence} for the local objectives (keyness, interpretability, and distinctiveness). Here, while \emph{ClueNetwork} accepts the existence of gold keyphrases to maintain minimal links with reality, it circumvents elusive gold edge weights (and consequently, elusive gold CD partition). To achieve this circumvention, \emph{ClueNetwork} incorporates the MCC--based veracity pretest and the percolation--theory--based $\text{RI}$. Although MCC \cite{b26} and percolation theory \cite{b129} are certainly not our creations, the idea of integrating them into \emph{ClueNetwork} ultimately facilitates \emph{ranking} candidate SNs. Therefore, this integration constitutes a substantial methodological refinement of conventional SNC practices.

\medskip

\noindent\textbf{Scientific Transparency}. Domain researchers can report their exploratory research by disclosing, ``an SN was constructed from the dataset, \emph{and} its confidence was evaluated at $15.6\%$ by using \emph{ClueNetwork},'' instead of merely stating, ``an SN was constructed from the dataset.'' That is, \emph{ClueNetwork} can enhance scientific transparency and provide domain researchers better starting points for their research programs.

\medskip

\noindent\textbf{Composability}. Most components (including parameters, local evaluation criteria, and local methods) of \emph{ClueNetwork} are modular and substitutable. Indeed, we have never argued throughout this paper that they must be exclusively selected. That is, \emph{ClueNetwork} offers users who want to update SOTA SNC processes opportunities to embed technologically sound components into it. This high composability is an \emph{intentional design} aimed at sustaining \emph{ClueNetwork} as a framework for process optimization rather than a singular process.

\section{Conclusions}\label{sec:conclusions}

\medskip

This paper has reviewed the theoretical foundation of Semantic Networks (SNs) as clues and proposed \emph{ClueNetwork} that facilitates pinpointing an optimal SN among candidate SNs for a textual dataset. To accomplish both goals, the eight Research Questions (RQs) were formulated in Section \ref{sec:introduction}, and this paper has sequentially focused on them.

On \textbf{RQ1} (``How do SNs as clues represent knowledge?''), Subsection \ref{subsec:surrogates and clues} has clarified that such SNs represent knowledge by implying data producers' beliefs as \emph{non--propositional} triads. The definition of an SN as a clue has also been presented as a system of such triads.

On \textbf{RQ2} (``When are SNs as clues, rather than typical Knowledge Graphs (KGs), recommended to be built?''), Subsection \ref{subsec:surrogates and clues} has clarified that while KGs are recommended for works requiring verification--oriented Knowledge Representation (KR), SNs are recommended for works requiring hypothesis--generation--oriented KR. It has also been clarified that, unlike KGs, qualities of SNs are decoupled from the veracity of derived hypotheses concerning what is implied.

On \textbf{RQ3} (``What are the definitions of the SN Construction (SNC) stages and their objectives?''), Subsubsections \ref{subsubsec:what is ake, and why does it matter}--\ref{subsubsec:why does cd matter, what is it} have defined Automatic Keyphrase Extraction (AKE), Edge Weighting (EW), and Community Detection (CD) as shown \hyperlink{definition of ake}{here (link)}, \hyperlink{definition of ew}{here (link)}, and \hyperlink{definition of cd}{here (link)}, respectively. Then, Subsubsections \ref{subsubsec:what is ake, and why does it matter} and \ref{subsubsec:why does cd matter, what is it} have defined the objectives of AKE and CD as shown \hyperlink{definition of max keyness}{here (link)} and \hyperlink{definition of max distinctiveness}{here (link)}, respectively. The objective of EW has also been defined in Subsubsection \ref{subsubsec:robustness improvement} as shown \hyperlink{definition of max interpretability}{here (link)}.

On \textbf{RQs} \textbf{4} (``Which stage--specific methods have been selected for the PoC of \emph{ClueNetwork}?'') and \textbf{6} (``Which stage--specific selected methods yield high performance?''), Subsubsections \ref{subsubsec:how does AKE work}--\ref{subsubsec:how does cd work} have briefed the philosophies of the selected local methods, and Subsubsections \ref{subsubsec:results of ake}--\ref{subsubsec:results of cd} have reported their illustrative evaluation results, respectively.

Before that, on \textbf{RQ5} (``How are the local evaluation criteria defined and justified?''), Subsection \ref{subsec:local evaluation criteria} has ensured the clear definitions of $h\text{F}_{1}$ (Table \ref{table:Table 6}), $\text{RI}$ (\eqref{Eq 7}), and Modularity $Q$ (\eqref{Eq 4}). Their justifications have also been presented as shown \hyperlink{justification of hf1}{here (link)}, \hyperlink{justification of ri}{here (link)}, and \hyperlink{justification of q}{here (link)}.

On \textbf{RQ7} (``How are local evaluation results across the SNC stages integrated to identify the optimal SN?''), Subsection \ref{subsec:problem reformulation} has first reformulated SNC as a Process Optimization Problem (POP) and presented its objective function $J$ that integrates local evaluation results across AKE, EW, and CD.

Then, Subsection \ref{subsec:cluenetwork} has presented the reusable framework \emph{ClueNetwork} that incorporates not only $J$ but also the mechanisms of percolation--theory--based evaluation and Matthews--Correlation--Coefficient--(MCC)--based minimal veracity pretest for EW measures. This lightweight framework not only prevents false edges but also facilitates identifying optimal SNs for documents, provided that researchers only prepare much more accessible gold keyphrases without gold edge weights and gold community partitions.

On \textbf{RQ8} (``How is \emph{ClueNetwork}'s objective function $J$ defined and justified?''), Subsection \ref{subsec:justification of j} has first clarified that $J$ is ultimately the joint subjective probability representing a putative peer reviewer's belief that keyness, interpretability, and distinctiveness of a given SN have jointly been achieved.

Then, the subsection has justified the selected expression of $J$ in two ways. First, it has theoretically explained that the expression aligns with the concept of subjective probability in Bayesian statistics. Second, through the perturbation analysis, the subsection has empirically demonstrated that the expression reacts more strongly against untruth, thereby moving more directly toward truth than the alternatives do.

Consequently, by addressing \textbf{RQs} \textbf{1} to \textbf{4}, this paper has confirmed the scientific legitimacy of SNs as clues and reviewed the key concepts and methods associated with the fundamental stages for constructing them. Thereby, their theoretical foundation has been clarified. Furthermore, by addressing \textbf{RQs} \textbf{5} to \textbf{8}, this paper has demonstrated that optimizing SNC is feasible and presented \emph{ClueNetwork} for it.

\appendices

\section{S. Korean One--Way Traffic Sign}\label{Appendix A}

\medskip

Non--propositionally structured data can manifest anywhere. For example, the one--way traffic sign in the Republic of Korea (South Korea) is structured as a combination of an up--arrow ($\uparrow$) and the word \texttt{일방통행} (one--way):

\begin{figure}[h!]
    \centering
    \includegraphics[width=0.5\linewidth]{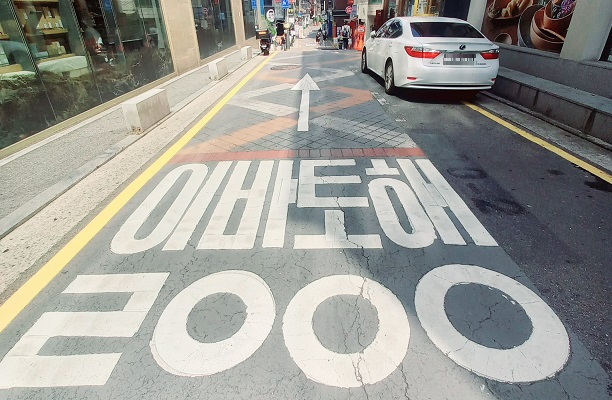}
    \caption{\textbf{One--way Traffic Sign in S. Korea.}}
    \label{Figure 99}
\end{figure}

\noindent This sign is certainly information, yet not a proposition. On public roads, where split--second judgments determine life or death, there is no time to convert the sign into a normative proposition. Nonetheless, the sign can manifest even \emph{knowledge}, as qualified Korean drivers share a \emph{belief} that they \textbf{MUST NOT} drive in the opposite direction.

In contrast, for those who have not yet embodied the belief, the sign remains \emph{some information}, yet not knowledge. Some foreigners unfamiliar with Korean have interpreted \texttt{일방통행} as \texttt{이바토해 2000}, as the final consonants \texttt{ㄹㅇㅇㅇ} resemble \texttt{2000} to them (\href{https://www.reddit.com/r/korea/comments/1ntn4c5/foreigners_are_obsessed_with_this_korean_street/}{link}). SNs as clues are also most useful when interpreted by \emph{experts} who can derive knowledge from such clues. So, Section \ref{sec:introduction} states that ``the lower bound of the expectation of SNC is the extraction of meaningful information, and the upper bound is knowledge representation.''

\section{Semantic Relatedness/Similarity}\label{Appendix B}

\medskip

Humans can perceive two words as semantically related if there exists at least one semantic or mere lexical relation between them \cite{b92, b93}. Such relations are diverse, for instance, antonymy, hypernymy \& hyponymy, IS--A, meronymy \& holonymy, troponymy, derivation, entailment, and others \cite{b95, b96}. Criteria for distinguishing semantic similarity from semantic relatedness are presented in Table \ref{table:Table 19}.

\begin{table}[h!]
\caption{Possible Criteria for Determining Whether Similarity beyond Relatedness Exists between Two Words}\label{table:Table 19}
\centering
\begin{tabular}{>{\arraybackslash}m{1.6cm}|>
{\arraybackslash}m{6.0cm}}
Criterion & Description \\
\hline\hline
Sharing many properties & Whether two words share many properties \cite{b91, b92}, \cite{b96}. \emph{Kangaroo} and \emph{wallaroo} are similar because both live in Australia, have pouches, move by hopping, can even attack innocent zoologists, etc. In contrast, \emph{kangaroo} and \emph{Australia} are related but not similar. \\ \hline
Substitution & Whether two words can be substituted for each other ``without changing the underlying semantics.'' \cite{b95} Given that \emph{mad} can be substituted with \emph{crazy}, they are similar. While \emph{free} belongs to the same ``semantic field'' \cite{b94} (i.e., possible mental states of graduate students), it is related to but not similar to the former. \\ \hline
Presence of an IS--A relation & \cite{b92} considers two words similar if a direct or indirect IS--A relation exists between them. Given that \emph{pineapple on pizza} and \emph{Americano coffee} can be unforgivable sins to many Italians, they are similar. In such a case, the direct relation between two words does not have to be IS--A. \cite{b93} lists possible relations as troponymy, synonymy, hypernymy \& hyponymy, or antonymy. \\ \hline
\end{tabular}
\end{table}

\section{Why undirected SNs are popular?}\label{Appendix C}

\medskip

When a directed edge from ``Simpson'' to ``family'' is derived by using CF, it can be inferred that they frequently co--occur as a collocation. However, when the edge is derived by using Cosine, the directional meaning becomes elusive. Moreover, directed edges between multigrams are scarcer than those between unigrams. Furthermore, even if multigrams can be decomposed, it remains unclear how their original scores should be distributed among their constituent unigrams.

\begin{table*}[b!]
\caption{Overview of Representative Network Types}\label{table:Table 20}
\centering
\begin{tabular}{>{\arraybackslash}m{1.6cm}|>{\arraybackslash}m{4.4cm}|>
{\arraybackslash}m{3.7cm}|>
{\arraybackslash}m{3.1cm}|>
{\arraybackslash}m{2.7cm}}
Type & Intuition & Key Features & Base Degree Distribution & Base Model \\
\hline\hline
Random & ``\emph{The world, ruled by uncertainty...}'' & Low clustering and logarithmic scaling of average path length & Single scale (Poisson) & Erdős--Rényi \cite{b136}
\\ \hline
Small--world & ``\emph{My friend's friend is likely my friend!}'' & High clustering and \newline short average path length & Single scale (Exponential) & Watts--Strogatz \cite{b137}
\\ \hline
Scale--free & ``\emph{The richer get richer.}'' & Dominant hubs and ultra--short average path length & Scale--free (Power Law) & Barabási--Albert \cite{b138}
\\ \hline
Generalized
random & ``\emph{Preserve a given degree distribution, but rewire edges at random.}'' & Preserved degree distribution & User--defined & Configuration \cite{b135}
\\ \hline
\end{tabular}
\end{table*}

\section{Loss of Discriminative Power}\label{Appendix D}

\medskip

According to \cite{b109}, when most entries of vectors are populated wiht meaningless zeros, the zero vector loses its discriminative power; Minkowski then becomes unstable even under small noise; as the dimensionality or the norm parameter $p$ increases, the noise scaling effect is amplified; for a vector, the distance difference between its farthest neighbor and its closest neighbor cannot increase as rapidly as the expected distance between any vector and its closest neighbor; consequently, such differences lose their discriminative power, rendering distances between vectors meaningless and unstable.

In contrast, given that Cosine is driven by the dot product of two keyphrases and the discrete measures focus on the set relations between the two, these measures remain free from the problematic $|t_{i,k}-t_{j,k}|^{p}$. Indeed, for applications such as clustering in high--dimensional sparse spaces, Cosine and Jaccard yield superior performances to even Euclidean \cite{b110}.

\section{Analytical Approximations for $f_{c}$}\label{Appendix E}

\medskip

Tracing back the idea that $f_{c}$ can be purely mathematically derived, there are Molloy and Reed \cite{b135} who proved that any random network with $\langle k^{2} \rangle / \langle k \rangle > 2$ has a $\mathbf{G}_{\mathbf{GC}}$. Herein, $\langle k \rangle$ and $\langle k^{2} \rangle$ are the first and second moments of the degree distribution of the $\mathbf{G}$. Cohen et al. \cite{b132} derived a closed--form expression for $f_{c}$ from the Molloy--Reed criterion:

\begin{equation} \label{Eq 13}
\begin{aligned}
f_{c}^{\text{Cohen}}=1-\frac{1}{\bigg(\frac{\langle k^{2} \rangle}{\langle k \rangle}-1\bigg)}
\end{aligned}
\end{equation}

\medskip

In fact, numerous networks in the world deviate from purely random networks (Table \ref{table:Table 20}), and the Molloy--Reed criterion does not account for triangles in empirical networks. Hence, under the intuition that percolation is hindered as $T_{\triangle}$ increases, Berchenko et al. \cite{b139} proposed the heuristic criterion $(1-T_{\triangle})(\langle k^{2} \rangle/\langle k \rangle - 1) > 1$ for the existence of $\mathbf{G}_{\mathbf{GC}}$ by incorporating the correction term $(1-T_{\triangle})$ into the Molloy--Reed criterion. Thereby, a new closed--form expression for $f_{c}$ \cite{b140} was also derived from Berchenko's heuristic:

\begin{equation} \label{Eq 14}
\begin{aligned}
f_{c}^{\text{Berchenko}}=1-\frac{1}{\bigg(\frac{\langle k^{2} \rangle}{\langle k \rangle}-1\bigg)(1-T_\triangle)}
\end{aligned}
\end{equation}

\medskip

In this manner, analytical approximations for $f_{c}$ have progressively incorporated additional properties exhibited by empirical networks. Nevertheless, deriving a universal closed--form expression for $f_{c}$ across diverse networks continues to be an open challenge. Should more comprehensive expressions become available, researchers could estimate $f_{c}$ directly from network statistics without repeated simulations.

\section{Preprocessing Types}\label{Appendix F}

\medskip

Please refer to Table \ref{table:Table 21}.

\begin{table}[b!]
\caption{Employed Preprocessing Types}\label{table:Table 21}
\centering
\begin{tabular}{>{\arraybackslash}m{1.1cm}|>
{\arraybackslash}m{6.5cm}}
Type & Description \\
\hline\hline
Common & Gold keyphrases are tokenized by using spaCy (v3.8.11) \cite{b151} with a custom infix finditer. Tokens in each gold keyphrase are stemmed by using the Porter stemmer in NLTK (v3.9.1) \cite{b152}, and the stems are joined with whitespace. Thereby, stemmed gold keyphrases are obtained.
\\ \hline
Type 1 & Each document is tokenized by using spaCy (v3.8.11) with the infix finditer. Then, $1$-- to $n$--grams are identified, and irrelevant $n$--grams are filtered out by using the \emph{candidate\_filtering} function in PKE \cite{b150}. Additional filtering can be applied by certain algorithms. For each $n$--gram, its surface forms, offsets, and other details are stored in a \emph{container} by using its lexical form (i.e., the stemmed form produced by the Porter stemmer in NLTK (v3.9.1)) as the key. This key is treated as a candidate, although the stored information may also be referenced by certain algorithms.
\\ \hline
Type 2 & Each document is tokenized and POS--tagged by using spaCy (v3.8.11) with the infix finditer. A graph is then built only based on nouns, proper nouns, and adjectives. Here, PositionRank adopts a stricter filter. Using the RegexParser in NLTK (v3.9.1) with the pattern \{<ADJ>*<NOUN|PROPN>+\}, PositionRank first identifies NPs. Then, a graph is built for the NPs by using only nouns, proper nouns, and adjectives.
\\ \hline
Type 3 & It is designed for KeyBERT, which cannot obtain appropriate candidates and their contextual embeddings from $n$--grams or stems. Type 3 exploits a candidate \emph{container} provided by PKE, as in Types $1$ and $2$. Here, each document is first tokenized and POS--tagged by using spaCy (v3.8.11) with the infix finditer. Then, among all surface forms of each NP identified by the RegexParser with the pattern, the one with the earliest offset is treated as a candidate.
\\ \hline
Type 4 & Each document is tokenized and POS--tagged by using StanfordCoreNLP (v3.9.1.1) \cite{b153}. Next, NP chunking is applied to the tokens by the RegexParser with the pattern {<NN.|JJ><NN.*>}, but any NP containing one or more stopwords from NLTK (v3.9.1), special characters, or punctuation is excluded from candidate consideration. The resulting NPs are then lemmatized by using the WordNetLemmatizer in NLTK (v3.9.1). When multiple lemmas share the same stemmed form, only the one with the highest keyness score is treated as a candidate.
\\ \hline
Type 5 & Each document is tokenized, POS--tagged, and NP--chunked by using spaCy (v3.8.11). Next, phrases that are stopwords, begin with trivial elements (e.g., pronouns, particles, or digits), are shorter than two characters, or contain URL terms or email terms are not regarded as candidates. The optional filters of LMRank can also be used. When multiple NPs share the same stemmed form, only the one with the highest keyness score is treated as a candidate.
\\ \hline
\end{tabular}
\end{table}

\section{About Normality Test}\label{Appendix G}

\medskip

When observations are sparse, and when all observations are zero (in which case SciPy \cite{b98} mechanically judges there is no evidence against normality), the K--S test \cite{b155} and the S--W test \cite{b160} are prone to Type II errors, respectively. When magnitudes of observations are extremely small, variance compression also misleads the S--W test to Type II errors.

\section*{Acknowledgment}\label{sec:acknowledgment}

\medskip

Please refer to the following items.

\medskip

\begin{enumerate}
    \item This research was supported by the G-LAMP Program of the National Research Foundation of Korea (NRF) grant funded by the Ministry of Education (No. RS-2025-25441317).
    \item The regression analysis class taught by Prof. ChanWoo Yoo, working in the Division of Advanced Engineering, Korea National Open University, enhanced the first author's statistical foundation.
    \item Yongjae Oh, a Ph.D. candidate in the Department of Physics and Astronomy, Seoul National University, provided insightful advice on percolation theory.
    \item We express our gratitude to the reviewers for thoroughly reading this lengthy (yet inevitably so) paper.
    \item Several figures in this paper were created using Gemini 3.5 Flash \cite{b41} or purchased with commercial licenses. Specifically, Fig. \ref{Figure 3} as a whole, the Einstein icon in Fig. \ref{Figure 2}, and the diamond icon in Fig. \ref{Figure 19} were generated by Gemini 3.5 Flash \cite{b41}. Most of the remaining icons in Fig. \ref{Figure 19} were commercially acquired.
    \item Except for the mentioned figures, we have created all content related to this work. Although Gemini 3.5 Flash \cite{b41}, Claude Sonnet 4.6 \cite{b183}, and GPT--5 \cite{b184} were used for reviewing the content, the final decision to incorporate any suggested improvements remained entirely at our discretion and responsibility.
\end{enumerate}

\begin{IEEEbiography}[{\includegraphics[width=1in,height=1.25in,clip,keepaspectratio]{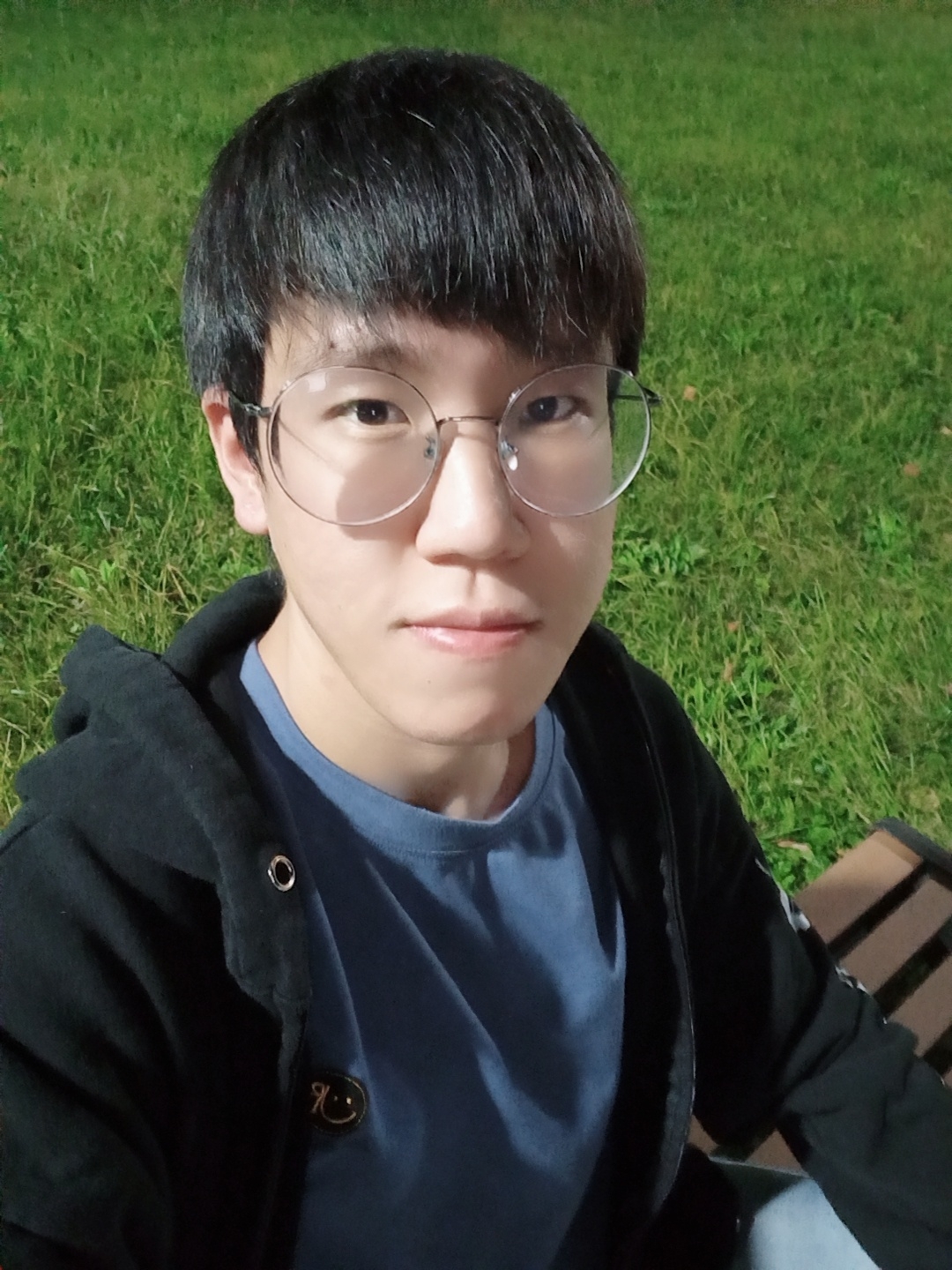}}]{JINWOO HA} will join the Department of Industrial and Information Systems Engineering at Soongsil University, Seoul, Republic of Korea, in September 2026 as an incoming Ph.D. student. He will receive the B.Eng. degree in AI from Korea National Open University in August 2026. Previously, he received the B.A. degree in philosophy with cum laude distinction from Chung--Ang University, Seoul, Republic of Korea, in 2020, and the M.S. degree in IT distribution and logistics from Soongsil University in 2024. His research interests include network science, knowledge engineering, process science, AI, and philosophy of AI.
\end{IEEEbiography}

\begin{IEEEbiography}[{\includegraphics[width=1in,height=1.25in,clip,keepaspectratio]{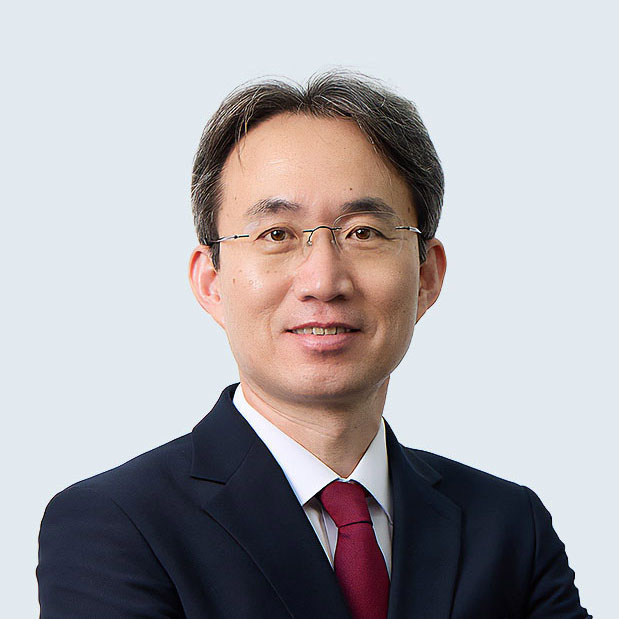}}]{DONGSOO KIM} received the B.S., M.S., and Ph.D. degrees in industrial engineering from Seoul National University, Seoul, Republic of Korea, in 1994, 1996, and 2001.

From 2003 to 2005 and 2005 to 2006, he was an Instructor and then an Assistant Professor at the Graduate School of Healthcare Management and Policy at The Catholic University of Korea. From 2006 to 2009 and from 2009 to 2015, he was an Assistant Professor, and Associate Professor at the Department of Industrial and Information Systems Engineering, Soongsil University, Seoul, Republic of Korea. Since 2015, he has been a Full Professor at the Department of Industrial and Information Systems Engineering, Soongsil University. He serves as an Associate Editor of the \textit{ICIC Express Letters} journal and the Dean of the College of Engineering at Soongsil University. His research interests include business process management, process mining for process optimization, industrial intelligent systems, medical informatics and u--health, convergence technologies of logistics, distribution, and information, and information security management systems.

Prof. Dongsoo Kim was a recipient of the International Conference on Innovative Computing, Information, and Control Contribution Award in 2015 and the International Symposium on Information and Knowledge Management Best Poster Award in 2020.
\end{IEEEbiography}

\EOD

\end{document}